\documentclass[11pt]{article}
\usepackage{amsmath,amssymb}
\usepackage{amsthm,url, booktabs}
\usepackage{multirow}
\usepackage[authoryear]{natbib}
\usepackage{cancel}
\usepackage{graphicx}
\usepackage{color}
\usepackage{caption}
\usepackage{subcaption}

\newtheorem{thm}{Theorem}[section] 
\newtheorem{lemma}{Lemma}[section] 

\newtheorem{prop}{Proposition}[section]
\newtheorem{definition}{Definition}[section]

\newcommand{\bed}{\begin{definition}}
\newcommand{\eed}{\end{definition}}

\newcommand{\eps}{\epsilon}

\newcommand{\bitem}{\begin{itemize}}
\newcommand{\eitem}{\end{itemize}}

\newcommand{\goto}{\rightarrow}

\newcommand{\beqn}{\begin{equation}}
\newcommand{\eeqn}{\end{equation}}
\newcommand{\balign}{\begin{align}}
\newcommand{\ealign}{\end{align}}

\newcommand{\diag}{\mathrm{diag}}

\newcommand{\beq}{\begin{equation}}
\newcommand{\eeq}{\end{equation}}

\allowdisplaybreaks
\usepackage{makecell}

\renewcommand{\baselinestretch}{1}

\begin{document}

\title{Mixed Membership Estimation for Social Networks}
\author{Jiashun Jin$^*$, Zheng Tracy Ke$^\dagger$ and Shengming Luo$^*$ \\\\
Carnegie Mellon University$^*$\\
and Harvard University$^\dagger$} 
\maketitle
\begin{abstract}    
In economics and social science, network data are regularly observed, and a thorough 
understanding of the network community structure facilitates the comprehension of economic patterns and activities. 
Consider an undirected network with $n$ nodes and $K$ communities. 
We model the network using the Degree-Corrected Mixed-Membership (DCMM) model, where for each node $i=1,2, \ldots,n$, there exists a membership vector $\pi_i = (\pi_i(1), \pi_i(2), \ldots, \pi_i(K))'$, where $\pi_i(k)$ is the weight that node $i$ puts in community $k$, $1 \leq k \leq K$.  
In comparison to the well-known stochastic block model (SBM), the DCMM permits both severe degree heterogeneity and mixed memberships, making it considerably more realistic and general. We present an efficient approach, Mixed-SCORE, for estimating the mixed membership vectors of all nodes and the other DCMM parameters. This approach is inspired by the discovery of a delicate simplex structure in the spectral domain. We derive explicit error rates for the Mixed-SCORE algorithm and demonstrate that it is rate-optimal over a broad parameter space. Our findings provide a novel statistical tool for network community analysis, which can be used to understand network formations, extract nodal features, identify unobserved covariates in dyadic regressions, and estimate peer effects. We applied Mixed-SCORE to a political blog network, two trade networks, a co-authorship network, and a citee network, and obtained interpretable results. 

\end{abstract}


 
\def\spacingset#1{\renewcommand{\baselinestretch}%
	{#1}\small\normalsize} \spacingset{1}

\spacingset{1.45}

\addtocontents{toc}{\setcounter{tocdepth}{2}}
\tableofcontents

\section{Introduction}  \label{sec:Intro}
Many economic activities happen on networks. Some examples of economic networks are the international trade networks, high-school friendship networks, stock co-jump networks, and job information networks. 
We denote a network with $n$ nodes by its adjacency matrix $A\in\mathbb{R}^{n\times n}$, with $A_{ij}=1$ if there is an edge between nodes $i$ and $j$ and $A_{ij}=0$ otherwise. 

In network econometrics, there is a surge of interests in understanding the interplay between network topology and economic activities \citep{graham2020network}. The literature can be divided into two categories, {\it formation} and {\it consequence}. Research in {\it formation} treats the network itself as the object of interest and studies the mechanism of forming the network. One popular model is the dyadic regression model, including the famous gravity model for bilateral trade \citep{tinbergen1962shaping} as a special example. In this model, $\mathbb{E}[A_{ij}]$ is a function of the dyadic covariates ${\boldsymbol X}_{ij}$ and nodal covariates ${\boldsymbol Y}_i$ and ${\boldsymbol Y}_j$, and the main goal is estimation and inference of parameters of this function. Another popular model is the strategic model of network formation \citep{jackson1996strategic}. In this model, each node has a utility function $u_i(A)$ that depends on the whole network, so deletion/addition of an edge affects the utilities of all nodes. Given these utility functions $\{u_i\}_{i=1}^n$, the network is in equilibrium if no node wishes to delete an edge and no pair of nodes wish to add an edge. The problems of interest include estimation and inference of these utility functions, e.g., by using network moment statistics \citep{miyauchi2016structural}.
Research in {\it consequence} treats the network as given information and aims to study influence of network structure on economic outcomes. 
There is a line of literature on estimation of the linear-in-means models \citep{manski1993identification, bramoulle2009identification}. In the simplest case of no covariates, let $y_i$ be the response of node $i$ and $d_i$ be the degree of node $i$; the linear-in-means model assumes $y_i=\alpha+\beta \sum_j (d_i^{-1}A_{ij})y_j+\epsilon_i$, with $\epsilon_i$'s being i.i.d. noise. The parameter $\beta$ captures the `peer effect' and is of main interest. 

Independent of the econometric literature, there is also a body of statistical literature on network data analysis, where the main interest is fitting a probabilistic, easy-to-interpret model for an observed network. Pioneered by \cite{bickel2009nonparametric}, the stochastic block model (SBM) has attracted much attention.  SBM assumes that nodes are divided into a few communities, and $\mathbb{E}[A_{ij}]$ is determined by community memberships of two nodes. 
Different from the {\it formation} literature of network econometrics, there are usually no observed covariates and the adjacency matrix $A$ is the only available data. 
Many methods have been proposed for estimating the underlying community structure from $A$.

Recently, the two lines of literature have crossed. There are many interests in applying statistical network models in econometrics. \cite{auerbach2022identification} proposed a joint regression and network formation model, where the goal is learning latent nodal features from the network and using these features in the regression. \cite{chen2020community} used network modeling to estimate the Bernoulli probability matrix $\mathbb{E}[A]$. They replaced $A$ by $\widehat{\mathbb{E}[A]}$ in fitting a network auto-regression model, in hopes of improving the estimation of peer effects. \cite{graham2015methods} combined the dyadic regression in econometrics and the latent space model in statistics to account for both observed and unobserved covariates in network formation.

Unfortunately, despite these encouraging progresses, we note two problems. First,  both the statistical literature and the econometric literature have been largely focused on some classical and idealized network models, such as the stochastic block model (SBM) and the graphon  \citep{lovasz2006limits}.  Second, recent developments in 
statistical network analysis have suggested new ideas in network modeling, 
but such ideas are largely unknown in the area of network econometrics. 
The SBM and graphon models are often too idealized for real networks. Many real networks have the so-called {\it severe degree heterogeneity}, meaning that the degree of one node is 
higher than another by $10$ or even $100$ times \cite[Table 1]{SCORE+}.  
Also, many networks have the so-called {\it mixed-membership}, meaning that different network 
communities overlap with each, and a node may belong to multiple communities \citep{airoldi2009mixed};
for such networks,  the SBM is too idealized, which does not model either mixed-membership or severe degree heterogeneity.  The graphon model is also too idealized.  It does not model severe degree heterogeneity and requires that the nodes are exchangeable (an assumption that is hard to check and is too strong for many real networks).  It is therefore desirable to (a) develop more realistic network models and new algorithms, and (b) introduce the most recent developments in statistical network analysis to the area of network econometrics.

We propose the Degree-Corrected Mixed-Membership (DCMM) model as a more suitable network model.  Compared with SBM, DCMM allows for both severe degree heterogeneity and mixed membership and it is much broader.  Compared with graphon, DCMM accommodates severe degree heterogeneity and does not require node exchangeability.  Since many real networks have strong mixed-membership,  an interesting problem is how to estimate the mixed-memberships of nodes. We propose a fast spectral method, Mixed-SCORE, for estimating network mixed-memberships,  and show that it is rate-optimal in a decision theory framework.  Given the interesting connections between the two areas (statistical network analysis and network econometrics)  we discuss above, our model and method not only provide new contributions to the former but also provide new opportunities to the latter.  
For example, for many existing works in network econometrics that used SBM or graphon 
as the network model, we may improve the results by using the more realistic DCMM model.   Also, our method is useful in several problems of network econometrics. For example,  one can use the output of our method to understand network formation, create nodal features, estimate the Bernoulli probability matrix, and learn the unobserved dyadic covariates.

In what follows, we first present a motivating example. In this example, DCMM has a relatively simple form. We use this example to illustrate why DCMM is a reasonable model and how to use the output of our method to answer real questions of interest.

\spacingset{1}
\begin{figure}[tb]
\centering
\includegraphics[height=.39\textwidth, width=.35\textwidth]{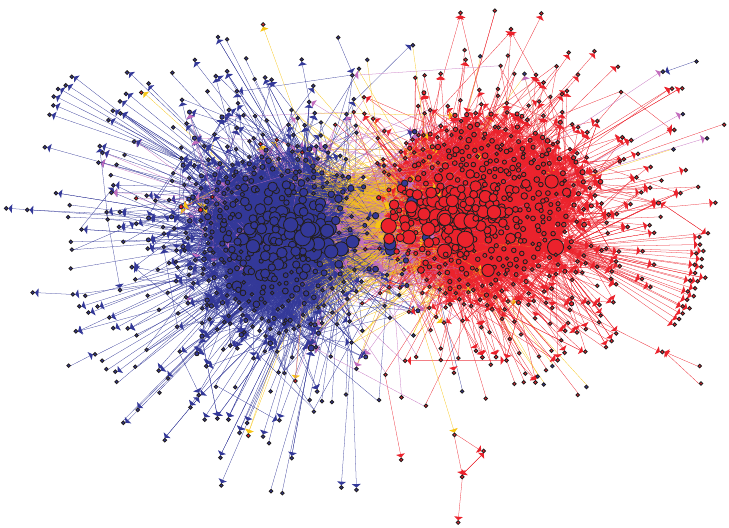} $\quad$
\includegraphics[height=.38\textwidth, width=.5\textwidth]{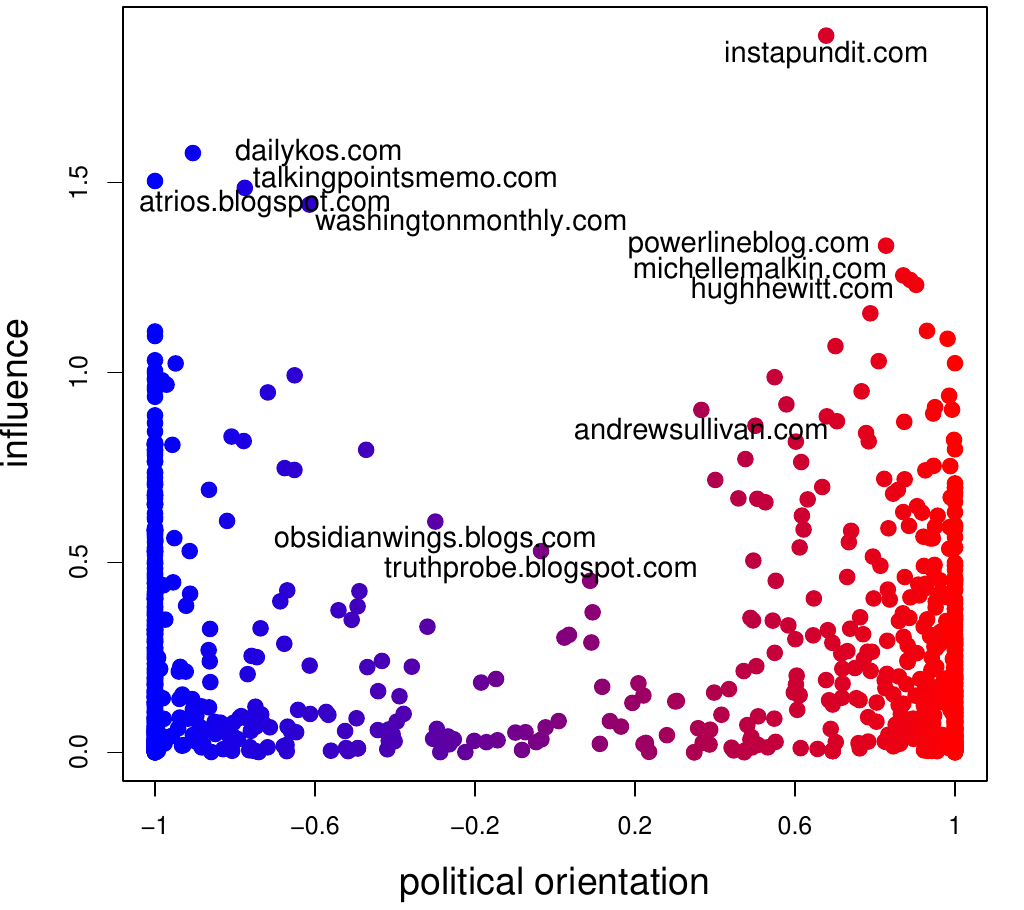}
\caption{The political blog network and the output of Mixed-SCORE. Left: A visualization of the network (figure source: \cite{adamic2005political}), where blue/red colors indicate the manually assigned community labels by \cite{adamic2005political}, and yellow/purple colors indicate the edges between two commmunities. Right: The estimated $p_i$ (x-axis) and $\theta_i$ (y-axis) by the Mixed-SCORE algorithm. }\label{fig:polblogs}
\end{figure}
\spacingset{1.45}

\subsection{A motivating example: Political blog network} \label{subsec:polblog}
The 2004 U.S. Presidential Election was the first presidential election in the United States in which blogging played an important role. 
\cite{adamic2005political} recorded the linkages of political blogs in a single day snapshot before the election. 
We use the data to construct an undirected network, where each node is a blog and two blogs are connected by an edge if they have links between them (one-way or reciprocal). 
The giant component of the network has $n=1222$ nodes. We assume each blog has a political orientation parameter $p_i\in [-1,1]$, 
where $p_i>0$, $p_i=0$ and $p_i<0$ corresponds to liberal, neutral and conservative. A node with $p_i=1$ is extremely conservative, while a node with $p_i=0.2$ is only mildly conservative. 
We also assume each blog has a popularity score $\theta_i>0$. The larger $\theta_i$, the more influence of the blog. 
Suppose the edges are independently generated.  We model the edge probability between two nodes as a function of their political orientations and popularities: 
\spacingset{1}
\footnote{Model \eqref{model-polblog} is not identifiable, as we can multiple $(\alpha,\beta)$ by a scalar $c$ and divide each $\theta_i$ by $\sqrt{c}$ to make the edge probabilities invariant.  
For identifiability, we let $\alpha+\beta=1$. This is the same as the identifiability condition we use for a general DCMM model (see Section~\ref{sec:Methods}). }\spacingset{1.45}
\beq \label{model-polblog}
\mathbb{P}(A_{ij}=1) = \theta_i\theta_j\cdot (\alpha+\beta p_ip_j), \qquad 1\leq i < j\leq n. 
\eeq
Here,  $\alpha>0$ is the baseline effect, and $\beta>0$ captures the effect of political orientations on linkage probabilities. When two blogs are both liberal or both conservative,  $\beta p_ip_j>0$, so they are more likely to be linked. When one blog is liberal and the other is conservative, $\beta p_ip_j<0$, so they are less likely to be linked. The more extreme of political orientations of two nodes, the larger $|\beta p_ip_j|$ and the stronger effect on linkage probability. 
Besides political orientations, the linkage probability is also affected by the popularity of nodes. Suppose two blogs $i$ and $j$ have exactly the same political orientation, but blog $i$ has a larger influence in the internet. It is more likely for other blogs to link to blog $i$ than blog $j$. 

We propose a fast spectral method, Mixed-SCORE, for estimating $(p_i, \theta_i)$ of each node and the global parameters $(\alpha,\beta)$. 
The details of this method will be deferred to Section~\ref{sec:Methods}. 
Figure~\ref{fig:polblogs} plots $(\hat{p}_i,  \hat{\theta}_i)$ of political blogs. The points in the top left regions correspond to {\it influential and liberal} blogs, and those in the top right region are {\it influential and conservative} blogs. Some of these influential blogs are more `extreme' than others in political orientation, such as the liberal blog \texttt{atrios.blogspot.com} and the conservative blog \texttt{hughhewitt.com}. Blogs with large $\hat{\theta}_i$ typically have clear political orientations and are far away from being neutral, with some exceptions like \texttt{truthprobe.blogspot.com}. 

When \cite{adamic2005political} collected this data set, they assigned a manual label $\ell_i\in\{\text{liberal}, \text{conservative}\}$ to each blog $i$ by checking the host website directory or reading blog posts. 
Our method does not need any manual efforts to label the blogs; using the sign of $\hat{p}_i$, we can recover their manual labels with an accuracy of $95.5\%$. 
Meanwhile, people are interested in not only the label of a blog but also the extremity of its political orientation, as an extremely conservative blog and a mildly conservative blog can have different opinions on issues such as abortion, gun control, and death penalties \citep{hindman2003googlearchy}. 
The $\hat{p}_i$'s from our method help reveal such information that is not seen in manual labels. 

We can use the output of Mixed-SCORE in several different ways. First, it is useful for understanding the formation of links between blogs. Our method obtains  
$\hat{\beta}=1-\hat{\alpha}=0.471$. It captures the effect of political orientation on link formation.\spacingset{1}\footnote{We focus on estimation in this paper. In a companion paper \cite{JKL2021}, we also provide a test for testing against the null hypothesis $\beta=0$. The p-value is $<10^{-7}$ for this political blog network, suggesting a significant effect of political orientation on link formation.} \spacingset{1.45}Second, our method creates two covariates, $\hat{p}_i$ (`political orientation') and $\hat{\theta}_i$ (`influence'), for each blog. These covariates will be useful in other tasks such as predicting the opinion of a blogger on a given topic. Third, we obtain $\widehat{\mathbb{E}[A_{ij}]}=\hat{\theta}_i\hat{\theta}_j(\hat{\alpha}+\hat{\beta}\hat{p}_i\hat{p}_j)$,
 which can be plugged into the linear-in-means model to improve the estimation of peer effect. Let $y_i$ be an outcome of interest (e.g., the frequency of a key word in blog posts). We fit a model $y_i=\gamma+\delta  \hat{q}_i^{-1}\sum_j\hat{\theta}_j(\hat{\alpha}+\hat{\beta}\hat{p}_i\hat{p}_j)y_j+\epsilon_i$, where $\hat{q}_i=\sum_{k=1}^n \hat{\theta}_k(\hat{\alpha}+\hat{\beta}\hat{p}_i\hat{p}_k)$. Compared with the standard linear-in-means model, this one better deals with measurement errors on the network itself.

\subsection{Main results and contributions}
Model~\eqref{model-polblog} is a special case of the Degree-Corrected Mixed Membership (DCMM) model to be introduced in Section~\ref{sec:Methods}. In the DCMM model, the network has $K$ perceivable communities. 
Each node has a mixed membership vector $\pi_i\in \mathbb{R}^K$, where $\pi_i(k)\geq 0$ is the weight that node $i$ puts on community $k$, satisfying $\sum_{k=1}^K\pi_i(k)=1$. When $\pi_i$ is degenerate (i.e., $\pi_i$ has only one nonzero entry which is equal to $1$, and the other entries are zero), we call node $i$ a {\it pure node}; otherwise, we call it a {\it mixed node}.  
In Model~\eqref{model-polblog} for the political blog network, $K=2$, $\pi_i=(\frac{1-p_i}{2}, \frac{1+p_i}{2})'$, and a node is pure if and only if $p_i\in \{\pm 1\}$. Each node also has a degree heterogeneity parameter $\theta_i>0$. The probability of forming an edge between nodes $i$ and $j$ is determined jointly by their mixed membership vectors and degree heterogeneity parameters (see Section~\ref{subsec:DCMM}). 
Given the adjacency matrix $A$, we are interested in estimating parameters of DCMM, especially the membership matrix $\Pi:=[\pi_1,\pi_2,\ldots,\pi_n]'$. Estimation of $\Pi$ is known as the problem of mixed membership estimation \citep{airoldi2009mixed}. 

In the statistical literature of network data analysis, many works focused on community detection, which clusters nodes into $K$ non-overlapping communities. Overlapping community detection \citep{COPRA}
allows the assignment of a node to more than one community. It is equivalent to a community detection problem with $2^K$ non-overlapping communities. 
Community detection is a clustering problem, so the methods and theory do not apply to mixed membership estimation. \cite{airoldi2009mixed} is a pioneer work on mixed membership estimation. They considered a special setting of DCMM with $\theta_1=\theta_2=\ldots=\theta_n$ (i.e., no degree heterogeneity) and assumed that $\pi_i$'s  are i.i.d. generated from a Dirichlet prior. They proposed a variational Bayes approach to computing the posterior of $\pi_1,\ldots,\pi_n$. 
However, in many real networks, degree heterogeneity is severe \citep{newman2003structure}, so we must assume unequal $\theta_i$'s. 
\cite{JiZhuMM} proposed the OCCAM algorithm for mixed membership estimation. OCCAM has the nice property of accommodating degree heterogeneity, but it requires a condition that the fraction of mixed nodes must be properly small, and so it does not work for networks with a large fraction of mixed nodes. 

We propose a new method Mixed-SCORE for network mixed membership estimation. 
It is inspired by our discovery of a low-dimensional simplex geometry associated with the leading eigenvectors of $A$.
Using linear algebra, we establish an explicit connection between this simplex and the target quantity $\Pi$. It leads to a fast spectral algorithm for estimating $\Pi$. 
Compared with the existing methods of mixed membership estimation \citep{airoldi2009mixed,JiZhuMM}, 
Mixed-SCORE successfully deals with degree heterogeneity and allows for an arbitrary fraction of mixed nodes. 
Furthermore, we also give a characterization of the error rate of Mixed-SCORE and show that it is rate-optimal for a wide range of settings. In comparison, the competitors either have no theoretical guarantees \citep{airoldi2009mixed} or have non-optimal error rates \citep{JiZhuMM}. 
Given $\hat{\Pi}$ from Mixed-SCORE, we also propose estimates of other parameters of DCMM. 


\subsection{Applications in network econometrics}\label{subsec:3EconApplications}

We give a few examples of using our model and method in network econometrics.

{\bf Example 1}: Economic outcomes are often affected by social influence. For example, a high school student's academic performance might depend on the attitudes and expectations of his/her friends and family. Such a social influence is not directly observed, and a popular solution is to collect network data and hope the unobserved social influence is revealed by linking behavior in the network (e.g., students with similar reported friendships may have similar family expectations \citep{auerbach2022identification}). 
Let $y_i\in\mathbb{R}$ be the outcome (e.g., academic performance of a student) and $X_i\in\mathbb{R}^p$ the observed features (e.g., school rating,  family income, etc.). Consider an unobserved social influence such as the family expectation. We assume there are $K$ extreme types of family expectation and the family expectation of a student is represented by a mixed membership vector $\pi_i\in\mathbb{R}^K$.  
We model the network by DCMM and the outcome by a regression $y_i=X_i'\beta + f(\pi_i)+\epsilon_i$. This model is similar to the model in \cite{auerbach2022identification}, except that he models the network by graphon but we model it by DCMM. We can apply Mixed-SCORE to obtain $\hat{\pi}_i$ and plug them into the regression. 
Compared with the method in \cite{auerbach2022identification}, our approach has some advantages: First, we allow the social feature $\pi_i$ to have an arbitrary dimension $K$, but in a graphon, $\pi_i$ is a scalar in $[0,1]$. 
Second, our approach deals with severe degree heterogeneity and guarantees that the estimated social feature is not biased by the student's own friendship popularity.

{\bf Example 2}: Understanding the social interactions or `peer effects' in decision making is of great interest in economics. To estimate the peer effect, we propose a new linear-in-means model based on DCMM: Given a network generated from DCMM, let $y=(y_1,y_2,\ldots,y_n)'$ store the response at each node and $X=[X_1,X_2,\ldots,X_n]'\in\mathbb{R}^{n\times p}$ store the feature vectors. Define $G\in\mathbb{R}^{n\times n}$ by $G_{ij}=\pi_i'\pi_j/(\sum_{k:k\neq i}\pi_i'\pi_k)$, for $i\neq j$, and $G_{ii}=0$. For some parameters $\alpha,\beta\in\mathbb{R}$ and $\gamma,\delta\in\mathbb{R}^p$, we model that $y=\alpha{\bf 1}_n + \beta Gy + X\gamma + GX\delta + \epsilon$, where $\epsilon$ is the noise vector.  
This model differs from the standard linear-in-means model \citep{manski1993identification} in the definition of $G$. In the standard form, $G$ is chosen as the normalized adjacency matrix.  
However, the adjacency matrix itself has stochastic errors. For example, two friends in real life may or may not be each other's Facebook friend. Our $G$ allows for a possibly nonzero peer effect between two nodes even when they are not directly connected by an edge. Under this model, we can apply Mixed-SCORE to obtain $\hat{\pi}_i$'s and then plug them into the model for $y_i$. 
A similar idea has been considered by \cite{chen2020community} for vector autoregression. They model the network with SBM, but we use the more general DCMM model.

{\bf Example 3}: The dyadic regression model \citep{graham2020network} is a popular network model. When there are unobserved covariates, how to make accurate parameter estimation is not fully understood. Inspired by \cite{graham2015methods}, we assume that an unobserved dyadic covariate is a function of unobserved nodal covariates and propose a dyadic regression model with a DCMM-like structure. 
Let $X\in\mathbb{R}^{n\times n}$ be the adjacency matrix of a weighted network (e.g., in the international trade network, $X_{ij}$ is the trade flow from country $i$ to country $j$). 
Suppose $X_{ij}\sim \mathrm{Poisson}(\lambda_{ij})$, with $\ln(\lambda_{ij})=  \sum_{m=1}^{M}\gamma_m\ln(Z_{m,i,j}) + \beta \ln(\pi_i'P\pi_j) + c_i + c_j$. Here $Z_1,\ldots, Z_M$ are the observed dyadic covariates,  $c_i$ is the fixed effect of node $i$, and $U_{ij}:=\pi_i'P\pi_j$ is an unobserved dyadic covariate, with $(\pi_i, P)$ similar to those in DCMM (to be introduced in Section~\ref{subsec:DCMM}). 
This model is connected to the model in  \cite{graham2015methods}: In his model, $U_{ij}=g(\xi_i, \xi_j; \delta_0)$, where $\xi_i\in\mathbb{R}$ is an unobserved nodal covariate and $g(\cdot,\cdot;\delta_0)$ is a symmetric distance function; in our model, the latent covariate $\pi_i$ can take an arbitrary dimension $K$. 
We introduce a practical algorithm in Section~\ref{subsec:data-tradeGoods}: We first construct a network from the residuals of fitting a dyadic regression with only observed covariates; we then apply Mixed-SCORE to obtain $\hat{U}_{ij}=\hat{\pi}_i'\hat{P}\hat{\pi}_j$; last, we plug in $\hat{U}_{ij}$ and re-fit the dyadic regression. 
Although this approach is mainly from a practical perspective, it points out a new direction, that is, using spectral algorithms to learn unobserved covariates. 
Compared with the existing approaches such as Markov Chain Monte Carlo  and triad probit \citep{graham2020network}, the spectral approach is computationally fast and allows for multidimensional $\pi_i$'s.

Since the main focus of this paper is estimation of $\pi_i$, we leave a careful study of these examples to future work. 
One of the key requirements for plugging $\hat{\pi}_i$ into a downstream economic model is that the error on $\hat{\pi}_i$ can be well-controlled. In this paper, we provide not only a method for estimating $\pi_i$ but also the explicit error bounds. In the case that the network is properly dense, the error bound reduces to $\mathbb{E}[n^{-1}\sum_{i=1}^n\|\hat{\pi}_i-\pi_i\|^2]=O(n^{-1}K^3)$, suggesting that the errors on $\hat{\pi}_i$ are negligible for downstream tasks (please see the discussions following Theorem~\ref{thm:main-rev}). 

The remaining of this paper is organized as follows. In Section~\ref{sec:Methods}, we formally introduce our model and method. In Section~\ref{sec:mainthm}, we state the theoretical results. In Sections~\ref{sec:simu}-\ref{sec:realdata}, we present the simulations and real data, respectively.  
We conclude the paper with discussions in Section~\ref{sec:Discu}. The technical proofs are relegated to the online supplementary material. 

\section{A spectral method for network membership estimation} \label{sec:Methods}

\subsection{The DCMM model} \label{subsec:DCMM}
Consider an undirected network with $n$ nodes. 
Suppose the network contains $K$ communities. Each node has a mixed membership vector $\pi_i=(\pi_i(1), \pi_i(2), \ldots,\pi_i(K))'$, where the entries of $\pi_i$ are nonnegative and sum to $1$. We interpret $\pi_i(k)$ as the fractional weight that node $i$ puts on community $k$. If node $i$ puts 100\% weight on community $k$, then $\pi_i(k)=1$ and $\pi_i(\ell)=0$ for all other $\ell\neq k$; we say that $\pi_i$ is degenerate and call node $i$ a pure node of community $k$. 
If node $i$ is not a pure node of any community, we call it a mixed node. Each node also has a degree heterogeneity parameter $\theta_i>0$. Let $P\in\mathbb{R}^{K,K}$ be a symmetric nonnegative matrix.
Recall that $A\in\mathbb{R}^{n\times n}$ is the adjacency matrix of the network. Since we do not allow for self-edges, the diagonal entries of $A$ are all zero. We assume that the upper triangle of $A$ (excluding the diagonal) contains independent Bernoulli variables,  where for any $1 \leq i, j \leq n$ and $i \neq j$,  
\begin{align} \label{def:DCMM} 
\mathbb{P}(A_{ij} = 1)  =  \theta_i \theta_j \times \sum_{k = 1}^K \sum_{\ell = 1}^K \pi_i(k) \pi_j(\ell) P_{k\ell}  
= \theta_i\theta_j\times \pi_i'P\pi_j. 
\end{align}  
Take Model~\eqref{model-polblog} for the political blog network for example. It is a special case with $K=2$, $\pi_i=(\frac{1-p_i}{2}, \frac{1+p_i}{2})'$ and $P$ being a $2\times 2$ matrix whose diagonal entries are equal to $\alpha+\beta$ and the off-diagonal entries are equal to $\alpha-\beta$. 
The parameters in \eqref{def:DCMM} are not identifiable. For identifiability, we assume that the diagonal entries of $P$ are equal to $1$ (see Section~\ref{subsec:proof-identifiability} of the supplementary material for a proof of model identifiability).

We call \eqref{def:DCMM} the degree-corrected mixed membership (DCMM) model. DCMM includes several popular network models as special cases. The stochastic block model (SBM) is a special DCMM where $\theta_i$'s are equal to each other (i.e., no degree heterogeneity) and all $\pi_i$'s are degenerate (i.e., no mixed membership). The MMSBM model \citep{airoldi2009mixed} is a special case with equal $\theta_i$'s (but $\pi_i$'s can be non-degenerate). The DCBM model \citep{DCBM} is a special case where all $\pi_i$'s are degenerate (but $\theta_i$'s can be unequal). DCMM can also be viewed as an equivalence to the OCCAM model \citep{JiZhuMM}, except that $\pi_i$'s are re-normalized by their $\ell^2$-norms in the OCCAM model.

It is convenient to express \eqref{def:DCMM} in a matrix form. Write $\Theta=\diag(\theta_1,\theta_2,\ldots,\theta_n)\in\mathbb{R}^{n,n}$ and  $\Pi=[\pi_1,\pi_2,\ldots,\pi_n]'\in\mathbb{R}^{n,K}$. Introduce an $n\times n$ matrix $\Omega=\Theta\Pi P\Pi'\Theta$. It is seen that $\Omega_{ij}=\theta_i\theta_j\cdot \pi_i'P\pi_j$. By Model \eqref{def:DCMM}, $\mathbb{E}[A_{ij}]=\Omega_{ij}$ for all $1\leq i\neq j\leq n$. It follows that 
\beq \label{model-MatrixForm}
A = \Omega - \diag(\Omega)    + W, \qquad\mbox{with}\quad W:=A-\mathbb{E}[A] \quad \mbox{and}\quad \Omega:=\Theta\Pi P\Pi'\Theta.      
\eeq
We call $\Omega$, $\diag(\Omega)$, and $W$ the ``main signal", ``secondary signal" and ``noise" respectively.

{\bf Remark 1}: DCMM distinguishes from the latent space models \citep{handcock2007model} or graphons \citep{lovasz2006limits,pensky2019dynamic} by not requiring exchangeability of nodes. In DCMM, we have no assumptions saying that $\theta_i$'s and $\pi_i$'s are i.i.d. drawn from some distributions. We treat all of them as unknown parameters. 

{\bf Remark 2}: DCMM has an interesting connection to the dyadic regression model. In DCMM, we can view $\theta_i$ and $\theta_i$ as nodal covariates, and $\pi_i'P\pi_j$ as a dyadic covariate, but a major difference is that these covariates are unobserved.

\subsection{The simplex structure in the spectral domain}

We first consider an oracle case where we observe the ``main signal" matrix $\Omega$ in \eqref{model-MatrixForm}. We would like to construct an estimate of $\Pi$ from $\Omega$. Note that $\Omega$ is a rank-$K$ matrix. For each $1\leq k\leq K$, let $\lambda_k$ be the $k$th largest eigenvalue of $\Omega$ in magnitude, and let $\xi_k\in\mathbb{R}^n$ be the associated eigenvector. Write $\Lambda=\diag(\lambda_1,\ldots,\lambda_K)$ and $\Xi=[\xi_1,\xi_2,\ldots,\xi_K]$. \cite{SCORE} proposed a  normalization of eigenvectors called the SCORE normalization. It constructs a matrix $R\in\mathbb{R}^{n\times (K-1)}$ containing the entry-wise ratios of eigenvectors, where
\begin{equation} \label{defineR}
R(i,k)=\xi_{k+1}(i)/\xi_1(i), \qquad 1\leq i\leq n, \;\;   1\leq k\leq K-1.
\end{equation} 
Let $r_i\in\mathbb{R}^{K-1}$ denote the $i$-th row of $R$. 
Viewing each $r_i$ as a point in the $(K-1)$-dimension 
Euclidean space, there is a simplex structure for the point cloud $\{r_i\}_{1\leq i\leq n}$: \spacingset{1}
\footnote{By definition, the simplex ${\cal S}$ spanned by $v_1,v_2,\ldots,v_K$ is the set of  points $r$ such that $r=\sum_{k=1}^K \beta_kv_k$ for some nonnegative vector $\beta$ with $\|\beta\|_1=1$. If $v_1,v_2,\ldots,v_K$ are affinely independent,  ${\cal S}$ is non-degenerate; and we call $v_1,\ldots,v_K$ the vertices of ${\cal S}$ and $\beta$ the barycentric coordinate vector of $r$.} \spacingset{1.45}

\begin{lemma}[The simplex geometry in $R$]  \label{lem:ideal} 
Consider Model~\eqref{def:DCMM} and assume that $P$ is non-singular, $P(\Pi'\Theta^2\Pi)$ is irreducible, and each community has at least one pure node. The following statements are true: (1) All entries of $\xi_1$ are strictly positive, so that the matrix $R$ in \eqref{defineR} is well-defined. 
(2) There exists a $K$-vertex simplex ${\cal S}\subset\mathbb{R}^{K-1}$, whose vertices are denoted by $v_1,v_2,\ldots,v_K$, such that each $r_i$ is contained in ${\cal S}$ and that $r_i$ falls on one vertex of ${\cal S}$ if and only if node $i$ is a pure node.  
(3) Let $w_i\in\mathbb{R}_+^K$ contain the barycentric coordinates of $r_i$ in ${\cal S}$. 
The vector $w_i$ is connected to $\pi_i$ through the equation $w_i = (\pi_i\circ b_1)/\|\pi_i\circ b_1\|_1$, where  $b_1\in\mathbb{R}^K$ is the vector defined by $b_1(k) = [\lambda_1 + v_k' \diag( \lambda_2, \ldots, \lambda_K) v_k]^{-1/2}$, $\lambda_1,\lambda_2, \ldots,\lambda_K$ are the nonzero eigenvalues of $\Omega$, and $\circ$ denotes the entrywise product between two vectors.
\end{lemma}

We call ${\cal S}$ the {\it Ideal Simplex}. Lemma~\ref{lem:ideal} inspires a method to recover $\Pi$ from $\Omega$. 
Step 1: Obtain $R$ from \eqref{defineR}. Step 2: By the second claim of Lemma~\ref{lem:ideal}, we can retrieve the vertices $v_1,\ldots,v_K$ by computing the convex hull of the point cloud $\{r_i\}_{1\leq i\leq n}$. Step 3: Given the vertices, we obtain the barycentric coordinate vector $w_i$ for each node $i$ (by solving a simple linear equation); we also compute the vector $b_1$ using the definition in Lemma~\ref{lem:ideal}; by the third claim of Lemma~\ref{lem:ideal}, we can recover $\pi_i$ from $w_i\propto \pi_i\circ b_1$ and $\|\pi_i\|_1=1$.

\spacingset{1}
\begin{figure}[t!]
\centering
\includegraphics[width=.325\textwidth]{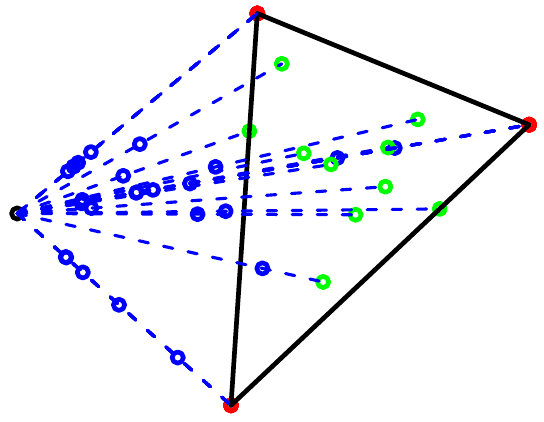}
\includegraphics[width=.29\textwidth]{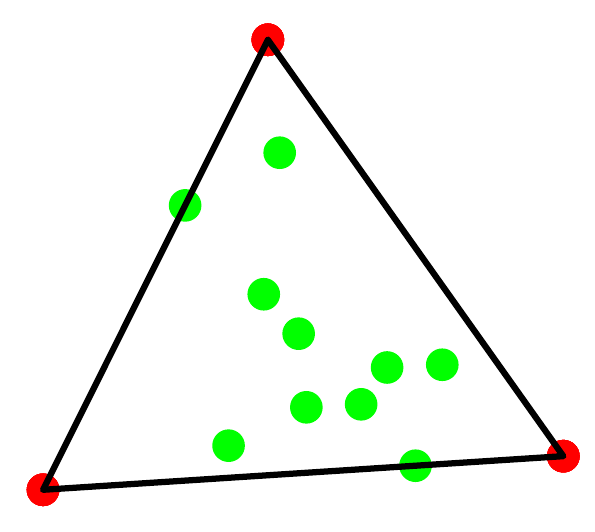}
\includegraphics[width=.29\textwidth]{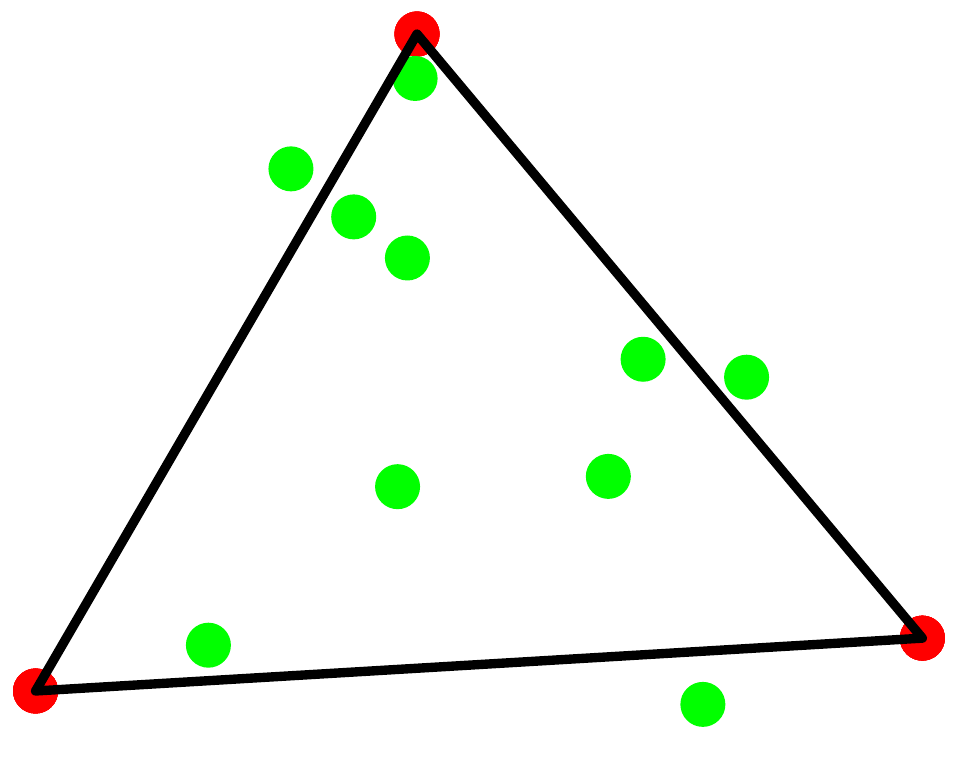} 
\caption{\small Illustration for why the simplex exists and the role of SCORE normalization ($K=3$). Left: rows of $\Xi$ (blue points). The point cloud is contained in a simplicial cone, and it is desirable to normalize the cone to a simplex. Middle: rows of $R$ (red: pure nodes; green: mixed nodes). It shows that the SCORE normalization successfully produces a simplex. Right: rows of $\Xi$ normalized by row-wise $\ell^1$-norm (for visualization, we have projected these points to $\mathbb{R}^2$). This normalization fails to produce a simplex.} \label{fig:WhySimplex}
\end{figure}
\spacingset{1.45}

{\bf Remark 3} {\it (Why the simplex exists and the crucial role of the SCORE normalization)}. 
In the proof of Lemma~\ref{lem:ideal}, we will see that the rows of $\Xi$ are contained in a simplicial cone with $K$ supporting rays, where all the pure nodes in one community are on one supporting ray, and the mixed nodes are in the interior of the cone. The SCORE normalization \eqref{defineR} transforms the simplicial cone to a simplex 
and provides a direct link between the simplex and $\Pi$. 
Interestingly, other normalizations of eigenvectors (e.g., to normalize each row of $\Xi$ by its own $\ell^1$-norm) fail to produce a simplex structure. See Figure \ref{fig:WhySimplex}.

\subsection{The Mixed-SCORE algorithm for estimating $\Pi$} \label{subsec:MixedSCORE} 
We extend the aforementioned method of recovering $\Pi$ to the real case where $A$, instead of $\Omega$, is observed.  For $1\leq k\leq K$, let  $\hat{\lambda}_k$  be the $k$th largest eigenvalue of $A$ in magnitude,  and let $\hat{\xi}_k\in\mathbb{R}^n$ be the associated eigenvectors. Write $\hat{\Lambda}=\diag(\hat{\lambda}_1,\ldots,\hat{\lambda}_K)$ and $\hat{\Xi}=[\hat{\xi}_1,\ldots,\hat{\xi}_K]$. We propose the following algorithm:

\medskip
\noindent 
{\it Mixed-SCORE} algorithm for estimating $\Pi$.  Input: $A, K$. Output: $\hat{\pi}_i$, $1 \leq i \leq n$.   
\begin{itemize} 
\itemsep -.5em 
\item {\it SCORE step}.  Fix a threshold $T > 0$ ($T = \log(n)$ by default).  Obtain $(\hat{\lambda}_1, \hat{\xi}_1), \ldots, (\hat{\lambda}_K, \hat{\xi}_K)$ and define $\hat{R} = [\hat{r}_1, \hat{r}_2, \ldots, \hat{r}_n]'$ as the matrix where for $1 \leq i \leq n$ and $1 \leq k \leq K-1$,   
\begin{equation} \label{defineRhat}
\hat{R}(i,k)= 
  \mathrm{sign}(\hat{\xi}_{k+1}(i)/\hat{\xi}_1(i)) \cdot \min\bigl\{|\hat{\xi}_{k+1}(i)/\hat{\xi}_1(i)|, \; T\bigr\}. \end{equation} 
\item {\it VH (vertex hunting) step}. Use  the rows of $\hat{R}$ to estimate the vertices of Ideal Simplex (details below). Denote the estimated vertices by $\hat{v}_1, \hat{v}_2, \ldots, \hat{v}_K$. 
\item {\it MR (membership reconstruction) step}.  Obtain an estimate of $b_1$ by 
\begin{equation} \label{estimateb1} 
\hat{b}_1(k) = [\hat{\lambda}_1 + \hat{v}_k' \diag(\hat{\lambda}_2, \ldots, \hat{\lambda}_K) \hat{v}_k]^{-1/2}, \qquad 1 \leq k \leq K.  
\end{equation} 
For each $1 \leq i \leq n$, solve $\hat{w}_i\in\mathbb{R}^K$ from the linear equations: $\hat{r}_i = \sum_{k = 1}^K \hat{w}_i(k) \hat{v}_k$, $\sum_{k=1}^K\hat{w}_i(k)=1$. Define a vector  $\hat{\pi}_i^* \in \mathbb{R}^{K}$ by 
$\hat{\pi}_i^*(k) = \max\{0, \hat{w}_i(k)/\hat{b}_1(k)\}$, $1 \leq k \leq K$. Estimate $\pi_i$ by $\hat{\pi}_i  = \hat{\pi}_i^*/\|\hat{\pi}_i^*\|_1$, $1\leq i\leq n$.  
\end{itemize} 

In Step 1, $\hat{R}$ is an estimate of the matrix $R$ in \eqref{defineR}. In Step 3, $\hat{b}_1$ is an estimate of $b_1$ in Lemma~\ref{lem:ideal}. These two steps are similar to those in the oracle case.   
Step 2 is however very different from in the oracle case:  The point cloud $\{\hat{r}_i\}_{1\leq i\leq n}$ is noisy. It is no longer possible to retrieve the vertices of the Ideal Simplex by simply computing the convex hull of these points. We call the estimation of $v_1,v_2,\ldots, v_K$ the vertex hunting (VH) problem. We introduce several VH algorithms. A summary of these algorithms is in Table~\ref{tab:CompSVS}.

The first possible VH approach is to use Successive Projection (SP)  \citep{araujo2001successive}.  
SP is a greedy algorithm. It starts by setting $\hat{v}_1$ as the data point $\hat{r}_i$ that has the largest Euclidean norm among $\hat{r}_1,\hat{r}_2,\ldots,\hat{r}_n$. Then, for $2\leq k\leq K$ successively,  it projects $\hat{r}_i$'s to the orthogonal complement of $\mathrm{Span}(\hat{v}_1,\ldots,\hat{v}_{k-1})$ and finds the data point with the largest Euclidean norm after projection; the estimated $k$th vertex $\hat{v}_k$  is set as the corresponding $\hat{r}_i$.

\spacingset{1}
\begin{figure}[tb] 
\centering
\hspace{-15mm}
\includegraphics[width=.329\textwidth, trim = 40mm 35mm 25mm 30mm, clip=true]{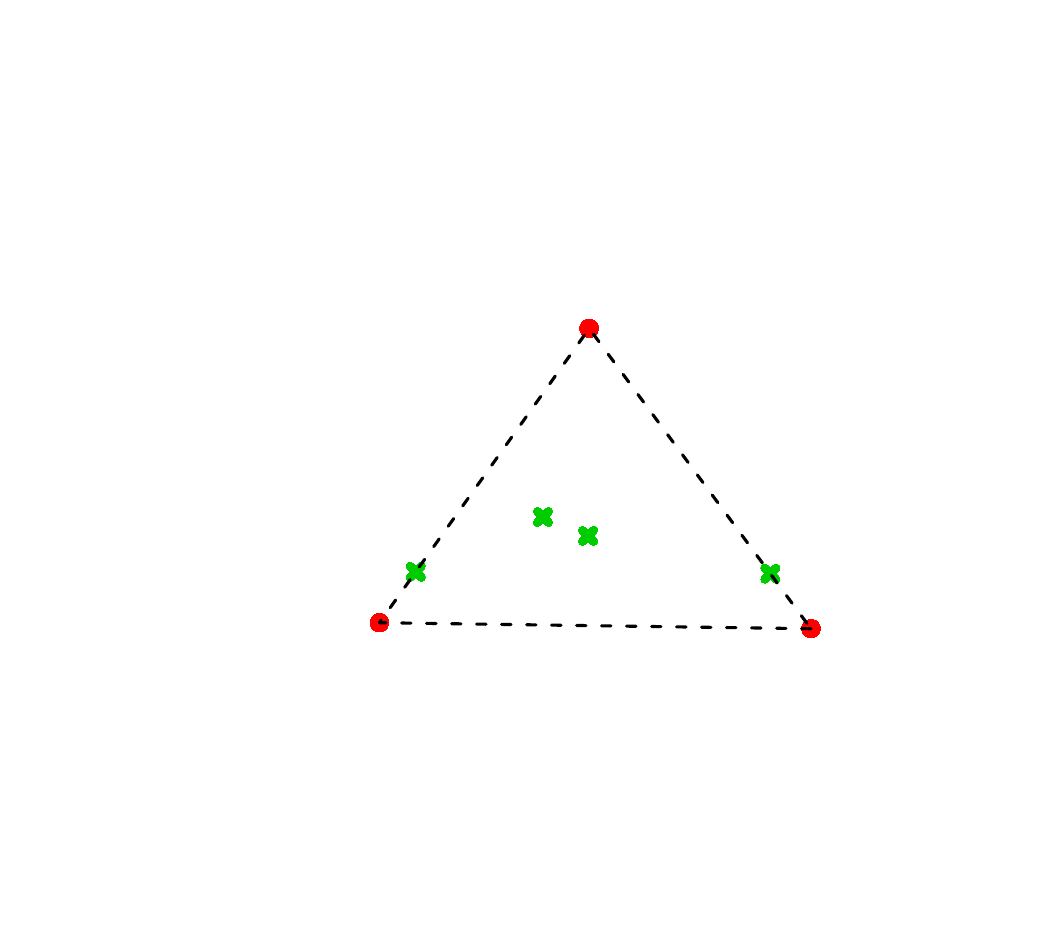}
\includegraphics[width=.329\textwidth, trim = 40mm 35mm 25mm 30mm, clip=true]{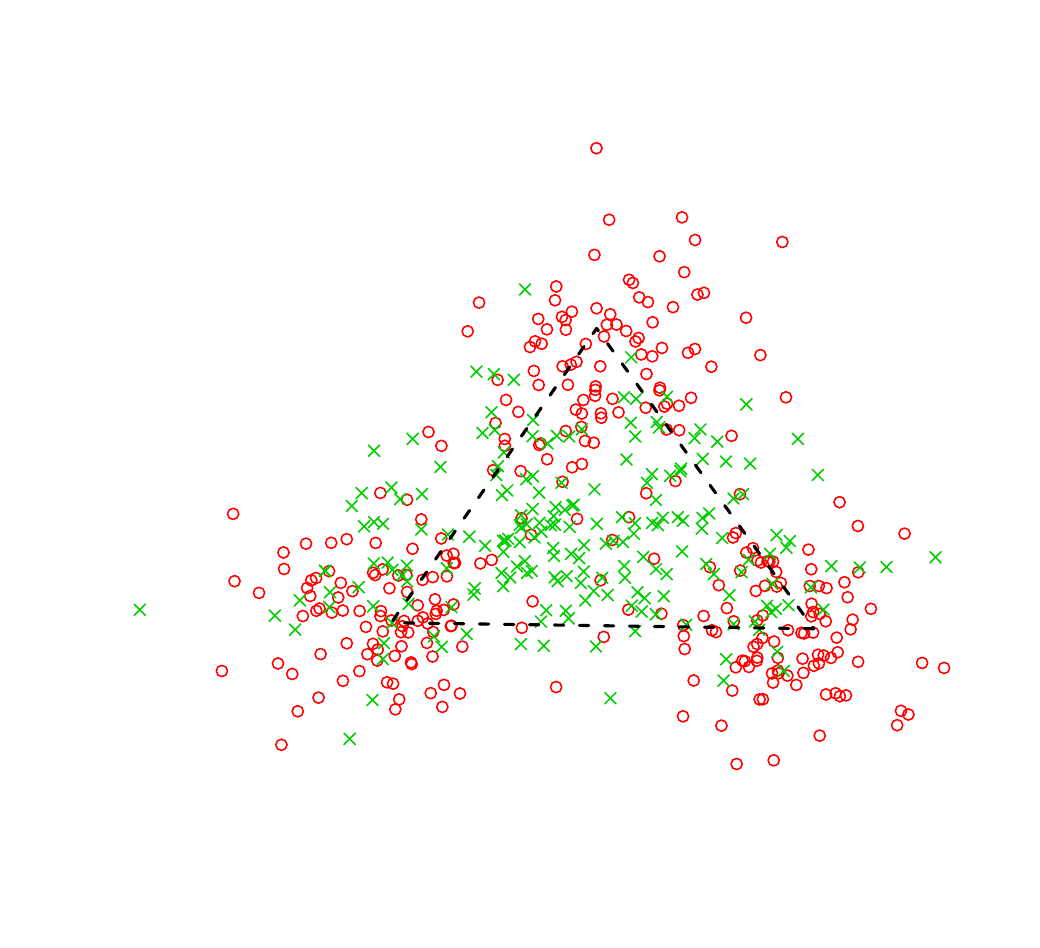}
\includegraphics[width=.329\textwidth, trim = 40mm 35mm 25mm 30mm, clip=true]{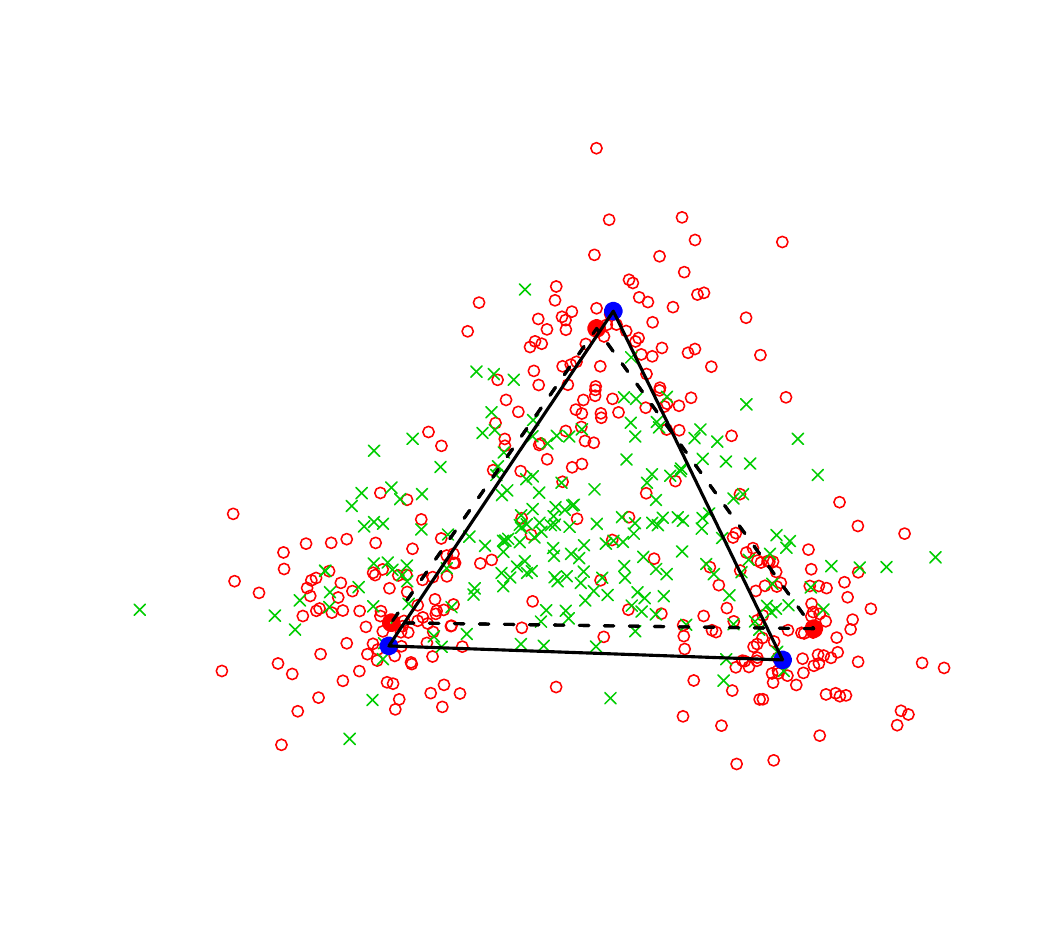}
\caption{\small Left: rows of $R$ (many rows are equal so a point may represent many rows). Middle: each point is a row of  $\hat{R}$ (it is seen that we have strong noise and many outliers, so we may have poor results if we hunt for vertices directly).  
Right: same as the middle panel except that a triangle (solid blue) estimated by  SVS  is added. In all panels, dashed triangle is the Ideal Simplex, and red/green points correspond to pure/mixed nodes respectively. The figure suggests (a) the rows of $\hat{R}$ are quite noisy, with many outliers, and (b) SVS works reasonably well.} \label{fig:mixedmember}
\end{figure}
\spacingset{1.45}

\spacingset{1}
\begin{table}[tb!] 
\centering 
\caption{\small Comparison of four versions of SVS  (for completeness, we analyze all versions theoretically. 
Numerically, we recommend SVS and SVS* for they have better performances).}  \label{tab:CompSVS}
\vspace{2pt}
\scalebox{0.85}{ 
\begin{tabular}{c|c|c}
\hline
& Using exhaustive search in 2nd stage  &   Using SP in 2nd stage  \\
\hline 
$L < n$ & SVS   & SVS* \\ 
$L = n$ & CVS  & SP  \\
\hline
\end{tabular}
} 
\end{table}
\spacingset{1.45}

However, the SP algorithm frequently underperforms numerically.  The Ideal Simplex is highly corrupted by noise and outliers (see Figure~\ref{fig:mixedmember}), but SP is well-known to be sensitive to outliers. To overcome the challenge, we propose {\it Sketched Vertex Search (SVS)}. SVS is a two-stage algorithm. 
In the denoise stage, we cluster $n$ points into $L$ clusters by $k$-means, for a tuning integer $K\ll L\ll n$. The center of each cluster (called a ``local center") is the average of many nearby points and thus robust to outliers. 
In the second stage, we estimate $K$ vertices from these $L$ ``local centers". The full algorithm is as follows:

{\it Sketched Vertex Search (SVS)} for vertex hunting. 
Input: $K$, a tuning integer $L\geq K$, the point cloud $\hat{r}_1,\hat{r}_2, \ldots,\hat{r}_n$.  Output: vertices $\hat{v}_1, \hat{v}_2, \ldots, \hat{v}_K$.  
\begin{itemize} 
\itemsep -.5em 
\item {\it Denoise}. Apply the classical $k$-means algorithm to $\{\hat{r}_i\}_{1\leq i\leq n}$ assuming there are  $L$ clusters.  
Denote the centers of the clusters by  $\hat{m}_1, \hat{m}_2, \ldots, \hat{m}_{L} \in \mathbb{R}^{K-1}$.   
\item {\it Vertex search}.  For any $K$ distinct indices $1 \leq j_1 < \ldots < j_K  \leq   L$,  
let ${\cal H}(\hat{m}_{j_1}, \ldots, \hat{m}_{j_K})$ be the convex hull of $\hat{m}_{j_1}, \ldots, \hat{m}_{j_K}$, and 
\beq \label{d_L}
d_{L}(j_1,\cdots, j_K) = \max_{1\leq j\leq L} \mathrm{distance}\big(\hat{m}_j, \; \mathcal{H}\{ \hat{m}_{j_1},\cdots, \hat{m}_{j_K} \}\big).  \spacingset{1}\footnote{For a point $v$ and a set $H$, $\mathrm{distance}(v, H)$ is the Euclidean distance from $v$ to $H$. When $H$ is a simplex, this distance can be easily computed via a standard quadratic programming.}\spacingset{1.45}
\eeq
Find $1\leq \hat{j}_1   < \hat{j}_2< \ldots < \hat{j}_K\leq L$ that minimizes \eqref{d_L}.   
Output $\hat{v}_k = \hat{m}_{\hat{j}_k}$,  $1 \leq k \leq K$.   
\end{itemize} 
The tuning integer $L$ can be chosen in a data-driven fashion. 
For each $L\in [K+1, 3K]$, let $d_L(\hat{R})=d_L(\hat{j}_1,\cdots,\hat{j}_K)$ be the same as in \eqref{d_L} and 
$\delta_L(\hat{R})  =  \min_{\{j_1,...,j_K\}}\bigl( 
\max_{1\leq k\leq K} \{\|\hat{v}_{j_k}^{(L)}- \hat{v}_k^{(L-1)}\|\}\bigr)$, where the minimum is taken over all permutations of $\{1,2,\ldots,K\}$. 
The quantity $\delta_L(\hat{R})$ tracks the change of estimated vertices when we increase the tuning parameter from $(L-1)$ to $L$. We select $L$ by (if there is a tie, pick the largest integer): 
\beq  \label{chooseL}
\hat{L}^*_n(A) = \mathrm{argmin}_{K+1 \leq L\leq 3K}\{\delta_L(\hat{R})/(1 +  d_L(\hat{R}))\}.  
\eeq
We also consider three variants of SVS.  
The first is SVS*, where in the second stage we apply SP to the $L$ ``local centers".  
The second is {\it Combinatorial Vertex Search (CVS)}, where we take $L = n$ in SVS (i.e., the denoise stage is skipped, so in the second stage, each $\hat{r}_i$ is viewed as a local center). 
In the last variant, we take $L = n$ in SVS*,  so it reduces to SP.    
For practical use, we recommend SVS and SVS*; they have the denoise step by k-means, which is crucial for good numerical performance. 

We view Mixed-SCORE a generic algorithm and treat VH as a ``plug-in" step. For each VH approach, we can plug it in and obtain a different version of Mixed-SCORE. We denote them by  Mixed-SCORE-X, e.g., for  
X $\in\{\mbox{SVS, SVS*, CVS, SP}\}$. Mixed-SCORE can also be used with other possible VH approaches.

The complexity of Mixed-SCORE mainly comes from obtaining the first $K$ eigenvalues and eigenvectors of $A$, which is $O(nK^2)$, and the VH step, which is $O(nK^2)$ if we use the SP algorithm. Hence, 
Mixed-SCORE-SP is a polynomial-time algorithm. 
Mixed-SCORE-SVS is also a polynomial-time algorithm if $(K,L)$ are both finite. 

{\bf Remark 4} {\it (Comparison with the standard PCA)}. The standard PCA approach creates a $K$-dimensional vector  $x_i=\hat{\Xi}'e_i$ for each node $i$. 
These vectors do not have real meanings and are hard to interpret; moreover, each $x_i$ is determined by all the parameters of DCMM and cannot faithfully represent the community structure among nodes.  In comparison, the $\hat{\pi}_i$'s from Mixed-SCORE have clear interpretations.

\subsection{Estimation of $\Theta$ and $P$} \label{subsec:Refitting} 
We are also interested in estimating the other parameters of DCMM. 
Among all the parameters, $\Pi$ is the hardest to estimate. Once $\hat{\Pi}$ is obtained, estimation of $(\Theta,P)$ is comparably easy. 
Therefore, as a byproduct, we use the output of Mixed-SCORE to construct estimates of $(\Theta,P)$. Recall that $\lambda_1,\ldots,\lambda_K$ are the nonzero eigenvalues of $\Omega$ and $\xi_1,\ldots,\xi_K$ are the associated eigenvectors. Let $v_1,v_2,\ldots,v_K$ be the vertices of the Ideal Simplex  and $b_1$ be  as in Lemma~\ref{lem:ideal}. The next lemma is proved in the supplementary material. 
\begin{lemma}  \label{lem:refitting} 
Let $\Lambda=\diag(\lambda_1, \ldots,\lambda_K)$, $V=[v_1,\ldots,v_K]$, and $B=\diag(b_1)[{\bf 1}_K, V']$.  If the conditions of Lemma~\ref{lem:ideal} hold,   then $P=B\Lambda B'$ and $\theta_i = \xi_1(i) / (\pi_i' b_1)$, $1 \leq i \leq n$.    
\end{lemma} 

After running Mixed-SCORE, we collect the following quantities: (i) the leading eigenvector $\hat{\xi}_1$; (ii) the estimated vertices $\hat{V}=[\hat{v}_1,\hat{v}_2,\ldots,\hat{v}_K]$; (iii) a vector $\hat{b}_1$; (v) the estimated mixed membership vectors in $\hat{\Pi}=[\hat{\pi}_1,\hat{\pi}_2,\ldots,\hat{\pi}_n]'$. 
Inspired by Lemma~\ref{lem:refitting}, we let
\beq \label{estimator-P-theta}
\hat{P}=\hat{B}\hat{\Lambda}\hat{B}', \qquad \mbox{and}\qquad \hat{\theta}_i = \hat{\xi}_1(i)/(\hat{\pi}_i'\hat{b}_1), \quad 1\leq i\leq n. 
\eeq

\section{Theoretical properties}  \label{sec:mainthm} 
We state some regularity conditions. Recall that $\theta_i$'s are the degree parameters in Model~\eqref{def:DCMM}. Let $\theta_{\max}=\max_i \theta_i$, $\theta_{\min}=\min_i \theta_i$, $\bar{\theta} = n^{-1}\sum_{i = 1}^n \theta_i$, and $\bar{\theta}_*=\sqrt{n^{-1} \sum_{i = 1}^n \theta^2_i}$. Define  
\beq \label{def:err}
err_n=err_n(\Theta)  =  [(\theta^{3/2}_{\max}\bar{\theta}^{3/2})/(\theta_{\min} \bar{\theta}_*^2)]   \cdot    
\sqrt{\log(n)/(n \bar{\theta}^2)}.   
\eeq

\noindent
{\bf Assumption 1}. $\theta_{\max} \leq C$, and  $err_n \goto 0$. 
 
 Here, the interesting range for $\theta_i$ is from 
$O(n^{-1/2})$ (up to a 
multi-$\log(n)$ term) to $O(1)$, so the first condition is mild.  To appreciate the second condition, note that when $\theta_{\max} \leq C \theta_{\min}$,
$err_n \asymp \sqrt{\log(n)/(n \bar{\theta}^2)}$, where $n\bar{\theta}^2$ is the order of the expected average node degree. Therefore, the condition of $err_n\to 0$ is the same as that the average node degree grows to $\infty$   faster than $\log(n)$, which is mild.   Introduce a $K\times K$ matrix $G=K\|\theta\|^{-2} (\Pi' \Theta^2 \Pi)$. 

\smallskip
\noindent
{\bf Assumption 2}. $\|P\|_{\max}\leq C$, $\|G\|\leq C$, and $\|G^{-1}\|\leq C$. 
\smallskip

The first one is seen to be mild. For the other two conditions, it is instructive to consider a special case where  
all nodes are pure. In this case,   $G = K  \|\theta\|^{-2} \cdot  \diag(\|\theta^{(1)}\|^2,  \ldots, \|\theta^{(K)}\|^2)$, where $\|\theta^{(k)}\|^2 = \sum_{i \in {\cal C}_k} \theta_i^2$. Therefore,  the two conditions reduce to that of $\max_{k}\|\theta^{(k)}\|^2\leq C\min_{k}\|\theta^{(k)}\|^2$, which is only  mild.  
Denote by $\lambda_k(PG)$ the $k$-th largest right eigenvalue of $PG$, and by $\eta_k\in\mathbb{R}^K$ the associated right eigenvector, $1\leq k\leq K$. 

\smallskip
\noindent
{\bf Assumption 3}. $|\lambda_2(PG)|\leq (1-c_1)\lambda_1(PG)$, and $c_1\beta_n\leq |\lambda_K(PG)|\leq |\lambda_2(PG)| \leq c^{-1}_1\beta_n$, where $\beta_n\in (0,1)$ and $c_1\in (0,1)$ is a constant.  
\smallskip 

The first item is a mild eigen-gap condition. In the second item, the quantity $\beta_n$ captures the `distinction' between communities and can be interpreted as the ``signal strength" of the DCMM model, 
where $\beta_n = O(1)$ is the case of ``strong signal" and $\beta_n = o(1)$ is the case of ``weak signal" ($\beta_n$ is a component in the error rate to be introduced).   
We assume $\lambda_2,\ldots,\lambda_K$ are at the same order. This is only for convenience and can be relaxed  (e.g., $\lambda_2,\ldots,\lambda_K$ split into several groups and those in the same group are at the same order).   

\smallskip
\noindent
{\bf Assumption 4}. $\min_{1\leq k\leq K}\eta_1(k)>0$, and $\frac{\max_{1\leq k\leq K}\eta_1(k)}{\min_{1\leq k\leq K}\eta_1(k)}\leq C$.  
\smallskip

In Section~\ref{subsec:condition-check} of  the supplementary material, we show that this assumption is satisfied in either of the following cases: As $n\to\infty$, (a) all entries of $PG$ are lower bounded by a constant, (b) $K$ is fixed and $P$ tends to a fixed irreducible matrix $P_0$, (c) $K$ is fixed and $G$ tends to a fixed irreducible matrix $G_0$, and (d) the maximum and minimum row sums of $P$ are at the same order and $\pi_i$'s are i.i.d. generated from a Dirichlet distribution.

\subsection{Large-deviation bounds for $\hat{R}$}
\label{subsec:hatR} 
The following entry-wise large-deviation bounds for matrix $\hat{R}$ plays a key role in our analysis. Let $\hat{R} = [\hat{r}_1, \hat{r}_2, \ldots, \hat{r}_n]'$ be as in \eqref{defineRhat}. Let $R = [r_1, r_2, \ldots, r_n]'$ be as in (\ref{defineR}). 
\begin{thm}[Large-deviation bounds for $\hat{R}$] \label{thm:r-entrywise-bound}
Consider the DCMM model where Assumptions 1-4 hold. Suppose $\sqrt{K\log(n)}\leq T\leq\infty$ for $T$ in \eqref{defineRhat}.   Let $err_n$ be as in \eqref{def:err} and $\beta_n$ as in Assumption 3.  With probability $1 - o(n^{-3})$, there exists an orthogonal matrix $H \in \mathbb{R}^{K-1, K-1}$ such that 
$\max_{1\leq i\leq n}\|H\hat{r}_i - r_i \| \leq CK^{3/2}\beta_n^{-1}err_n$.  
If, additionally, $\theta_{\max} \leq C \theta_{\min}$, then with probability $1 - o(n^{-3})$,
$\max_{1\leq i\leq n}\|H\hat{r}_i - r_i \| \leq  CK^{3/2}(n\bar{\theta}^2\beta_n^2)^{-1/2}\sqrt{\log(n)}$.   
\end{thm} 

In Theorem~\ref{thm:r-entrywise-bound}, $(K, \beta_n, \bar{\theta})$ may all vary with $n$. Among them, $\beta_n$ captures the ``strength of community signals", where we either have $\beta_n = O(1)$ or $\beta_n \goto 0$ reasonably fast, so the claims applies to both the cases of 
``strong signals" and ``weak signals".  

The proof of Theorem~\ref{thm:r-entrywise-bound} is based on a row-wise large deviation bound for the eigenvectors of the adjacency matrix (Lemma~\ref{lem:Aeigenvecs} in the supplement). 
In the literature, there were few results about row-wise deviation bounds for eigenvectors of a network adjacency matrix \citep{abbe2017entrywise,fan2019simple,fan2020asymptotic}. 
They focused on moderate degree heterogeneity and assumed that the nonzero population eigenvalues are at the same order, so they do not apply to our setting.
We need non-trivial efforts to prove Lemma~\ref{lem:Aeigenvecs} and Theorem~\ref{thm:r-entrywise-bound}.

\subsection{Rates of Mixed-SCORE with a generic but efficient VH step} \label{subsec:thm-generic}
Mixed-SCORE has a plug-in VH step, and the goal of the VH step is to estimate the vertices $v_1, \ldots, v_K$ of the ideal simplex. In this section, we present the rate of Mixed-SCORE for a {\it generic but efficient} VH step. 
Next in Section \ref{subsec:thm-4versions}, we discuss the 
 rate of Mixed-SCORE for all $4$ proposed VH step in Table~\ref{tab:CompSVS} (where the rate can be much faster in some cases).

\medskip
\noindent
{\bf Definition 1} ({\it Efficient VH}).  We call a VH step efficient if it satisfies that $\max_{1\leq k\leq K}\|H\hat{v}_k - v_k\|\leq C \max_{1\leq i\leq n}\|H\hat{r}_i - r_i\|$, where  $H$ is the orthogonal matrix in Theorem~\ref{thm:r-entrywise-bound}. 
\smallskip 
 
For our proposed VH methods in Table~\ref{tab:CompSVS},  CVS and SP are efficient under Assumptions 1-4,  and SVS and SVS* are efficient if some additional conditions hold; see Section \ref{subsec:thm-4versions}.

For any estimate $\hat{\Pi}=[\hat{\pi}_1,\hat{\pi}_2,\ldots,\hat{\pi}_n]'$ for $\Pi$, we measure the error by  
the mean squared error (MSE) $\mathbb{E}[\frac{1}{n} \sum_{i = 1}^n \|\hat{\pi}_i - \pi_i\|^2]$. 
Recall that $err_n$ is defined in (\ref{def:err}).  
\begin{thm}[Error of Mixed-SCORE] \label{thm:main-rev}  
Consider the DCMM model where Assumptions 1-4 hold. Let $\hat{\Pi}$ be the estimate of $\Pi$ by 
Mixed-SCORE with a generic but efficient VH step. 
Then, $\mathbb{E}[\frac{1}{n} \sum_{i = 1}^n \|\hat{\pi}_i - \pi_i\|^2 ] \leq CK^3 \beta_n^{-2}err_n^2+o(n^{-2})$. 
If additionally $\theta_{\max} \leq C \theta_{\min}$, 
then $\mathbb{E}[\frac{1}{n} \sum_{i = 1}^n \|\hat{\pi}_i - \pi_i\|^2]  \leq CK^{3}(n\bar{\theta}^2\beta_n^2)^{-1}\log(n)+o(n^{-2})$. 
\end{thm}

We now discuss the implication of Theorem~\ref{thm:main-rev} on economic applications. For simplicity, we consider a case where $\theta_{\max}\asymp\theta_{\min}$, $K=O(1)$ and $\beta_n\geq C$. By Theorem~\ref{thm:main-rev}, the MSE is $O((n\bar{\theta}^2)^{-1}\log(n))$. For a dense network, $\bar{\theta}\asymp 1$, and the MSE becomes $O(n^{-1}\log(n))$, which is quite negligible. Suppose we have a downstream economic model $y_i=\alpha + \pi_{i(-1)}'\beta+\epsilon_i$, where $y_i$ is an outcome of interest and $\pi_{i(-1)}$ is the sub-vector of $\pi_i$ by dropping the last coordinate (to remove co-linearity). We plug in the $\hat{\pi}_i$'s from Mixed-SCORE and let $\hat{\beta}$ be the least-squares coefficient. It can be shown that $|\hat{\beta}-\beta|^2=O\bigl(n^{-1}\sum_{i=1}^n \|\hat{\pi}_i-\pi_i\|^2\bigr)+O_{\mathbb{P}}(n^{-1})$. Therefore, as long as $n\bar{\theta}^2\gg\log(n)$, we have consistency on $\hat{\beta}$. Furthermore, using the faster rates in Section~\ref{subsec:thm-4versions}, we can further remove the $\log(n)$ factor in MSE; as a result, when the network is dense, we also have root-$n$ consistency of $\hat{\beta}$.

{\bf Remark 5} {\it (Rate optimality)}. \cite{jin2017sharp} derived a  minimax lower bound  for the case where $K$ is finite and that $\theta_i$'s are equal. They showed that for any estimate $\hat{\Pi}$, there is a constant $c_0 > 0$ such that $
\frac{1}{n} 
\sum_{i = 1}^n \|\hat{\pi}_i - \pi_i\|^2  \geq C  / (n \bar{\theta}^2 \beta_n^2)$ with probability $\geq c_0$. 
Comparing it with Theorem~\ref{thm:main-rev}, the error rate of Mixed-SCORE is optimal (up to a $\log(n)$ factor) for DCMM with  $\theta_{\max} \leq C \theta_{\min}$. 

{\bf Remark 6} {\it (Comparison with the rate of the OCCAM algorithm \citep{JiZhuMM})}. 
Since the theory of OCCAM does not allow $\beta_n = o(1)$ or  $K$ diverging with $n$, we compare two methods only in the case that $K \leq C$ and $\beta_n \geq C$.    
The rate of Mixed-SCORE reduces to $(n\bar{\theta}^2)^{-1/2}$, but the rate of OCCAM cannot be faster than $(n\bar{\theta}^2)^{-1/5}$, which is strictly slower. 
Also, OCCAM works only if the fraction of mixed nodes is properly small (hinged in Assumption-B of \cite{JiZhuMM}). For example, when $K=3$, $P=0.9I_3+0.1{\bf 1}_3{\bf 1}_3'$, and $\pi_i=\frac{1}{\sqrt{3}}{\bf 1}_3$ for all mixed nodes, the fraction of mixed nodes has to be $<1/4$.

{\bf Remark 7} {\it (Comparison with theory of community detection)}.  Community detection is a less challenging problem,  where $\pi_i$'s are known to be degenerate. It has exponential rates \citep{gao2016community}, but membership estimation only achieves polynomial rates \citep{jin2017sharp}. 
Consider an example with $K=2$, 
$\pi_i\overset{iid}{\sim}\mathrm{Dirichlet}(\alpha_0)$, and $P(A_{ij}=1) = n^{-1} \pi_i'  P  
\pi_j$, where $P_{km} = a\cdot 1\{k = m\}+b\cdot 1\{ k\neq m\}$.   
As $n\to\infty$, $\alpha_0$ is fixed but $(a,b)$ can depend on $n$.
This is equivalent to a DCMM with $\bar{\theta}\asymp n^{-1/2}\sqrt{a}$ and $\beta_n\asymp (a-b)/a$. Write $I=(a-b)^2/a$. The rate of Mixed-SCORE is $O(I^{-1/2}\sqrt{\log(n)})$, but when $\pi_i$'s are all degenerate, the rate of  community detection is $\exp(-O(I))$. 

Given the results for $\hat{\Pi}$,  we further study the estimates $(\hat{\Theta}, \hat{P})$ defined in Section~\ref{subsec:Refitting}.

\begin{thm}[Estimation of $(\Theta, P)$ in DCMM] \label{thm:other-params}  
Under the conditions of Theorem~\ref{thm:main-rev}, with probability $1-o(n^{-3})$,   $\|\hat{P}-P\|\leq C(K^2+K^{3/2}\beta_n^{-1})err_n$ and $\|\hat{\Theta}-\Theta\|_F^2\leq C\|\theta\|^2 K^3\beta_n^{-2}err_n^2$. 
\end{thm}

\subsection{Rates for Mixed-SCORE with proposed VH steps,  and faster rates} \label{subsec:thm-4versions}
Section~\ref{subsec:thm-generic} analyzes a generic Mixed-SCORE algorithm with an efficient VH step.
In this subsection, we discuss Mixed-SCORE with each specific VH approach in Table~\ref{tab:CompSVS}. First, we consider CVS and SP. 
The following theorem shows that CVS and SP are both efficient, and Mixed-SCORE-CVS and Mixed-SCORE-SP  attain the rate in Theorem~\ref{thm:main-rev}. 

\begin{thm} \label{thm:CVS+SP}  
Consider the DCMM model where Assumptions 1-4 hold and each community has at least one pure node. Let $H$ be the orthogonal matrix in Theorem~\ref{thm:r-entrywise-bound}. If we apply either CVS or SP to rows of $\hat{R}$, then with probability $1-o(n^{-3})$, $\max_{1\leq k\leq K}\|H\hat{v}_k - v_k\|\leq C \max_{1\leq i\leq n}\|H\hat{r}_i - r_i\|$, so both CVS and SP are efficient.  
Moreover, for Mixed-SCORE-CVS or Mixed-SCORE-SP, $\mathbb{E}[\frac{1}{n} \sum_{i = 1}^n \|\hat{\pi}_i - \pi_i\|^2 ] \leq CK^3 \beta_n^{-2}err_n^2+o(n^{-2})$. 
\end{thm}


Next, we consider Mixed-SCORE-SVS and Mixed-SCORE-SVS*. SVS and SVS* use a denoise stage, which provides a significant advantage in numerical performance, but also makes them harder to analyze. For this reason, we only consider two settings. In the first setting, we assume all $\pi_i$'s for mixed nodes are $iid$ drawn from a continuous distribution. In the second setting, $\pi_i$'s form several {\it loose clusters}. Owing to space limit, we only present Setting 1 here. Setting 2 is in Section~\ref{sec:Setting2} of the supplementary material.

\medskip
\noindent
{\bf Setting 1}. Let ${\cal S}_0 = {\cal S}_0(e_1, e_2, \ldots, e_K)$ be the standard simplex in $\mathbb{R}^K$, 
where the vertices $e_1, e_2, \ldots, e_K$ are the standard Euclidean basis vectors of $\mathbb{R}^{K}$.  Fix a density $g$ defined over ${\cal S}_0$. 
Let ${\cal R} = \{\pi \in {\cal S}_0: g(\pi) > 0\}$ be the support of $g$. Suppose there is a constant $c_0 > 0$ such that ${\cal R}$ is an open subset of ${\cal S}_0$, and $\mathrm{distance}(e_k, {\cal R}) \geq c_0$, $1\leq k\leq K$. Let $\delta_{v}(\pi)$ be the point mass at  $\pi = v$. Fixing constants $\eps_1,\ldots,\eps_K>0$ with $\sum_{k = 1}^K \eps_k < 1$,
we invoke a random design model where $\pi_i$'s are $iid$ drawn from $f(\pi) = \sum_{k=1}^K \eps_k \cdot \delta_{e_k}(\pi) + \bigl(1 - \sum_{k = 1}^K \eps_k\bigr) \cdot  g(\pi)$. 
The following is similar to $err_n$ in (\ref{def:err}), and quantifies the ``faster rate" aforementioned. 
\beq \label{err-rate-new}
err^*_n=err^*_n(\Theta)  =  [(\theta_{\max}^{1/2}\bar{\theta}^{3/2})/(\theta_{\min} \bar{\theta}_*)]   \cdot    
(n \bar{\theta}^2)^{-1/2}.   
\eeq

\begin{thm} \label{thm:SVS1}  
Consider the DCMM model where Assumptions 1-4 hold and $\pi_i$'s are as in Setting 1. Let $H$ be as in Theorem~\ref{thm:r-entrywise-bound}. 
There exists a constant $L_0(g,\epsilon_1,\ldots,\epsilon_K)>0$ such that, if we apply SVS or SVS* to rows of $\hat{R}$ with $L\geq L_0$,  then with probability $1-o(n^{-3})$, $\max_{1\leq k\leq K}\|H\hat{v}_k - v_k\|\leq C \bigl(n^{-1}\sum_{i=1}^n\|H\hat{r}_i - r_i\|^2\bigr)^{1/2}$.  
Moreover, for Mixed-SCORE-SVS or Mixed-SCORE-SVS*,  $\mathbb{E}[\frac{1}{n} \sum_{i = 1}^n \|\hat{\pi}_i - \pi_i\|^2 ] \leq CK^3 \beta_n^{-2}(err^*_n)^2+o(n^{-2})$. 
\end{thm}

By Theorem~\ref{thm:SVS1}, the rates of Mixed-SCORE-SVS and Mixed-SCORE-SVS* are faster than those of 
Mixed-SCORE-SP and Mixed-SCORE-CVS. 
In fact, by  \eqref{def:err}-\eqref{err-rate-new}, we have $
err_n^*/ err_n = [\bar{\theta}_*/(\theta_{\max}\sqrt{\log(n)})]$. 
Since $\bar{\theta}_* / \theta_{\max} \leq 1$ and $\bar{\theta}_* / \theta_{\max}$ may tend to $0$ rapidly, 
we have the following observations: 1) The rate here is faster than that of 
Theorem \ref{thm:main-rev} by at least a factor of $\log(n)$.
2) The rate here can be much 
faster than that of Theorem \ref{thm:main-rev}  if  $\bar{\theta}_* / \theta_{\max}\to 0$ rapidly. As an example, suppose   
$\theta_1 =  \ldots =\theta_{n-1} = \alpha_n$ and 
$\theta_n = n^{\gamma} \alpha_n$, where $0 < \gamma < 1/2$ is a constant; in this case,  $err_n^* / err_n = \bar{\theta}_* / (\theta_{max} \sqrt{\log(n)})  \leq n^{-\gamma} / \sqrt{\log(n)}$, and so the rate here is much faster than that of Theorem \ref{thm:main-rev}. 
Once we have a faster rate for $\hat{\Pi}$, we also enjoy a faster 
rate for the proposed $(\hat{\Theta}, \hat{P})$ in 
Section \ref{subsec:Refitting}  
(proof is omitted).

{\bf Remark 8}. The faster rates here are because SVS and SVS* use a denoise stage, which improves the accuracy in vertex hunting and so in membership estimation.  
The improved rate is not due to the more strict setting considered here (in fact, in Setting 1 and Setting 2,  if we use SP and CVS for VH in Mixed-SCORE,  then we do not have a much faster rate). 
For more general settings, Mixed-SCORE-SVS or Mixed-SCORE-SVS* continue to enjoy this faster rate, as supported by numerical experiments in Section~\ref{sec:simu}. 

%
%

\section{Simulations} \label{sec:simu}

\begin{figure}[tb!]
\centering
\includegraphics[width=.3\textwidth]{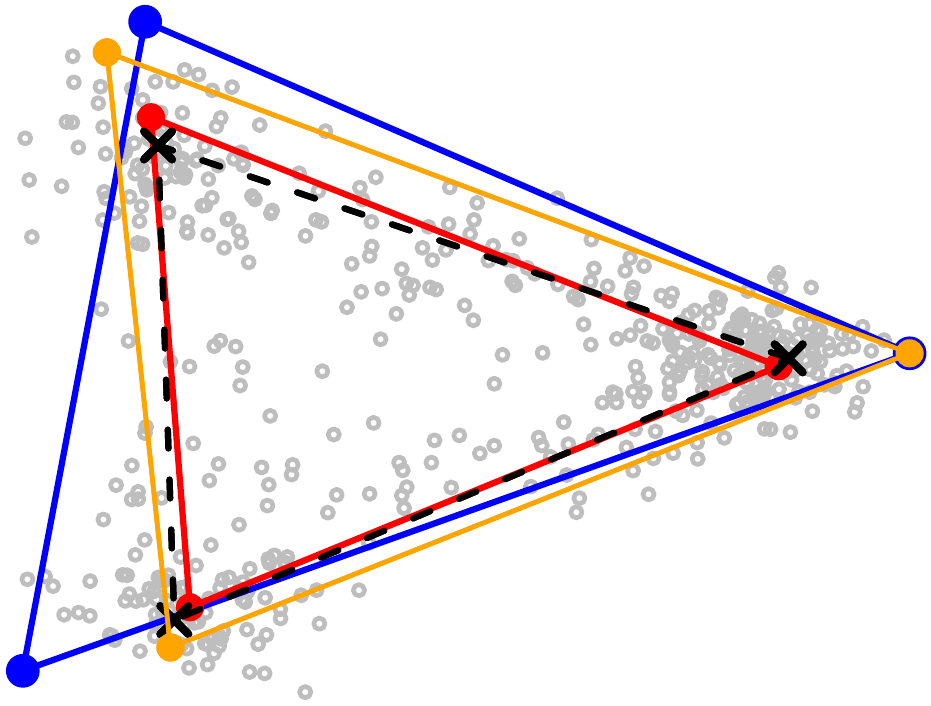}
\includegraphics[width=.3\textwidth]{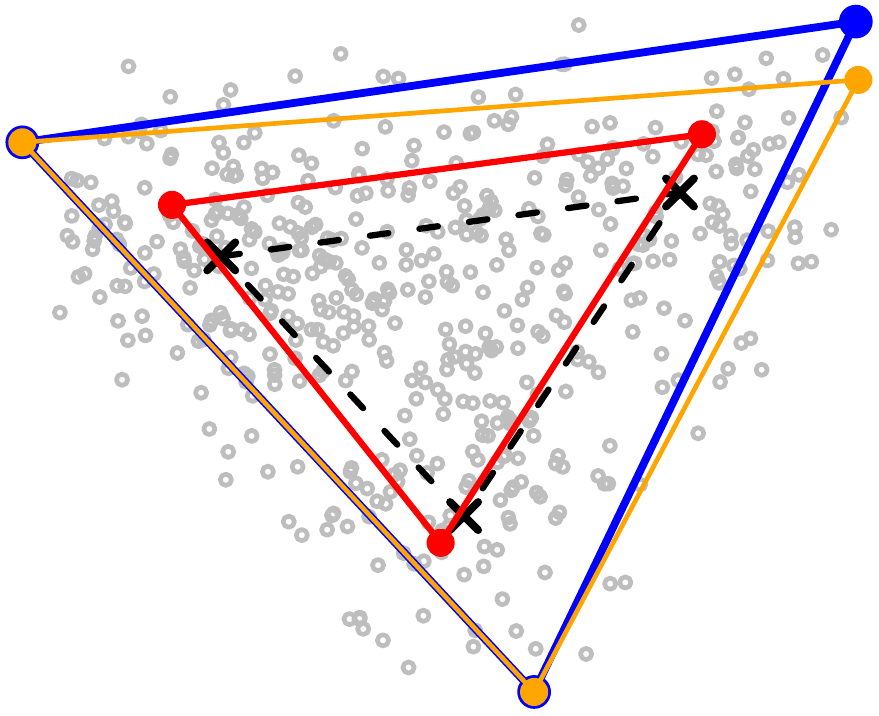} \hspace{1pt}
\includegraphics[width=.29\textwidth]{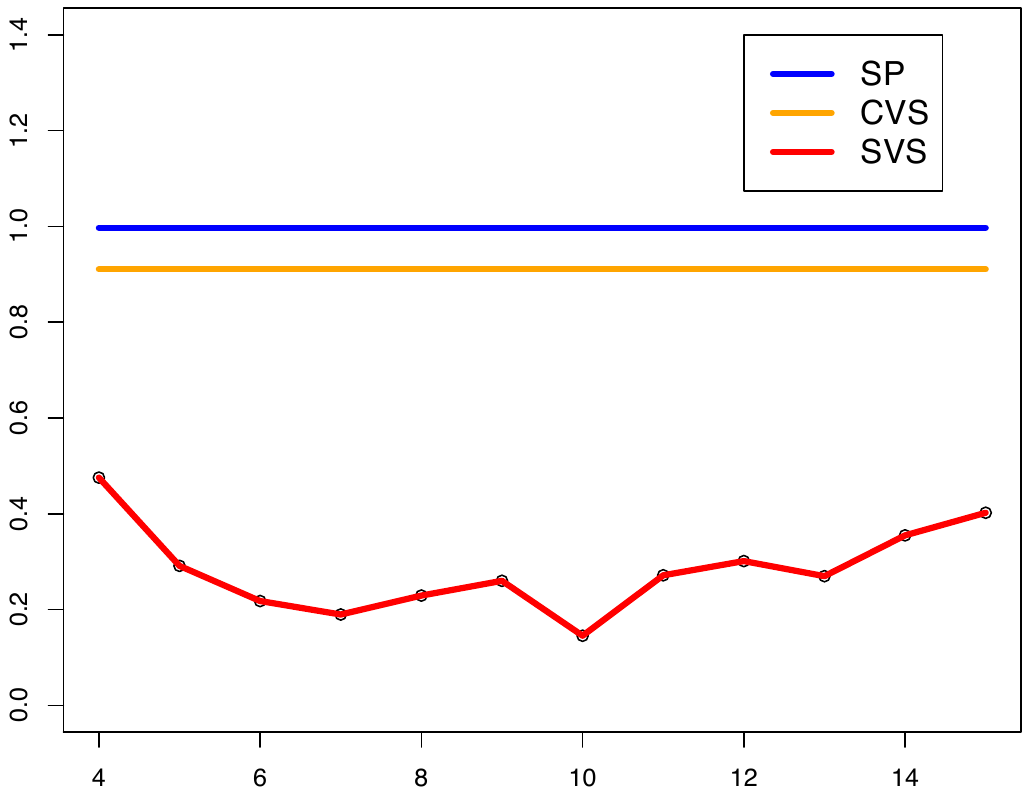}
\caption{\small Comparison of VH methods (black: truth; blue: SP; yellow: CVS; red: SVS). Left: The case of weak noise. CVS and SVS perform well, but SP performs less satisfactorily (possible reason: SP is a  greedy algorithm). Middle: The case of strong noise. SVS performs well, but SP and CVS perform unsatisfactorily. This is because SVS is much less sensitive to outliers. Right: Robustness of SVS to the choice of $L$ (y-axis is $\max_k\|H\hat{v}_k-v_k\|^2$).} \label{fig:VHcompare}
\end{figure}

{\bf Experiment 1} {\it (Comparison of VH approaches)}. We view Mixed-SCORE as a generic algorithm, where we can plug in any VH approach. In Table~\ref{tab:CompSVS}, we list four VH approaches. We now compare SP, CVS and SVS (the performance of SVS* is very similar to SVS, thus omitted). 
Fix $(n,  K) = (500, 3)$. $P$ is a matrix whose diagonals are $1$ and off-diagonals are $0.3$. Each community has $50$ pure nodes. For $\pi_i$'s of the remaining $350$ nodes, half of them are $iid$ drawn from $\mbox{Dirichlet}(0.6, 0.2, 0.2)$, and half are $iid$ drawn from $\mbox{Dirichlet}(0.3, 0.4, 0.3)$. We consider two cases: (a) Weak noise ($\theta_i\equiv 0.7$, and the network is denser) (b) Strong noise ($\theta_i\equiv 0.4$, and the network is sparser). We choose  $L$ as in (\ref{chooseL}), but we also investigate SVS for all $L\in\{4,5,6,\ldots,15\}$. 
We report the average of $\max_k\|H\hat{v}_k-v_k\|^2$ over 100 repetitions. The results are in Figure~\ref{fig:VHcompare}. We observe the following:
(i) In the strong signal case, three methods perform similarly. (ii) In the weak signal case, CVS and SP are significantly worse than SVS. (iii) The performance of SVS is insensitive to the choice of $L$. The results confirm our claims in Section~\ref{subsec:MixedSCORE} and Section~\ref{subsec:thm-4versions} that the de-noise stage in SVS plays a crucial role in improving the numerical performance. 

\spacingset{1}
\spacingset{1.45}

{\bf Experiments 2-4} {\it (Performance of Mixed-SCORE-SVS)}. 
From now on, we fix the VH approach as SVS. The tuning integer $L$ is chosen from data using (\ref{chooseL}). In the literature, other mixed membership estimation approaches only work for MMSBM. The only exception is OCCAM \cite{JiZhuMM}. OCCAM assigns to each node a non-negative ``membership" vector with unit $\ell_2$-norm; we renormalize them by 
 their $\ell_1$-norms and use them as the estimated $\pi_i$. Fix $n=500$ and $K=3$. For $0\leq n_0\leq 160$, let each community have $n_0$ number of pure nodes. Fixing $x\in (0,1/2)$, let the mixed nodes have four different memberships $(x, x, 1-2x)$, $(x, 1-2x, x)$, $(1-2x, x, x)$ and $(1/3, 1/3, 1/3)$, each with $(500-3n_0)/4$ number of nodes. Given $\rho\in (0,1)$, $P$ has diagonals $1$ and off-diagonals $\rho$. Fixing $z\geq 1$, we generate the degree parameters such that $1/\theta_i\overset{iid}{\sim} U(1,z)$, where $U(1,z)$ denotes the uniform distribution on $[1,z]$. The tuning parameter $L$ is selected as in \eqref{chooseL}. For each parameter setting, we report $n^{-1}\sum_{i=1}^n\|\hat{\pi}_i-\pi_i\|^2$ averaged over $100$ repetitions. 

\begin{figure}[tb!] 
\centering
\includegraphics[height=.28\textwidth]{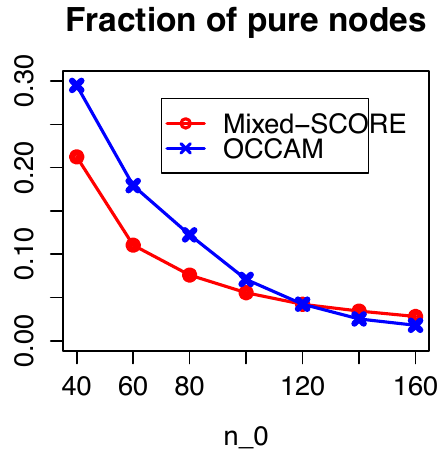}
\includegraphics[height=.285\textwidth]{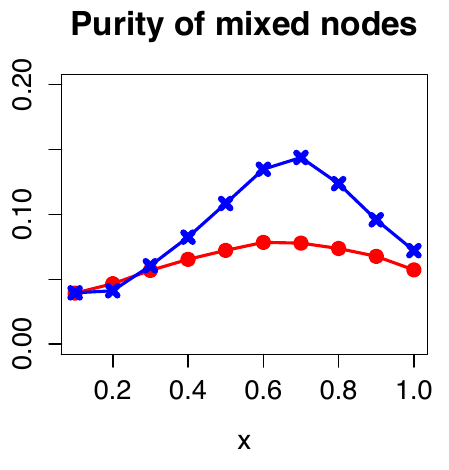}
\includegraphics[height=.28\textwidth]{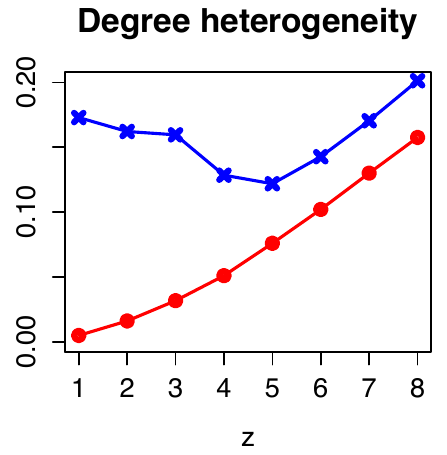}
\caption{\small Estimation errors of Mixed-SCORE and OCCAM ($y$-axis: $n^{-1}\sum_{i=1}^n\|\hat{\pi}_i - \pi_i\|^2$). 
} \label{fig:comparison}
\end{figure}

Experiment 2 (fraction of pure nodes).  Fix $(x, \rho, z)=(0.4, 0.1, 5)$ and let $n_0$ range in $\{40, 60, 80, 100, 120, 160\}$. As $n_0$ increases, the fraction of pure nodes increases from around $25\%$ to around $95\%$. See Figure~\ref{fig:comparison} (left). When the fraction of pure nodes is $<70\%$, Mixed-SCORE significantly outperforms OCCAM; when the fraction of pure nodes is $>70\%$, the two methods have similar performance.  

Experiment 3 (purity of mixed nodes). 
We call $\max_{1 \leq k \leq K} \{\pi_i(k) \}$ the ``purity" of node $i$.   
Fix $(n_0, \rho, z)=(80, 0.1, 5)$ and let $x$ range in $\{0.05, 0.1, 0.15, \cdots, 0.5\}$. 
In our settings, there are four types of mixed nodes. For the first three types, their purity is $(1-2x)1\{x\leq 1/3\}+x 1\{x>1/3\}$. Therefore, as $x$ increases to $1/3$, these nodes become less pure; then, as $x$ further increases, these nodes become more pure. 
See Figure~\ref{fig:comparison} (middle). It suggests that membership estimation is harder as the purity of mixed nodes decreases. Mixed-SCORE outperforms OCCAM in almost all settings, especially when $x$ is close to $1/3$. 

Experiment 4 (degree heterogeneity). Fix $(x, n_0, \rho)=(0.4, 80, 0.1)$ and let $z$ range in $\{1, 2, \cdots, 8\}$. Since $1/\theta_i\overset{iid}{\sim}U(1,z)$, a larger $z$ means the lower average degree and more severe degree heterogeneity (so the problem is harder). See Figure~\ref{fig:comparison} (right). Mixed-SCORE uniformly outperforms OCCAM.  Interestingly, when $z$ is small (so the problem is ``easy"), Mixed-SCORE is very accurate, but the performance of OCCAM is unsatisfactory.  

\spacingset{1}
\spacingset{1.45}

{\bf Experiments 5-8}. For space limit, we have relegated them to the supplement. Experiment 5 studies settings where the matrix $P$ varies. Experiment 6 studies settings where $\pi_i$'s drawn from a continuous distribution. Experiment 7 further investigates robustness of Mixed-SCORE-SVS to the choice of $L$. Experiment 8 compares Mixed-SCORE with the latent space modeling of networks \cite{handcock2007model}.

\section{Real data applications} \label{sec:realdata}

\subsection{The international trade networks and the trade triangles} \label{subsec:data-tradeGoods}
There are two lines of literature on the analysis of international trade networks. The first is the gravity model  \citep{anderson2003gravity}. It fits a generalized linear model for trade flows using countrywise `size' covariates and pairwise `trading cost' covariates. The second is in physics, which studies the topology of trade networks \citep{serrano2003topology}. 
Mixed-SCORE is useful in both approaches.

{\bf Combination of  Mixed-SCORE and gravity models}. Let $X(i,j)$ be the trade flow from country $i$ to country $j$. 
The (general) gravity model assumes $X(i,j)\sim \mathrm{Poisson}(\lambda(i,j))$, with $\ln(\lambda(i,j))=\sum_{m=1}^{M}\alpha_m G_{m}(i)+\sum_{m=1}^M\beta_m G_{m}(j) +  \sum_{s=1}^S\gamma_sD_{s}(i,j) + c_i+ c_j$, where $G_1, \ldots, G_M$ are the (log) `size' covariates, $D_1, \ldots, D_S$ are the (log) `trading cost' covariates, and $c_i$'s are the fixed effects of countries. We fit this model using Poisson pseudo maximum likelihood and let $\hat{\lambda}(i,j)$ denote the fitted value.  We define two `p-values' for each country pair: $Q_1(i,j)=\mathbb{P}(\mathrm{Poisson}(\hat{\lambda}(i,j))>X(i,j))$ and $Q_2(i,j)=\mathbb{P}(\mathrm{Poisson}(\hat{\lambda}(i,j))<X(i,j))$. 
A small value of $Q_1(i,j)$ implies that the observed trade flow is significantly higher than the fitted one, and a small value of $Q_2(i,j)$ indicates the opposite. 
We construct two undirected networks. In the first one,  there is an edge between nodes $i$ and $j$ if $\min\{Q_1(i,j), Q_1(j,i)\}<0.05$. In the second network, edges are defined similarly except that $Q_1$ is replaced by $Q_2$. We call them the {\it gravity-under-shooting (GUS)} network and {\it gravity-over-shooting (GOS)} network, respectively. For each network, we apply Mixed-SCORE to obtain $(\hat{\Pi}, \hat{\Theta}, \hat{P})$
and then construct a new nodal covariate, $U(i) = \ln(\hat{\theta}(i))$, and a new dyadic covariate, $H(i, j)=\ln(\hat{\pi}_i'\hat{P}\hat{\pi}_j)$.
We use them as surrogates of those unobserved covariates in the gravity model and plug them back to re-fit the gravity model. 
As explained in Example 3 of Section~\ref{subsec:3EconApplications}, we assume here that the unobserved covariates have a DCMM-like structure, which has the same spirit as the model in \cite{graham2015methods}. 
Our proposed `Mixed-SCORE + refitting' is a proxy approach to fitting the model we introduce there.  

To test the performance of our approach, we use an edited version of the gravity data set in \cite{head2010erosion} (available in the R package \texttt{gravity}).  The original data set contains the bilateral trade flows for 166 countries in 1948-2006. We only use the data in 2006. This edited version includes a nodal covariate, {\it gdp}, and five dyadic covariates, {\it distw}, {\it rta}, {\it contig}, {\it comlang\_off} and {\it comcur} 
(their meanings are in Column 2 of Table~\ref{tab:gravity-fitting}). Compared with the original gravity model fitting in \cite{head2010erosion}, this edited version does not provide all covariates, so it serves as a good example of {\it unobserved covariates}. 
Since there is only one year of data, we did not include any nodal covariate, because their effects will be absorbed into the fixed effect $c_i$;  all five dyadic covariates were included. We constructed the GUS and GOS networks as above and ran Mixed-SCORE separately on these two networks. We set $K=3$ for both networks.\spacingset{1}\footnote{We also tried other values of $K$. For different $K$, the networks and Mixed-SCORE output are different, but the newly created covariates and the subsequent gravity model fitting are similar.}\spacingset{1.45} 
It gave rise to two new dyadic covariates $H^{\text{GUS}}$ and $H^{\text{GOS}}$ (again, we did not include the new nodal covariates because of the fixed effects $c_i$). The results are in Table~\ref{tab:gravity-fitting}, where both new covariates created by Mixed-SCORE are significant. 
The other coefficients have mild changes and slightly smaller standard errors after re-fitting, except the coefficient of {\it comcur}. 
Initially, the coefficient of {\it comcur} is negative, with a very small p-value. This contradicts our common sense: sharing common currency should not have a significantly negative impact on trading.  
After adding the Mixed-SCORE covariates, the coefficient of {\it cumcur} becomes positive and insignificant. It suggests that our proposed approach is potentially useful in correcting the bias caused by unobserved covariates. 

\spacingset{1}
\begin{table}[tb!]
\centering
\caption{\small Combination of Mixed-SCORE and gravity model. The bigger model has two new covariates created by Mixed-SCORE. The F statistic for model comparison is 928.56 (p-value $<$ 2.2e-16). We note that these coefficients are not supposed to be directly compared with the fitted coefficients in Column 2 of Table 2 in \cite{head2010erosion}, because they use panel data but we only use one year's data (this also explains why our standard errors are considerably smaller).}\label{tab:gravity-fitting}
\scalebox{.7}{
\begin{tabular}{ll | rr | rr}
\toprule
 & & \multicolumn{2}{c|}{Before} &  \multicolumn{2}{c}{After}\\
 \cline{3-6}
Covariate & Meaning & Coef.  & Pval & Coef. & Pval \\
\hline
{\it distw} & weighted distance & -.832 (.012) & $<$2e-16 *** & -.722 (.011) & $<$2e-16 ***  \\
{\it rta} &   regional trade agreement dummy & .429 (.026) & $<$2e-16 ***&  .429 (.022) &  $<$2e-16 ***  \\
{\it contig} &   contiguity dummy & .415 (.022)  & $<$2e-16 *** &   .403 (.019)  & $<$2e-16 *** \\
{\it comlang\_off} & common official language dummy & .242 (.022) & $<$2e-16 *** &  .181 (.019) &  $<$2e-16 *** \\
{\it comcur} &   common currency dummy & -.167 (.031) & 7e-08 *** &  .005 (.027)  & .852 $\;\;\;\;\;\,$\\
{\it dyadic\_GUS} & new trade cost covariate (GUS) & & &  1.294 (.033) &  $<$2e-16 ***  \\
{\it dyadic\_GOS} & new trade cost covariate (GOS) &  &  & -.337 (.037) & $<$2e-16 ***  \\
 \bottomrule
\end{tabular}
}
\end{table} 
\spacingset{1.45}

\begin{figure}[tb!] 
\centering
\begin{subfigure}[t]{0.495\linewidth}
\vspace{0pt}
    \includegraphics[width=1\textwidth, height=.85\textwidth]{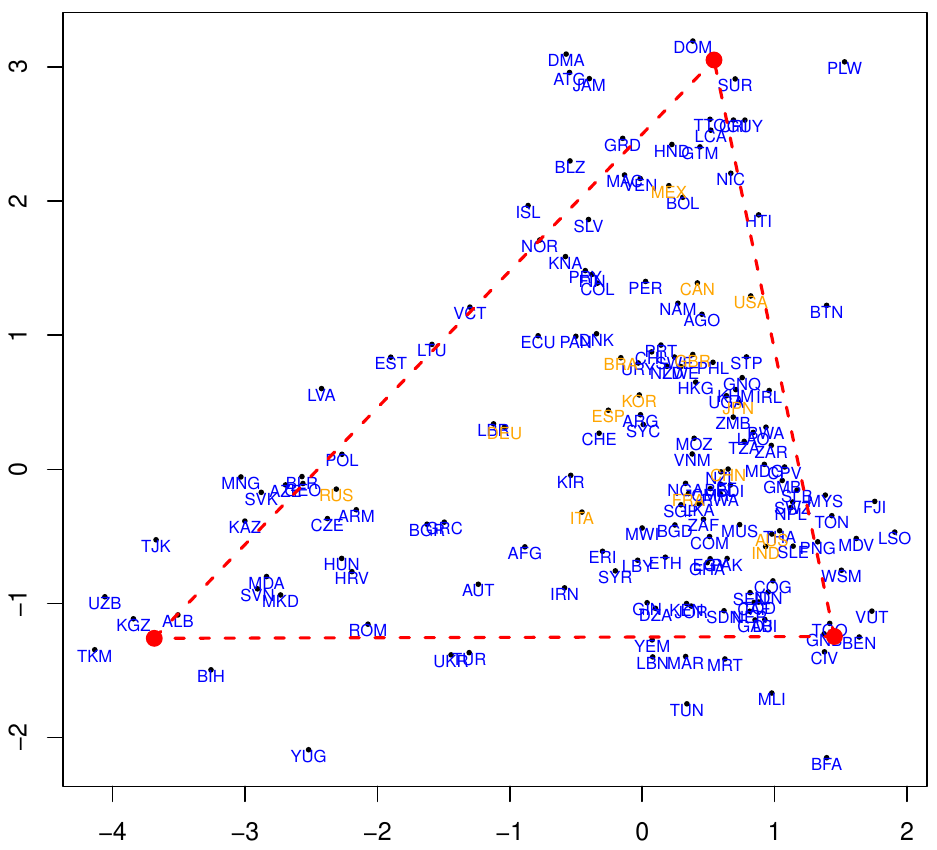}
    \caption{The GUS network after gravity model fitting.} \label{subfig:GUS}
\end{subfigure}
\begin{subfigure}[t]{0.49\linewidth}
\vspace{6pt}
    \includegraphics[width=1\textwidth, height=.8\textwidth]{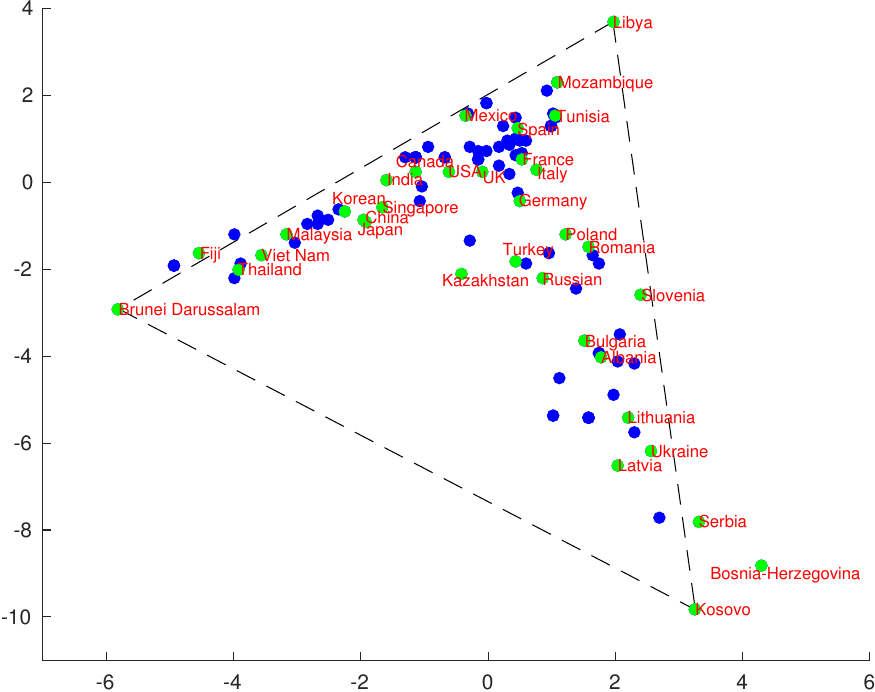}
    \caption{The trade in service (TIS) network} \label{subfig:TSI}
\end{subfigure}
\caption{\small Rows of $\hat{R}$ and the estimated simplex ($K=3$, so the simplex is a triangle). Left: Orange dots are top 15 countries with highest GDPs.  Right: Green dots are $35$ manually-picked economies.}  
\label{fig:trade-network-new} 
\end{figure}

To appreciate what information Mixed-SCORE captures, we check the rows of $\hat{R}$ for the GUS and GOS networks. Owing to space limit, we only discuss the GUS network here but relegate the results of the GOS network to the supplementary material (see Section~\ref{sec:Real-moreResults}).  
The edges in the GUS network indicate significant under-estimation of trade flows in the initial gravity model. Therefore, if $\hat{r}_i$ and $\hat{r}_j$ are close, the two countries may have unmodeled connections that benefit trade.  
The rows of $\hat{R}$ and the estimated simplex (which is a triangle since $K=3$) for GUS are shown in Figure~\ref{subfig:GUS}. We have some observations: (a) The 3 vertices may be interpreted as {\it Caribbean} (top), {\it Former Soviet Union} (bottom left), and {\it Western African} (bottom right). 
(b) United States, Canada and Mexico are close. These countries are in the North American Free Trade Agreement (NAFTA). The benefit of NAFTA cannot be fully captured by the regional trade agreement dummy {\it rta} \citep{anderson2016terms} and is further revealed in the covariates created by Mixed-SCORE.   
(c) United States and Russia are far away from each other - a consequence of the historical confrontation between two countries \citep{hufbauer2003impact}. 
(d) High-GDP countries tend to be in the interior of the triangle (i.e., they have low `trading costs' with many countries). This is consistent with economic theory that good `tradability' can boost economic growth \citep{waugh2010international}. (e) United States (with the highest GDP)  is not in the deep  interior of the triangle but on an edge. Interestingly, this position is farthest from the {\it Former Soviet Union} vertex. 

{\bf Remark 9}. In re-fitting the gravity model, an alternative approach is replacing $H(i,j)$ by $\ln(\widehat{\Omega}_{ij})$, where $\widehat{\Omega}$ is an arbitrary estimate of $\Omega$. Using the output of Mixed-SCORE, we can obtain an estimate $\widehat{\Omega}^{\text{MS}}$ by $\widehat{\Omega}^{\text{MS}}_{ij}=\hat{\theta}_i\hat{\theta}_j\cdot \hat{\pi}_i'\hat{P}\hat{\pi}_j$. Since $\hat{\theta}_i$ and $\hat{\theta}_j$ will be absorbed into the fixed effects, this approach is equivalent to the approach we have used above. 
However, we may plug in a different estimate of $\Omega$, such as $\widehat{\Omega}^{\text{PCA}}=\sum_{k=1}^K\hat{\lambda}_k\hat{\xi}_k\hat{\xi}_k$, where $\hat{\lambda}_k$ and $\hat{\xi}_k$ are the $k$th eigenvalue and eigenvector of $A$. In Section~\ref{sec:EstimateOmega} of the supplementary material, we compare the two estimates of $\Omega$ and find that $\widehat{\Omega}^{\text{MS}}$ has much better numerical performance. The reason is that $\widehat{\Omega}^{\text{MS}}$ utilizes the DCMM model structure, not just low-rankness of $\Omega$. 

{\bf Remark 10}. In the recent literature of gravity modeling of trade data, it has become common to use panel data and to include the importer-year and exporter-year fixed effects \citep{weidner2021bias}. We did not use panel data because Mixed-SCORE only applies to static networks. In a working paper, we extend Mixed-SCORE to dynamic networks. It will be useful for analysis of panel data. We leave this to future work. 

{\bf Remark 11}. In the analysis of panel data, an interesting approach is using the pairwise fixed effects \citep{weidner2021bias} to account for unobserved covariates. However, for our example here where we only use one year's data, this approach will introduce $n(n-1)/2$ free parameters, but we only have $n(n-1)$ observed trading flows; therefore, this approach will have the issue of over-fitting. In comparison, our Mixed-SCORE approach only allows for $O(nK)$ free parameters and does not have this over-fitting issue.

\spacingset{1}
\spacingset{1.45}


{\bf Using Mixed-SCORE for network analysis of the world trade web}. 
Studying the network topology of the world trade web is a problem of interest \citep{serrano2003topology}.  These works do not require observing any covariates. They build networks directly from trade flows and study the topology of these networks (e.g., power law degree distribution, latent community structure, centrality metric, clustering coefficient, etc.). We will show that Mixed-SCORE is useful for creating low-dimensional embeddings of countries in these networks. 
We downloaded the trade in services data from \url{https://data.wto.org/}. For each pair of economies $(i,j)$, we aggregated the total service export from economy $i$ to economy $j$ during 2014-2018 (we used the numbers reported by economy $i$). There are 202 economies in total, but we removed {\it European Union} and {\it Extra EU Trade}, as their data partially overlap with the data of individual countries. This gave rise to a $200\times 200$ weight matrix $X$. 
We symmetrize $X$ to $Y=(X+X')/2$. 
Let $u =(u_1,u_2,\ldots,u_{200})'$ contain the row sums of $Y$. Define $Z=[\diag(u)]^{-1/2}Y[\diag(u)]^{-1/2}$, where each entry of $Z$ is in $[0,1]$. \spacingset{1}\footnote{One may use GDP or population to normalize, but here we are primarily interested in the case with no observed covariates. We follow the literature to use total trade flows to normalize.}\spacingset{1.45}
Let $\mu$ and $\sigma$ be the mean and standard deviation of all nonzero entries of $Z$. 
We construct an undirected network, where each economy is a node and there is an edge between $i$ and $j$ if and only if $Z(i,j) \geq \mu+\sigma$.
We restrict it to the giant component, which has $n=116$ nodes. We call this network the trade-in-service (TIS) network. 
We applied Mixed-SCORE with $K=3$.\spacingset{1}\footnote{For the adjacency matrix,  the scree plot shows the elbow point is either at $K = 3$ or $K = 4$. We  applied Mixed-SCORE with both $K = 3$ and $K = 4$. It turns out that for $K = 3$,   the plot of the rows of $\hat{R}$ (see (\ref{defineRhat})) fits better with the simplex structure, and the results are easier to interpret,  so we choose $K = 3$. Furthermore, we set $T=2\log(n)$ and $L=25$ in Mixed-SCORE.}\spacingset{1.45}
The rows of $\hat{R}$ are displayed in Figure~\ref{subfig:TSI}.\spacingset{1}\footnote{The point associated with {\it Montenegro} is far away from the data cloud, which we treat as an outlier and do not show in the figure.}\spacingset{1.45} 
This creates an embedding of all economies into a 2-dimensional latent space. 
We have some noteworthy observations. 
(a) The point cloud fits well with a triangle, which we call the `trade triangle'. The three vertices may be interpreted as three different regions: `North Africa' (top vertex in Figure~\ref{subfig:TSI}),  `Southeast Asia' (bottom left vertex), and `Central/South Europe' (bottom left vertex).
(b) It agrees to economic theory that 
geographic proximity plays a key role in trade. In Figure~\ref{subfig:TSI}, countries that are geographically close tend to cluster together; e.g., countries in Southeast Asia ({\it Thailand}, {\it Vet Nam}, {\it Malaysia}, etc.), East Asia ({\it China}, {\it Japan}, {\it Korea}, etc.), North America ({\it USA}, {\it Canada}, {\it Mexico}, etc.), West Europe ({\it UK}, {\it France}, {\it Germany}, etc.), East Europe and West/Central Asia ({\it Russian}, {\it Kazakhstan}, {\it Turkey}, {\it Bulgaria}, etc.) and so on. 
(c) The node embedding contains more information than geographical proximity. 
For example, {\it Singapore} is geographically close to Southeast Asian countries, but it is closer to East Asian countries in the trade triangle; 
West European countries are geographically closer to East European countries, but they are closer to North American countries in the trade triangle. 
These can be explained by trading agreements and historical trading relationships. 
The above supports that Mixed-SCORE is useful for node embedding. Imagine that we are given the trade flows of a new product or service, with little known information; we can apply Mixed-SCORE to visualize the locations of countries in the embedded space and gain useful insights for next-step modeling.

\subsection{The coauthor and citee network of statisticians, and Fan's group} \label{application-stat}   
The study of coauthorship networks and citation networks is common in applied social science \citep{barabasi2002evolution}. 
The goal is using scientific publications in a field to study the development of the field itself. 
It is useful for discovering whether all sub-areas (`communities') are developed in a healthy and balanced way and whether any particular sub-area is under-developed and needs more allocation of resources \citep{foster2015tradition}.  For example, \cite{andrikopoulos2016four} studied the coauthorship network for {\it Journal of Econometrics}. 
In this subsection, we use a data set from \cite{SCC-JiJin}. It consists of bibtex 
and citation data of $3,248$ papers published in four top-tier statistics journals, {\it Annals of Statistics}, {\it Biometrika}, {\it Journal of American Statistical Association}, and {\it Journal of Royal Statistical Society -Series B}, during 2003--2012. 

{\bf The coauthorship network}.  \cite{SCC-JiJin} defined a coauthorship network, where each node is an author, and two authors have an edge 
if they coauthored $2$ or more papers in the data range. The giant component of the network contains 236 authors. 
\cite{SCC-JiJin} suggest that this is the ``High Dimensional Data Analysis" group, which has a ``Carroll-Hall" sub-group (including researchers in nonparametric and semi-parametric statistics and functional estimation) and a ``North Carolina" sub-group (including researchers from Duke, North Carolina, and NCSU).  In light of this, we consider a DCMM model assuming (a) there are $K=2$ communities called  ``Carroll-Hall"  and ``North Carolina"  respectively, and (b) some nodes have mixed memberships in two communities.   We applied Mixed-SCORE, and the results are in Table~\ref{tb:coauthor}. 
It was argued in \cite{SCC-JiJin} that the ``Fan" group (Jianqing Fan and collaborators) has strong ties to both communities.   Our results confirm such a finding but shed new light on the ``Fan" group: many of the nodes (e.g., Yingying Fan, Rui Song, Yichao Wu,    Chunming Zhang, Wenyang Zhang) have highly mixed memberships, and for each mixed node, we can quantify its weights in two communities.  For example, both Runze Li (former graduate of UNC-Chapel Hill)  
and Jiancheng Jiang (former post-doc at UNC-Chapel Hill and current faculty member at UNC-Charlotte)  have 
mixed memberships, but Runze Li is more on the ``Carroll-Hall" community (weight: $73\%$) and Jiancheng Jiang  is more on the ``North Carolina" community (weight: $62\%$).

\spacingset{1} 
\begin{table}[htb!]  
\caption{\small Left and Middle: high-degree pure nodes in the ``Carroll-Hall" community  and the ``North Carolina" community. Right: highly mixed nodes (data: Coauthorship network).} \label{tb:coauthor}
\centering
\scalebox{.58}{
\begin{tabular}{lc|| lc|| lcl}
\hline
Name & Deg. & Name & Deg. & Name & Deg. & Estimated PMF \\
\hline
Peter Hall & 21 & Joseph G Ibrahim & 14 & Jianqing Fan & 16 & 54\% of Carroll-Hall \\
Raymond J Carroll & 18 & David Dunson	& 8 & Jason P Fine & 5 & 54\% of Carroll-Hall\\
T Tony Cai & 10 & Donglin Zeng & 7 & Michael R Kosorok	 & 5 & 57\% of Carroll-Hall\\
Hans-Georg Muller & 7 & Hongtu Zhu & 7 & J S Marron & 4 & 55\% of North Carolina \\ 
Enno Mammen	& 6 & Alan E Gelfand & 5 & Hao Helen Zhang & 4 & 51\% of North Carolina \\
Jian Huang & 6 & Ming-Hui Chen & 5 & Yufeng Liu & 4 & 52\% of North Carolina \\
Yanyuan Ma & 5 & Bing-Yi Jing	 & 4 & Xiaotong Shen & 4 & 55\% of North Carolina \\
Bani Mallick & 4 & Dan Yu Lin & 4 & Kung-Sik Chan & 4 & 55\% of North Carolina  \\
Jens Perch Nielsen	& 4 & Guosheng Yin & 4 & Yichao Wu & 3	& 51\%  of Carroll-Hall\\
Marc G Genton	& 4 & Heping Zhang	& 4 & Yacine Ait-Sahalia & 3 & 51\%  of Carroll-Hall\\
Xihong Lin & 4  & Qi-Man Shao	 & 4 & Wenyang Zhang & 3 & 51\% of Carroll-Hall\\
Aurore Delaigle	 & 3 & Sudipto Banerjee & 4 & Howell Tong & 2	 & 52\% of North Carolina  \\
Bin Nan & 3 & Amy H Herring & 3 & Chunming Zhang & 2 &  51\% of Carroll-Hall\\
Bo Li & 3 & Bradley S Peterson	 & 3 & Yingying Fan & 2 & 52\% of North Carolina  \\
Fang Yao	& 3 & Debajyoti Sinha & 3 & Rui Song & 2	 & 52\% of Carroll-Hall \\
Jane-Ling Wang & 3 & Kani Chen & 3 & Per Aslak Mykland & 2 &  52\% of North Carolina \\
Jiashun Jin & 3 & Weili Lin & 3 & Bee Leng Lee &  2 &  54\% of Carroll-Hall \\
\hline \\
\end{tabular}}
\end{table}
\spacingset{1.45}

\begin{figure}[tb] 
\centering
\includegraphics[width=.4\textwidth, height=.25\textwidth]{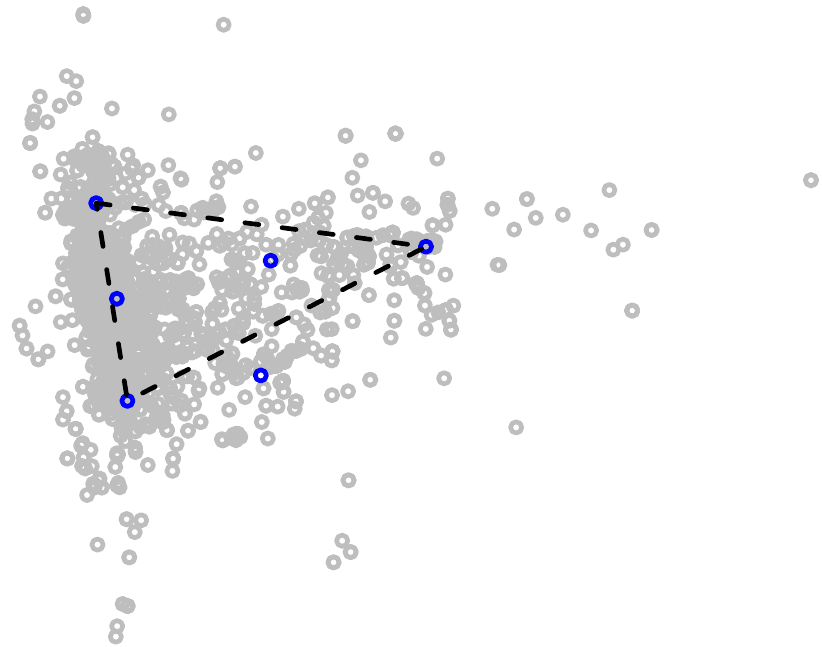}
\includegraphics[width=.4\textwidth, height=.25\textwidth]{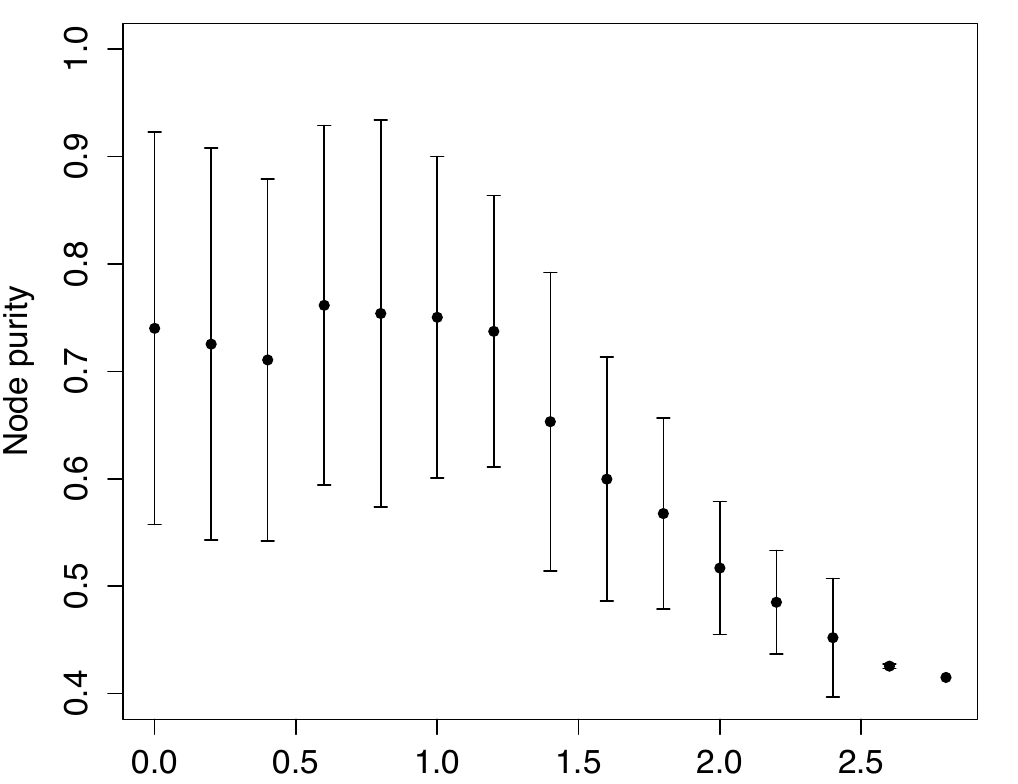}
\caption{\small Left:  rows of $\hat{R}$ and the estimated simplex. Right: node purity v.s. degree; $x$-axis is $\hat{\theta}(i)$ (grouped with an interval of $.2$; we plot the mean and standard deviation of $\|\hat{\pi}_i\|_\infty$ in each group).} \label{fig:citee}
\end{figure}

{\bf The citee network}. \cite{SCC-JiJin} also defined a citee network: there is an edge between two authors if they  have been cited at least once by the same author (other than themselves). 
The giant component of this network contains $1790$ authors. \cite{SCC-JiJin} suggested that the network has three meaningful  communities:  ``Large Scale Multiple Testing" (MulTest), ``Spatial and Nonparametric Statistics" (SpatNon) and ``Variable Selection" (VarSelect). We thereby set $K = 3$  and apply Mixed-SCORE.  Figure~\ref{fig:citee} (left) plots the rows of $\hat{R} \in \mathbb{R}^{n,2}$, where a simplex (triangle)  is clearly visible in the cloud.   
Table~\ref{tb:membership2} shows the estimated PMF of high degree nodes (please also see Table~\ref{tb:membership1} in the supplementary material). The results confirm 
those in \cite{SCC-JiJin} (especially on the existence of three communities aforementioned), but also 
shed new light on the network. 
First, high-degree nodes in VarSelect are frequently observed to have an interest in MulTest, 
and this is not true the other way around (e.g., 
compare {\it Jianqing Fan}, {\it Hui Zou} with {\it Yoav Benjamini}, {\it Joseph Romano}).  
Second, the citations from SpatNon to either MulTest or  VarSelect are comparably lower than those between MulTest and VarSelect.  This fits well with our impression. 
Conceivably,  a node with higher degree tends to be more senior and so tends to be more mixed. 
Figure~\ref{fig:citee} (right) 
is the plot of the node purity, $\max_{1 \leq k \leq K} \{  \hat{\pi}_i(k) \}$, versus the estimated degree heterogeneity parameter $\hat{\theta}(i)$. 
The results show a clear negative correlation between two quantities (especially on the right end, which corresponds to nodes with high degrees), 
which indicates that nodes with  higher degrees  tend to be more mixed.

\spacingset{1}
\spacingset{1.45}

\spacingset{1}
\begin{table}[tb!]  
\caption{\small Estimated PMF of the $12$ nodes with the highest degrees in the Citee network.} \label{tb:membership2} 
\centering
\scalebox{.7}{
\begin{tabular}{l|l| lll | l|l| lll}
\hline
Name & Deg. & MulTest & SpatNon & VarSelect & Name & Deg. & MulTest & SpatNon & VarSelect\\
\hline
Jianqing Fan & 977 & 0.365 & 0.220	& 0.415 &  Peter Buhlmann & 742 & 0.527 & 0.121 & 0.352\\
Raymond Carroll & 850 & 0.282 & 0.294 & 0.424 & Hans-Georg Muller & 714 & 0.413 & 0.237 & 0.350\\
Hui Zou & 824 & 0.348 & 0.225	& 0.427 & Yi Lin & 693 & 0.417 & 0.137 & 0.446\\
Peter Hall	& 780 & 0.501 & 0.032 & 0.467 &  Nocolai Meinshausen & 692 & 0.462 & 0.125 & 0.413\\
Runze Li & 778 & 0.282 & 0.226 & 0.491 &  Peter Bickel & 692 & 0.529 & 0.216 & 0.255\\
Ming Yuan & 748 & 0.391 & 0.166 & 0.444 & Jian Huang & 677 & 0.572 & 0 & 0.428\\
\hline
\end{tabular}
}
\end{table}
\spacingset{1.45}

\section{Discussion} \label{sec:Discu}
There have been independent interests on networks from both the econometric literature and the statistical literature. Recently, the use of statistical network models in economic problems has received increasingly more attention. However, the statistical models used in network econometrics are largely limited to the classical models, such as SBM and graphon. 
Recent developments in 
statistical network analysis have suggested new ideas in network modeling, 
but such ideas are largely unknown in the area of network econometrics. 
In this paper, we make two contributions: 1) We provide a new tool for estimating community structure and creating nodal features from network data. 2) We offer a new network model that accommodates severe degree heterogeneity and mixed memberships and is more suitable for real data; we also equip it with a fast spectral algorithm for estimating parameters of this model. For many existing works in network econometrics that use SBM or graphon 
as the network model, we may improve the results by using the more realistic DCMM model introduced here. This will inspire interesting future research.

The design of our algorithm includes several novel ideas, e.g., discovering the simplex structure in the spectral domain and the correct steps to estimate $\Pi$ from the simplex. We have also proposed new vertex hunting algorithms, which have much better numerical performance than the existing algorithms such as successive projection. Theoretically, we derive the explicit error bounds for $\hat{\Pi}$ and show that it is rate-optimal under some conditions. 

For future research, first, it is unclear how to estimate $K$ from data. \cite{jin2020estimating} proposed a stepwise goodness-of-fit procedure for estimating $K$ when there is no mixed membership (i.e., the DCMM model reduces to DCBM). It is an interesting question how to combine Mixed-SCORE with this approach for estimating $K$ under DCMM. 
Second, we mention several applications of our work in network econometrics (see Section~\ref{subsec:3EconApplications}). It is of great interest to study each application more carefully. For example, can we get a theoretical guarantee for using Mixed-SCORE in these problems? We briefly discuss it in the paragraph below Theorem~\ref{thm:main-rev}, but more rigorous theoretical studies are needed. We leave these open problems to future work.

\medskip
\noindent 
{\bf Data and code}: The code for implementing Mixed-SCORE and different VH algorithms is available at \url{https://github.com/ZhengTracyKe/MixedSCORE}. This link also contains all the real networks used in this paper.

\spacingset{0.9}
\small 

\bibliographystyle{jf}
\bibliography{network}

\newpage

\allowdisplaybreaks

\spacingset{1.45}

\appendix

\section{Identifiability and Regularity Conditions}
We prove the identifiability of DCMM and discuss Assumption 4 (where we give sufficient conditions for this assumption to hold).  

\subsection{The Identifiability of DCMM} \label{subsec:proof-identifiability} 

The following proposition shows that the DCMM model is identifiable if each community has at least one pure node. 
\begin{prop}[Identifiability] \label{prop:identifiability} 
Consider a DCMM model as in \eqref{def:DCMM}, where $P$ has unit diagonals.  
When each community has at least one pure node, the model is identifiable: For eligible $(\Theta,\Pi,P)$ and $(\widetilde{\Theta},\widetilde{\Pi}, \widetilde{P})$, if $\Theta\Pi P\Pi'\Theta=\widetilde{\Theta}\widetilde{\Pi}\widetilde{P}\widetilde{\Pi}'\widetilde{\Theta}$, then $\Theta=\widetilde{\Theta}$, $\Pi=\widetilde{\Pi}$, and $P = \widetilde{P}$. 
\end{prop}

\noindent
{\it Proof of Proposition~\ref{prop:identifiability}}: Let $G=K\|\theta\|^{-2}\Pi'\Theta^2\Pi$ be the same as in Section~\ref{sec:mainthm}. We consider two cases: (1) $PG$ is an irreducible matrix. (2) $PG$ is a reducible matrix.

First, we study Case (1). When $PG$ is irreducible, the matrix $R$ is well-defined (see Lemma~\ref{lem:ideal}). Additionally, by Lemma~\ref{lem:ideal}, there exists the Ideal Simplex, which is uniquely determined by the eigenvectors $\xi_1, \xi_2, \ldots,  \xi_K$ of $\Omega$.  
For either $(\Theta, \Pi, P)$ or $(\widetilde{\Theta}, \widetilde{\Pi}, \widetilde{P})$,  we have an Ideal Simplex. The two Ideal Simplexes can be different only when there are multiple choices of $\xi_1, \xi_2, \ldots,  \xi_K$. By Lemma~\ref{lem:B}, the first eigenvalue of $\Omega$ has a multiplicity $1$,  so by basic linear algebra,  $[\xi_1, \xi_2, \ldots,  \xi_K]$ are uniquely defined up to a rotation matrix of the form 
\[
\left[ 
\begin{array}{ll} 
a & 0 \\
0 & S
\end{array} 
\right], \quad \mbox{where $a \in \{-1, 1\}$ and $S \in \mathbb{R}^{K-1, K-1}$ is an orthogonal matrix}.   
\] 
Recalling $R = [\mathrm{diag}(\xi_1)]^{-1}[\xi_2, \xi_3, \ldots, \xi_K]$, it is seen that the property of ``a row of $R$ falls on one of the vertices of the Ideal Simplex" is invariant to the above rotation. Therefore, a row of $\Pi$ equals to the corresponding row of $\widetilde{\Pi}$, as long as one of them is pure.  

We now proceed to showing $(\Theta, \Pi, P) = (\widetilde{\Theta},\widetilde{\Pi}, \widetilde{P})$. By the above argument and that each community has at least one pure node, we assume without loss of generality that for $1 \leq k \leq K$, the $k$-th node is a pure node in community $k$. Comparing the first $K$ rows and the first $K$ columns of $\Theta \Pi P \Pi' \Theta$ with those of  $\widetilde{\Theta} \widetilde{\Pi} \widetilde{P} \widetilde{\Pi}' \widetilde{\Theta}'$, 
it follows that
\[
\diag(\theta_1,   \ldots, \theta_K) \cdot   P  \cdot    \diag(\theta_1,  \ldots, \theta_K)  = 
\diag(\tilde{\theta}_1,  \ldots, \tilde{\theta}_K)  \cdot  \widetilde{P}  \cdot    \diag(\tilde{\theta}_1,   \ldots, \tilde{\theta}_K).  
\] 
As both $P$ and $\widetilde{P}$ have unit diagonal entries,    $P = \widetilde{P}$ and $\theta_k = \tilde{\theta}_k$, $1 \leq k \leq K$. 
 
Moreover, note that $P\Pi'\Theta$ has a full row-rank. Since $\Theta \Pi P \Pi' \Theta = \widetilde{\Theta} \widetilde{\Pi} \widetilde{P} \widetilde{\Pi}' \widetilde{\Theta}$, it is seen that $\Theta \Pi = \widetilde{\Theta} \widetilde{\Pi} \Delta$, where $\Delta =  \widetilde{P} \widetilde{\Pi}' \widetilde{\Theta} X' (XX')^{-1}$, with $X = P \Pi' \Theta$ for short. We compare the first $K$ rows of $\Theta \Pi$ and $\widetilde{\Theta} \widetilde{\Pi} \Delta$, recalling that the first $K$ rows are pure and that $\theta_k = \tilde{\theta}_k$ for $1 \leq k \leq K$. It follows that $\Delta$ equals to the $K \times K$ identity matrix. Therefore, 
\[
\Theta \Pi = \widetilde{\Theta} \widetilde{\Pi}. 
\]  
Since each row of $\Pi$ or $\tilde{\Pi}$ is a PMF, $\Theta = \tilde{\Theta}$, $\Pi  = \tilde{\Pi}$, and the claim follows.  

Next, we study Case (2). By Lemma~\ref{lem:B}, 
\[
\Xi = \Theta\Pi B, \qquad\mbox{for a non-singular matrix $B$}. 
\]
Row $i$ of $\Xi$ equals to $\theta_i$ times a convex combination of rows of $B$. It follows that {\it all rows of $\Xi$ are contained in a simplicial cone with $K$ supporting rays, where a pure row falls on one supporting ray, and a mixed row falls in the interior of the simplicial cone}. Note that $\Xi$ is uniquely defined up to a $K\times K$ orthogonal matrix. The effect of this orthogonal matrix is to simultaneously rotate all rows of $\Xi$. Such a rotation does not change the property that ``a pure row falls on one supporting ray". Therefore, a row of $\Pi$ equals to the corresponding row of $\tilde{\Pi}$, provided that one of them is pure. The remaining of the proof is similar to that of Case (1).
\qed

\medskip 

\noindent
{\bf Remark} {\it (Comparison with the identifiability of other models)}. Compared to other models (e.g., MMSB, DCBM), DCMM has many more parameters (for degree heterogeneity and for mixed memberships). These parameters have more degrees of freedom than those in MMSB or DCBM, and so DCMM requires stronger conditions to be identifiable. 
\begin{itemize}
\item The assumption that $P$ has unit diagonals is not needed for identifiability of MMSB, but it is necessary for identifiability of DCMM. Consider a DCMM with parameters $(\Theta,\Pi, P)$. Given any $K\times K$ diagonal matrix $D$ with positive diagonals, let 
\[
\widetilde{P}=DPD, \quad \tilde{\pi}_i = (D^{-1}\pi_i)/\|D^{-1}\pi_i\|_1, \quad\mbox{and}\quad \tilde{\theta}_i=\|D^{-1}\pi_i\|_1\cdot\theta_i.
\]
It is seen that $\Theta\Pi P\Pi'\Theta=\widetilde{\Theta}\widetilde{\Pi}\widetilde{P}\widetilde{\Pi}'\widetilde{\Theta}$. This case will be eliminated by requiring $P$ to have unit diagonals.
\item The assumption that $P$ has a full rank is not needed for identifiability of DCBM, but it is necessary for identifiability of DCMM. If the rank of $P$ is $<K$, there exists a nonzero vector $\beta\in\mathbb{R}^K$ such that $P\beta=0$. 
As long as there is a $\pi_i$ such that $\pi_i(k)>0$ for all $k$, we can change $(\pi_i,\theta_i)$ to $(\tilde{\pi}_i, \tilde{\theta}_i)$ but keep $\Omega$ unchanged. To see this, let 
\[
\tilde{\pi}_i =  (\pi_i+\epsilon\beta)/\|\pi_i+\epsilon\beta\|_1, \quad \mbox{and}\quad \tilde{\theta}_i=\|\pi_i+\epsilon\beta\|_1\cdot \theta_i,
\]
for a sufficiently small $\epsilon>0$. Since the two vectors, $\theta_i\cdot P\pi_i$ and $\tilde{\theta}_i\cdot P\tilde{\pi}_i$, are equal, $\Omega$ remains unchanged.  
\end{itemize}

\subsection{Sufficient conditions for Assumption 4 to hold} \label{subsec:condition-check}
We give two propositions showing examples where  Assumption 4 is satisfied. Below, for a matrix $M$, let $\lambda_k(M)$ denote the $k$-th largest eigenvalue in magnitude.  
\begin{prop} \label{prop:condition-check}
Consider a DCMM model where $\Omega=\Theta\Pi P\Pi'\Theta$ and $\|P\|_{\max}\leq C$. Write $G=K\|\theta\|^{-2}(\Pi'\Theta^2\Pi)$. Let $\eta_1$ be the first (unit-norm) right singular vector of $PG$. As $n\to\infty$, suppose at least one of the following conditions hold, where $c>0$ is a constant:
\begin{itemize}
\item $\min_{1\leq k,\ell\leq K}P(k,\ell)\geq c$, and $\min_{k}\{\sum_{i=1}^n \theta_i^2\pi_i(k)\}\geq c\max_{k}\{\sum_{i=1}^n \theta_i^2\pi_i(k)\}$.  
\item $K$ is fixed, $\min_{k} G(k,k)\geq c$, and $|\lambda_1(PG)|\geq c+|\lambda_2(PG)|$. For a fixed irreducible matrix $P_0$, $\|P-P_0\|\to 0$. 
\item $K$ is fixed, and $|\lambda_1(PG)|\geq c+|\lambda_2(PG)|$. For a fixed irreducible matrix $G_0$, $\|G-G_0\|\to 0$. 
\end{itemize}
Then, we can select the sign of $\eta_1$ such that all its entries are strictly positive. Furthermore, $[\max_{1\leq k\leq K}\eta_1(k)]/[\min_{1\leq k\leq K}\eta_1(k)]\leq C$. 
\end{prop}

\begin{prop} \label{prop:condition-check2}
Consider a DCMM model where $\Omega=\Theta\Pi P\Pi'\Theta$. We assume that  $\max_{1\leq k\leq K}\{\sum_{\ell=1}^K P(k, \ell)\}\leq C\min_{k}\{\sum_{\ell=1}^K P(k, \ell)\}$. Suppose $\pi_i$'s are i.i.d. generated from Dirichlet($\alpha$), where $\alpha=(\alpha_1,\alpha_2,\ldots,\alpha_K)'$ satisfies $C_1 \leq \alpha_k\leq C_2$ for two constants $C_2>C_1>0$. Write $G=K\|\theta\|^{-2}(\Pi'\Theta^2\Pi)$. Let $\eta_1$ be the first (unit-norm) right singular vector of $PG$. As $n\to\infty$, $[\max_{1\leq k\leq K}\eta_1(k)]/[\min_{1\leq k\leq K}\eta_1(k)]\leq C$, with probability $1-o(1)$. 
\end{prop}

\noindent
{\it Proof of Propositions~\ref{prop:condition-check}-\ref{prop:condition-check2}:}
First, we prove Propositions~\ref{prop:condition-check}. 
Consider the first case. Let $x_k=K\|\theta\|^{-2}\sum_{i=1}^n\theta_i^2\pi_i(k)$. It is seen that
$\sum_{k=1}^K x_k=K$. The assumption says that $\min_k x_k\geq c\max_k x_k$. Therefore, $x_k\asymp 1$ for all $ k$. At the same time, $\sum_{\ell=1}^K G(\ell,k)=K\|\theta\|^{-2}\sum_{\ell=1}^K\sum_{i=1}^n\theta_i^2\pi_i(\ell)\pi_i(k)=x_k$. It follows that
\[
\max_k\bigl\{ \sum_{\ell}G(\ell,k)\bigr\}\asymp \min_k\bigl\{ \sum_{\ell}G(\ell,k)\bigr\}\asymp 1.  
\] 
For any $1\leq m,k\leq K$, the $(m,k)$-th entry of $PG$ equals to 
$\sum_\ell P(m,\ell)G(\ell,k)$, which is between $c\sum_\ell G(\ell,k)$ and $C\sum_\ell G(\ell,k)$ by the assumption on $P$. It follows that
\beq \label{prop-condcheck-1}
\max_{k,\ell}\bigl\{(PG)(k,\ell)\bigr\} \asymp \min_{k,\ell}\bigl\{(PG)(k,\ell)\bigr\}\asymp 1. 
\eeq
In particular, $PG$ is a positive matrix. By Perron's theorem \cite[Theorem 8.2.8]{HornJohnson}, the first right singular value $\lambda_1(PG)$ is positive and has a multiplicity of $1$, and the first eigenvector $\eta_1$ is a positive vector. Write $\lambda=\lambda_1(PG)$ for short. By definition,
\[
\lambda\eta_1 =(PG)\eta_1. 
\]
It follows that
\beq \label{prop-condcheck-2}
\max_k\eta_1(k)\leq \frac{\|\eta\|_1}{\lambda}\max_{k,\ell} \{(PG)(k,\ell)\}, \quad \min_k\eta_1(k)\geq \frac{\|\eta\|_1}{\lambda}\min_{k,\ell}\{(PG)(k,\ell)\}. 
\eeq
Combining \eqref{prop-condcheck-1}-\eqref{prop-condcheck-2} gives $\max_k\eta_1(k)\asymp \min_k\eta_1(k)$. The claim follows. 

Consider the second case. We first state and prove a useful result: 
\beq \label{prop-condcheck-3}
\begin{array}{l}
\mbox{Let $A$ and $B$ be two nonnegative matrices with strictly}\\
\mbox{positive diagonals. If $A$ is irreducible, then $AB$ is irreducible.} 
\end{array}
\eeq
The proof uses the definition of primitive matrices (a subclass of irreducible matrices; see \cite[Section 8.5]{HornJohnson}). We aim to show $AB$ is a primitive matrix. By \cite[Theorem 8.5.2]{HornJohnson}, it suffices to show that there exists $m\geq 1$, such that $(AB)^m$ is a strictly positive matrix. By the assumption, $A$ is an irreducible matrix with positive diagonals; it follows from \cite[Theorem 8.5.4]{HornJohnson} that $A$ is a primitive matrix. By \cite[Theorem 8.5.2]{HornJohnson} again, there exists $m\geq 1$ such that $A^m$ is a strictly positive matrix. Let $\alpha>0$ be the minimum diagonal entry of $B$. Since $A$ and $B$ are nonnegative matrices, each entry of $(AB)^m$ is lower bounded by $\alpha^m$ times the corresponding entry of $A^m$; hence, $(AB)^m$ is also a strictly positive matrix. It follows that $AB$ is a primitive matrix, which is also an irreducible matrix. 

We then show the claim. Note that $P$ and $G$ are both nonnegative matrices with positive entries. Since $\|P-P_0\|\to 0$, the support of $P$ has to be a superset of the support of $P_0$ for large enough $n$; as a result, when $P_0$ is an irreducible matrix, $P$ has to be an irreducible matrix for sufficiently large $n$. We apply \eqref{prop-condcheck-3} to obtain that $PG$ is an irreducible matrix. It follows that $\lambda_1(PG)>0$ and it has a multiplicity $1$; additionally, the first right eigenvector $\eta_1$ is a positive vector. 

It remains to show $\max_k\eta_1(k)\asymp \min_k\eta_1(k)$. We prove by contradiction. Write $\eta_1=\eta_1^{(n)}$, $P=P^{(n)}$ and $G=G^{(n)}$ to emphasize the dependence on $n$. If the claim is not true, then there is a subsequence $\{n_s\}_{s=1}^\infty$ such that  
\beq  \label{prop-condcheck-4}
\lim_{s\goto\infty}\biggl\{\frac{\min_{k} \eta_1^{(n_s)}(k)}{\max_{k}\eta_1^{(n_s)}(k)}\biggr\}\goto 0.
\eeq
Since $K$ is fixed, all the entries of $G^{(n_s)}$ are bounded. It follows that there exists a subsequence of $\{n_s\}_{s=1}^\infty$, which we still denote by $\{n_s\}_{s=1}^\infty$ for notation convenience, such that $G^{(n_s)}\goto G^*$ for a fixed matrix $G^*$. Therefore, 
\beq  \label{prop-condcheck-5}
\bigl\|(PG)^{(n_s)}- P_0G^*\bigr\|\to 0, \qquad \mbox{as }s\to\infty.  
\eeq
Let $\eta_1^*$ be the first right eigenvector of $P_0G^*$. 
Since $|\lambda_1(PG)|\geq c+|\lambda_2(PG)|$, by the sin-theta theorem (e.g., see Lemma~\ref{lem:sine-theta}), it follows from \eqref{prop-condcheck-5} that
\beq \label{prop-condcheck-6}
\|\eta_1^{(n_s)} -  \eta_1^*\|\to 0, \qquad \mbox{as }s\to\infty.
\eeq
We now derive a contradiction from \eqref{prop-condcheck-4}-\eqref{prop-condcheck-6}. On the one hand, combining \eqref{prop-condcheck-5}-\eqref{prop-condcheck-6} and noting that $\eta_1^*$ is a fixed vector, we conclude that the minimum entry of $\eta_1^*$ is zero. On the other hand, the assumption of $\min_kG(k,k)\geq c$ ensures that $G^*$ has strictly positive diagonals. We apply \eqref{prop-condcheck-3} to conclude that $P_0G^*$ is a fixed irreducible matrix. By Perron's theorem, $\eta_1^*$ should be a strictly positive vector. This yields a contradiction. 

Consider the third case. The proof is similar to that of the second case, except that we switch the roles of $P$ and $G$. Note that we do not need additional conditions on the diagonals of $P$, since $P$ always has unit diagonals.

Next, we prove Propositions~\ref{prop:condition-check2}. By \eqref{prop-condcheck-1} and \eqref{prop-condcheck-2}, we only need to show that
\[
\max_{k,\ell}\bigl\{(PG)(k,\ell)\bigr\} \asymp \min_{k,\ell}\bigl\{(PG)(k,\ell)\bigr\}. 
\]
Since the maximum row sum and minimum row sum of $P$ are at the same order, it suffices to show that the maximum and minimum entries of $G$ are at the same order. Let $G_0=\mathbb{E}_{\pi\sim \mathrm{Dirichlet}(\alpha)}[\pi\pi']$. As $n\to\infty$, it is easy to show that $\|G-G_0\|_F=o(1)$. Therefore, we only need to show that the maximum and minimum entries of $G_0$ are at the same order. By direct calculations, 
\begin{align*}
G_0 &=(\mathbb{E}[\pi])(\mathbb{E}[\pi])' + \mathrm{Cov}(\Pi)\cr
&= \frac{1}{\|\alpha\|_1^2}\alpha\alpha' + \frac{1}{1+\|\alpha\|_1}\biggl[ \frac{1}{\|\alpha\|_1}\diag(\alpha) - \frac{1}{\|\alpha\|^2_1}\alpha\alpha'\biggr]\cr
&= \frac{1}{\|\alpha\|_1(1+\|\alpha\|_1)}[\diag(\alpha)+\alpha\alpha']. 
\end{align*}
Since all entries of $\alpha$ are bounded above and below by constants, it is easy to see that the maximum and minimum entries of $G_0$ are at the same order. This completes the proof. 
\qed

\section{Faster Rates of Mixed-SCORE (Setting 2)} \label{sec:Setting2}

In Section~\ref{subsec:thm-4versions}, we discuss Mixed-SCORE with each specific VH approach in Table~\ref{tab:CompSVS}. For Mixed-SCORE-SVS and Mixed-SCORE-SVS*, we consider two settings where they enjoy faster rates than the generic Mixed-SCORE algorithm. Due to space limit, we only present Setting 1 in Section~\ref{subsec:thm-4versions}. We now present Setting 1. 

\bigskip
\noindent
{\bf Setting 2}.  Let ${\cal N}_k$ be the set of pure nodes of community $k$,  $1\leq k\leq K$,  
and let ${\cal M}$ be the set of all mixed nodes.  
Suppose there are constants $c_1, c_2\in (0,1)$ such that  $
\min_{1\leq k\leq K} |{\cal N}_k|\geq c_1n$ and  $\min_{1\leq k\leq K}\sum_{i \in {\cal N}_k}  \theta^2(i)  \geq c_2\|\theta\|^2$.  Furthermore, for a fixed integer $L_0 \geq 1$, we assume there is a partition of ${\cal M}$,  ${\cal M}={\cal M}_1\cup \cdots\cup{\cal M}_{L_0}$,  a set of PMF's $\gamma_1,\cdots, \gamma_{L_0}$, and constants   $c_3, c_4>0$  such that  ($e_k$:  $k$-th standard basis vector  of $\mathbb{R}^{K}$)  
$\big\{ \min_{1\leq j\neq \ell\leq L_0} \|\gamma_j-\gamma_\ell\|, \; \min_{1\leq \ell\leq L_0, 1\leq k\leq K } \|\gamma_\ell -e_k\| \big\}\geq c_3$, 
and for each $1 \leq \ell \leq L_0$ (note: $err_n$ is the same as that in \eqref{def:err}),  
$|{\cal M}_\ell| \geq c_4 |{\cal M}|\geq n\beta_n^{-2}err_n^2$ and $\max_{i\in{\cal M}_\ell}\|\pi_i - \gamma_\ell\| \leq 1/\log(n)$.   

\bigskip

In this setting, $\pi_i$'s form several {\it loose clusters}, where the $\pi_i$'s in the same cluster are within a distance of $O(\frac{1}{\log(n)})$ from each other. Since $\frac{1}{\log(n)}$ is much larger than the order of noise, $ \max_{1\leq i\leq n}\|H\hat{r}_i-r_i\|$,  the assumed clustering structure is indeed ``loose". \spacingset{1}\footnote{In fact, by a slight modification of the proof, we can replace $(1/\log(n))$ in Setting 2 by any $o(1)$ term, or an appropriately small constant $\tilde{c}_3>0$ (this constant $\tilde{c}_3$ will depend on the constants in Setting 2 in a quite complicated way). We present the current version for its convenience.} \spacingset{1.45}

\begin{thm} \label{thm:SVS2}  
Consider the DCMM model where Assumptions 1-4 hold and $\pi_i$'s are from Setting 2. Let $H$ be as in Theorem~\ref{thm:r-entrywise-bound}.  Suppose we apply SVS or SVS* to rows of $\hat{R}$ with 
\[
L=\hat{L}_n(A) := \min\bigl\{L\geq K+1:  \epsilon_{L}(\hat{R})<\epsilon_{L-1}(\hat{R})/\log(\log(n)) \bigr\}.
\]
With probability $1-o(n^{-3})$, 
\[
\max_{1\leq k\leq K}\|H\hat{v}_k - v_k\|\leq C \sqrt{n^{-1}\sum_{i=1}^n\|H\hat{r}_i - r_i\|^2}.
\]
Moreover, for Mixed-SCORE-SVS or Mixed-SCORE-SVS*, 
\[
\mathbb{E}\Bigl[\frac{1}{n} \sum_{i = 1}^n \|\hat{\pi}_i - \pi_i\|^2\Bigr] \leq CK^3 \beta_n^{-2}(err^*_n)^2+o(n^{-2}).
\] 
\end{thm}

\section{The Oracle Case and Ideal Mixed-SCORE}
We consider the oracle case where $\Omega$ is observed. In Section~\ref{subsec:prelim}, we state a useful lemma, which is the key for analysis of the oracle case. 
In Section~\ref{subsec:lemSec2proof}, we prove Lemmas~\ref{lem:ideal} in the paper, which inspire Ideal Mixed-SCORE. 
In Section~\ref{subsec:lemRefitproof}, we prove Lemma~\ref{lem:refitting}, which is about recovering $(P,\theta)$ from $\Pi$.  
In Section~\ref{subsec:app-Omega}, we study eigenvalues and eigenvectors of $\Omega$ and the matrix $R$; these results are useful for the proofs in Sections~\ref{sec:app-RealCase}-\ref{sec:app-Rates}.

\subsection{A useful lemma and its proof} \label{subsec:prelim}
Let $G=K\|\theta\|^{-2}(\Pi'\Theta^2\Pi)$ is as in Section~\ref{sec:mainthm}. Let  $\lambda_1,\lambda_2,\ldots,\lambda_K$ be the nonzero eigenvalues of $\Omega$, sorted in the descending order of magnitudes. Let $\xi_1,\xi_2,\ldots,\xi_K$ be the corresponding eigenvectors. We have the following lemma: 
\begin{lemma} \label{lem:B}
Consider the DCMM model, where $PG$ is an irreducible matrix and there is at least one pure node for each community. The following statements are true:
\begin{itemize} \itemsep +2pt
\item There is a non-singular matrix $B\in\mathbb{R}^{K,K}$ such that $\Theta\Pi B=\Xi$, and $B$ is unique once $\Xi$ is chosen. 
\item For $1\leq k\leq K$, denote by $a_k$ the $k$th largest (in magnitude) eigenvalue of $PG$. Then, $a_k$'s are real, and the nonzero eigenvalues of $\Omega$ are $\lambda_k=(K^{-1}\|\theta\|^2)a_k$, $1\leq k\leq K$. 
\item For $1\leq k\leq K$, denote by $b_k$ the $k$th column of $B$. Then, $b_k$ is a (right) eigenvector of $PG$ associated with $a_k$. 
\item $\lambda_1>0$ and it has a multiplicity $1$ (so $\xi_1$ is uniquely determined up to a factor of $\pm 1$). 
\item $\xi_1$ can be chosen such that all of its entries are positive. For this choice of $\xi_1$, all the entries of the associated $b_1$ are also positive. 
\end{itemize}
\end{lemma}

\noindent{\it Proof of Lemma~\ref{lem:B}}: Consider the first claim. Denote by $Span(M)$ the column space of any matrix $M$. It suffices to show that $Span(\Theta\Pi)=Span(\Xi)$. 
Then, since $\xi_1,\cdots,\xi_K$ form an orthonormal basis of this subspace, there is a unique, non-singular matrix $\tilde{B}$ such that $\Theta\Pi=\Xi\tilde{B}$. We then take $B=\tilde{B}^{-1}$. 

We now show $Span(\Theta\Pi)=Span(\Xi)$. 
By the assumption that there is at least one pure node in each community, we can find $K$ rows of $\Pi$ such that they form a $K\times K$ identity matrix. So $\Pi$ has a rank $K$. Since $\Theta$ and $P$ are both non-singular matrices, $\Omega$ also has a rank $K$. 
By definition, $\Omega\xi_k=\lambda_k\xi_k$, for $1\leq k\leq K$. It follows that
\[
\Theta\Pi (P \Pi'\Theta \xi_k) = \lambda_k \xi_k.  
\]
Hence, each $\xi_k$ is in the column space of $\Theta\Pi$. This means the column space of $\Xi$ is contained in the column space of $\Theta\Pi$. Since both matrices have a rank $K$, the two column spaces are the same. 

Consider the second claim. Note that $P$ is symmetric and $G$ is positive definite. Let $G^{1/2}$ be the unique square root of $G$. For any matrices $A\in\mathbb{R}^{m,n}$ and $B\in\mathbb{R}^{n,m}$, if $m\geq n$, then the nonzero eigenvalues of $AB$ are the same as the nonzero eigenvalues of $BA$ \cite[Theorem 1.3.22]{HornJohnson}.
As a result, eigenvalues of $PG$ are the same as eigenvalues of the symmetric matrix $G^{1/2}PG^{1/2}$. It implies that $a_1,a_2,\ldots,a_K$ are real. 
 
Furthermore, the nonzero eigenvalues of $\Omega=(\Theta\Pi)(P \Pi'\Theta)$ are the same as the nonzero eigenvalues of $(P \Pi'\Theta)(\Theta\Pi)=(K^{-1}\|\theta\|^2) (PG)$. Hence, the nonzero eigenvalues of $\Omega$ are
$(K^{-1}\|\theta\|^2)a_1, (K^{-1}\|\theta\|^2)a_2,\ldots, (K^{-1}\|\theta\|^2)a_K$.

Consider the third claim. Write $\tilde{G}\equiv K^{-1}\|\theta\|^2 G=\Pi'\Theta^2\Pi$. Note that $\Omega\xi_k=\lambda_k\xi_k$ and $\xi_k=\Theta\Pi b_k$. Hence, $(\Theta\Pi P \Pi'\Theta)(\Theta\Pi b_k) = \lambda_k(\Theta\Pi b_k)$. Multiplying both sides by $\Pi'\Theta$ from the left, we have
\[
\tilde{G}P\tilde{G}b_k = \lambda_k \tilde{G}b_k. 
\]
Since $\tilde{G}$ is non-singular, $P\tilde{G}b_k=\lambda_k b_k$. Plugging in $\tilde{G}=(K^{-1}\|\theta\|^{2})G$ and $\lambda_k=(K^{-1}\|\theta\|^2)a_k$, we obtain $PGb_k=a_kb_k$. This shows that $b_k$ is a (right) eigenvector of $PG$ associated with $a_k$. Additionally, since $\eta_1$ is the first unit-norm right singular vector of $PG$, it yields that $\eta_1=b_1/\|b_1\|$.  


Consider the fourth claim. Since $\lambda_1=(K^{-1}\|\theta\|^2)a_1$, it suffices to show that $a_1>0$ and that it has a multiplicity $1$. This follows immediately from the Perron-Frobenius theorem \cite[Theorem 8.4.4]{HornJohnson} and the assumption that $PG$ is an irreducible matrix. 


Consider the last claim. Note that $b_1$ is the eigenvalue of $PG$ associated with $a_1$. Since $a_1$ has a multiplicity $1$, $b_1/\|b_1\|$ is unique up to a factor of $\pm 1$ (depending on the choice of $\xi_1$). By Perron-Frobenius theorem again, $b_1/\|b_1\|$ can be chosen such that all the entries are positive. Recalling that $\Xi=\Theta \Pi B$, we immediately have $\xi_1=\Theta\Pi b_1$. 
Since $\Theta\Pi$ is a nonnegative matrix with positive row sums and $b_1$ has strictly positive entries, all the entries of $\xi_1$ are also positive. \qed

\subsection{Proofs of Lemma~\ref{lem:ideal}} \label{subsec:lemSec2proof}
Consider the first claim. We have shown in  Lemma~\ref{lem:B} that
\[
\Xi = \Theta \Pi B, \qquad \mbox{for a non-singular matrix }B=[b_1,\ldots,b_K]\in\mathbb{R}^{K,K}. 
\]
Furthermore, by the last two bullet points of Lemma~\ref{lem:B}, if we pick the sign of $\xi_1$ such that $\sum_{i=1}^n\xi_1(i)>0$, then $\xi_1$ and $b_1$ are uniquely determined and have strictly positive entries. 
This proves the first claim. 

Consider the other two claims. 
We first show there are $K$ affinely independent vectors $v_1,v_2,\ldots,v_K$ such that each $r_i$ is a convex combination of them. For $1\leq k\leq K$, define $v_k\in\mathbb{R}^{K-1}$ by 
\beq \label{DefineV}
v_k(\ell) = b_{\ell+1}(k)/b_1(k), \qquad 1\leq \ell\leq K-1. 
\eeq
The vectors $v_1,v_2,\ldots,v_K$ are affinely independent, if and only if the following matrix 
\[
Q = \begin{pmatrix} 1 &\cdots & 1\\v_1 & \cdots & v_K\end{pmatrix} 
\]
is non-singular. By \eqref{DefineV}, we observe that $Q'=\diag(b_1)B$. Since $B$ is non-singular and $b_1$ is a positive vector, $Q$ has to be a non-singular matrix. This proves that $v_1,v_2,\ldots,v_K$ are affinely independent. We then study each $r_i$. 
Since $\Xi =\Theta\Pi B$, we have 
\[
\xi_\ell(i)=\theta(i)\sum_{k=1}^K\pi_i(k)b_\ell(k)=\theta(i)\|b_\ell\circ \pi_i\|_1,  \qquad 1\leq \ell\leq K.
\]
By definition of $R$, $r_i(\ell)=\xi_{\ell+1}(i)/\xi_1(i)$. It follows that 
\begin{align*}
r_i(\ell) = \frac{\theta(i)\sum_{k=1}^K\pi_i(k)b_{\ell+1}(k)}{\theta(i)\|b_1\circ \pi_i\|_1}= \sum_{k=1}^K\frac{b_1(k)\pi_i(k)}{\|b_1\circ \pi_i\|_1}\cdot \frac{b_{\ell+1}(k)}{b_1(k)} = \sum_{k=1}^K w_i(k) v_k(\ell). 
\end{align*}
This proves that $r_i=\sum_{k=1}^K w_i(k)v_k$, with $w_i=(b_1\circ \pi_i)/\|b_1\circ \pi_i\|_1$. Since $b_1$ is a positive vector and $\pi_i$ is a nonnegative vector, we have that $w_i$ is a nonnegative vector and $\|w_i\|_1=1$. Therefore, $r_i$ is a convex combination of $v_1,v_2,\ldots,v_K$.  

We now show the second claim. Each $r_i$ is in the convex hull of $v_1,v_2,\ldots,v_K$. Since these $K$ vectors are affinely independent, their convex hull is a non-degenerate simplex with $K$ vertices. Recall that $w_i=(b_1\circ \pi_i)/\|b_1\circ \pi_i\|_1$, where $b_1$ is a strictly positive vector. Therefore, for each $1\leq k\leq K$, node $i$ is a pure node of community $k$ if and only if $\pi_i=e_k$, which happens if and only if $w_i=e_k$; and $w_i=e_k$ means $r_i$ is located at the vertex $v_k$. 

We then show the last claim, which is the formula for $b_1$. 
Write $\Lambda=\mathrm{diag}(\lambda_1,\cdots,\lambda_K)$. Then, $\Omega=\Xi\Lambda\Xi'$. First, plugging in $\Xi=\Theta\Pi B$, we find that $\Omega=\Theta\Pi (B\Lambda B')\Pi'\Theta$. Multiplying both sides by $\Pi'\Theta$ from the left and $\Theta\Pi$ from the right, we have $\Pi' \Theta \Omega\Theta \Pi=\tilde{G} (B\Lambda B')\tilde{G}$, where $\tilde{G}=\Pi'\Theta^2\Pi$ is a non-singular matrix. Second, since $\Omega=\Theta\Pi P\Pi'\Theta'$, we have $\Pi' \Theta \Omega\Theta \Pi=\tilde{G}P\tilde{G}$. Combining the above gives
\beq \label{proof-refitting-add}
\tilde{G}P\tilde{G} = \tilde{G} (B\Lambda B')\tilde{G} \quad \Longrightarrow \quad P = B\Lambda B'. 
\eeq
It follows that 
\[
1=P(k,k)=\sum_{\ell=1}^K \lambda_\ell b_\ell^2(k)=b^2_1(k)[\lambda_1 + \sum_{\ell=2}^K \lambda^2_\ell v_k (\ell-1)].
\]
Noting that $b_1(k)$ is positive, the above gives the formula for computing $b_1$. \qed

\subsection{Proof of Lemma~\ref{lem:refitting}} \label{subsec:lemRefitproof} 
Write $V=[v_1,v_2,\ldots,v_K]$. By \eqref{DefineV}, $B=\diag(b_1)[1, V']$. Moreover, by \eqref{proof-refitting-add}, $P=B\Lambda B'$. 
Combining them gives the formula of recovering $P$. Note that $\Xi=\Theta\Pi B$. It follows that $\xi_1(i) = \theta(i)\cdot \pi_i'b_1$. This gives the formula of recovering $\theta$. \qed

\subsection{Spectral analysis of $\Omega$} \label{subsec:app-Omega}
First, we study the leading eigenvalues of $\Omega$.  
Let $\lambda_1,  \ldots, \lambda_K$  be the nonzero eigenvalues of $\Omega$, listed in the descending order in magnitude. 
The following lemma is proved in Section~\ref{subsubsec:lem-eigenOmega}:
\begin{lemma} \label{lem:eigenOmega}
Under conditions of Theorem~\ref{thm:r-entrywise-bound}, the following statements are true:
\begin{itemize} \itemsep +3pt
\item $C^{-1}K^{-1}\|\theta\|^2\leq \lambda_1\leq C\|\theta\|^2$. If $\beta_n=o(1)$, then $\lambda_1\asymp \|\theta\|^2$. 
\item $\lambda_1-|\lambda_2|\asymp \lambda_1$.
\item $|\lambda_k|\asymp \beta_nK^{-1}\|\theta\|^2$, for $2\leq k\leq K$. 
\end{itemize}   
\end{lemma}

Next, we study the leading eigenvectors of $\Omega$. For $1\leq k\leq K$, let $\xi_k$ be the eigenvector of $\Omega$ associated with $\lambda_k$. 
Write $\Xi_0=[\xi_2,\xi_3,\cdots,\xi_K]\in\mathbb{R}^{n,K-1}$, and let $\Xi'_{0,i}$ be its $i$-th row, $1\leq i\leq n$.   
The following lemma is proved in Section~\ref{subsubsec:lem-xi1Omega}:
\begin{lemma} \label{lem:xi1}
Under conditions of Theorem~\ref{thm:r-entrywise-bound}, the following statements are true:
\begin{itemize} \itemsep +3pt
\item If we choose the sign of $\xi_1$ such that $\sum_{i=1}^n \xi_1(i)>0$, then the entries of $\xi_1$ are positive satisfying $C^{-1}\theta(i)/\|\theta\|\leq \xi_1(i)\leq C\theta(i)/\|\theta\|$, $1\leq i\leq n$.
\item $\|\Xi_{0,i}\|\leq C\sqrt{K}\theta(i)/\|\theta\|$, $1\leq i\leq n$. 
\end{itemize} 
\end{lemma}

Last, we study the entry-wise ratio matrix $R$. Recall that $w_i$ is the barycentric coordinate vector of $r_i$ in the Ideal Simplex. The following lemma is proved in Section~\ref{subsubsec:lem-R}. 
\begin{lemma} \label{lem:R}
Under conditions of Theorem~\ref{thm:r-entrywise-bound}, the following statements are true: 
\begin{itemize} \itemsep +3pt
\item The vertices of the Ideal Simplex satisfy that $\max_{1\leq k\leq K}\|v_k\|\leq C\sqrt{K}$ and $\min_{k\neq \ell}\|v_k-v_\ell\|\geq C^{-1}\sqrt{K}$. 
\item $C^{-1}\|\pi_i-\pi_j\|_1\leq \|w_i-w_j\|_1\leq C\|\pi_i-\pi_j\|_1$, for all $1\leq i,j\leq n$. 
\item $C^{-1}\sqrt{K}\|w_i-w_j\|\leq \|r_i-r_j\|\leq C\sqrt{K}\|w_i-w_j\|$, for all $1\leq i, j\leq n$. 
\end{itemize}
\end{lemma}

Lemmas~\ref{lem:eigenOmega}-\ref{lem:R} are useful for proofs in Sections~\ref{sec:app-RealCase}-\ref{sec:app-Rates}. Below, we prove these lemmas.

\subsubsection{Proof of Lemma~\ref{lem:eigenOmega}} \label{subsubsec:lem-eigenOmega}
By Lemma~\ref{lem:B}, all nonzero eigenvalues of $\Omega$ are $(K^{-1}\|\theta\|^2)a_1,\ldots, (K^{-1}\|\theta\|^2)a_K$, where $a_k$ is the $k$-th largest eigenvalue (in magnitude) of $PG$. By Assumption 3,
\[
a_1-|a_2|\geq C^{-1} a_1, \qquad C^{-1}\beta_n\leq |a_K|\leq |a_2|\leq C\beta_n. 
\]
The second and third claims follow immediately. 

It remains to show the first claim, which reduces to studying $a_1$. For any two matrices $A$ and $B$, the nonzero eigenvalues of $AB$ are the same as the nonzero eigenvalues of $BA$. Hence, 
\[
a_1=\lambda_1(PG)=\lambda_1(G^{1/2}PG^{1/2})=\max_{x\neq 0}\frac{x'G^{1/2}PG^{1/2}x}{\|x\|^2}.
\] 
By Assumption 2, $\|G\|\leq C$ and $\|G^{-1}\|\leq C$.
It is easy to see that $a_1\leq C\lambda_1(P)$. 
Additionally, $\lambda_1(P)=\max_{y\neq 0}\frac{y'Py}{\|y\|^2}=\max_{x\neq 0}\frac{x'G^{1/2}PG^{1/2}x}{\|G^{1/2}x\|^2}$. Since $\|G^{1/2}x\|^2=x'Gx\geq C^{-1}\|x\|^2$, it follows that $\lambda_1(P)\leq \max_{x\neq 0}\frac{x'G^{1/2}PG^{1/2}x}{C^{-1}\|x\|^2}\leq C\lambda_1(PG)$. Together,
\[
C^{-1}\lambda_1(P)\leq \lambda_1(PG)\leq C\lambda_1(P). 
\]
Note that $\lambda_1(P)\leq K\|P\|_{\max}=O(K)$ and $\lambda_1(P)\geq P(k,k)\geq 1$. We plug them into the above inequality to get
\beq \label{lemPopEigVal-1}
C^{-1} \leq a_1\leq CK. 
\eeq
This inequality holds in all cases. If, additionally, $\beta_n\to 0$ as $n\to\infty$, we can get a stronger result. Note that $P$ and $G$ are nonnegative matrices, and for each $1\leq k\leq K$, $P(k,k)=1$ and $G(k,k)\geq \lambda_{\min}(G)\geq C^{-1}$. It follows that $(PG)(k,k)\geq P(k,k)G(k,k)\geq C^{-1}$. We thus have 
\[
\mathrm{trace}(PG)\geq C^{-1}K.
\]
At the same time, 
$\mathrm{trace}(PG)=a_1+\sum_{k=2}^Ka_2=a_1+O(K\beta_n)=a_1+o(K)$. It follows that 
\beq \label{lemPopEigVal-2}
C^{-1}K \leq a_1\leq CK, \qquad \mbox{if }\beta_n = o(1). 
\eeq
The first claim follows from \eqref{lemPopEigVal-1}-\eqref{lemPopEigVal-2} and the equality $\lambda_1=(K^{-1}\|\theta\|^2)a_1$. 
\qed

\subsubsection{Proof of Lemma~\ref{lem:xi1}} \label{subsubsec:lem-xi1Omega}
Consider the first claim. From the last item of Lemma~\ref{lem:B}, we can choose the sign of $\xi_1$ such that both $(\xi_1, b_1)$ have strictly positive entries, where this choice of sign corresponds to $\sum_{i=1}^n\xi_1(i)>0$. 
Note that $\Xi=\Theta\Pi B$, which implies $\xi_1(i)=\theta(i)\sum_{k=1}^K\pi_i(k)b_1(k)$. Since each $\pi_i$ is a PMF (a nonnegative vector whose entries sum to 1), 
\[
\theta(i)\min_{1\leq k\leq K} b_1(k) \leq \xi_1(i) \leq \theta(i)\max_{1\leq k\leq K}b_1(k), \qquad 1\leq i\leq n. 
\]
Hence, to show the claim, it suffices to show that
\beq \label{lem-xi1-0}
C^{-1}\|\theta\|^{-1}\leq b_1(k)\leq C\|\theta\|^{-1}, \qquad \mbox{for all $1\leq k\leq K$}. 
\eeq
Write $\tilde{G} = K^{-1}\|\theta\|^2 G =\Pi'\Theta^2\Pi$. Since $\Xi=\Theta\Pi B$ and $X'X=I_K$, we have $B'\Pi'\Theta^2\Pi B= I_K$, or equivalently, $B'\tilde{G}B =I_K$. Multiplying both sides by $B$ from the left and $B'$ from the right, we obtain $BB'\tilde{G}BB'=BB'$. Since $BB'$ is non-singular, it implies
\beq \label{BB'}
BB' = \tilde{G}^{-1}=K\|\theta\|^{-2} G^{-1}. 
\eeq
We note that $BB'=\sum_{k=1}^K b_kb_k'\succeq b_1b_1'$. So, $\|b_1\|^2\leq \|B\|^2\leq K\|\theta\|^{-2}\|G^{-1}\|$. By our assumption of $\|G^{-1}\|\leq C$. It follows that
\[
\|b_1\|\leq C\|\theta\|^{-1}\sqrt{K}. 
\]  
At the same time, $1 = \|\xi_1\|^2=\|\Theta\Pi b_1\|^2$. By direct calculations, $\|\Theta\Pi b_1\|^2=\sum_i\theta_i^2(\pi_i'b_1)^2\leq \sum_i\theta_i^2 \|b_1\|_\infty^2\leq \|\theta\|^2\|b_1\|_\infty^2$. It follows that
\[
\|b_1\|_\infty \geq C^{-1}\|\theta\|^{-1}. 
\] 
In Lemma~\ref{lem:B}, we have seen that $b_1$ is the first right singular vector of $PG$. Hence, $b_1\propto\eta_1$, where $\eta_1$ is the same as in Assumption 4. By Assumption 4, all the entries of $\eta_1$ are at the same order. Hence, all the entries of $b_1$ are at the same order. It follows that 
\[
b_1(k)\asymp \|b_1\|_\infty\asymp (1/\sqrt{K})\|b_1\|.
\]
This gives \eqref{lem-xi1-0} and completes the proof of the first claim.

Consider the second claim. Since $\Xi=\Theta\Pi B$, for $1\leq i\leq n$, 
\[
\|\Xi_{0,i}\|\leq \theta(i) \|B\pi_i \|\leq C\theta(i) \sqrt{\lambda_{\max}(B'B)}\leq C\sqrt{K}\|\theta\|^{-1}\theta(i),
\]
where the last inequality is due to \eqref{BB'} and and the condition $\|G^{-1}\|\leq C$.  \qed

\subsubsection{Proof of Lemma~\ref{lem:R}} \label{subsubsec:lem-R}
First, we prove the claim about the connection between $\|w_i-w_j\|_1$ and $\|\pi_i-\pi_j\|_1$. 
Let ${\cal S}_0\subset\mathbb{R}^K$ be the standard simplex whose vertices are $e_1,e_2,\ldots,e_K$. Define a mapping
\[
T_1: {\cal S}_0\to {\cal S}_0, \qquad \mbox{where}\quad T_1(x) = \frac{x\circ b_1}{\|x\circ b_1\|_1}. 
\]
Then, $w_i=T_1(\pi_i)$, for $1\leq i\leq n$. To show the claim, it suffices to
show that $T_1$ and $T_1^{-1}$ are both Lipschitz with respect to the $\ell^1$-norm, i.e., for any $x,y\in {\cal S}_0$, 
\beq \label{lemR-goal1}
C^{-1}\|x-y\|_1\leq \|T_1(x)-T_1(y)\|_1\leq C\|x-y\|_1. 
\eeq 
We now show \eqref{lemR-goal1}. Fixing any $x,y\in {\cal S}_0$, write $x^*=T_1(x)$ and $y^*=T_1(y)$. By definition, $x^*(k)=x(k)b_1(k)/\|x\circ b_1\|_1$ and $y^*(k)=y(k)b_1(k)/\|y\circ b_1\|_1$.
We write
\begin{align*}
x^*(k) -& y^*(k) = \frac{[x(k)-y(k)]b_1(k)}{\|x\circ b_1\|_1} + y(k)b_1(k) \left[\frac{1}{\|x\circ b_1\|_1} - \frac{1}{\|y\circ b_1\|_1}\right]\cr
&=  \frac{b_1(k)}{\|x\circ b_1\|_1}[x(k)-y(k)] + \frac{y^*(k)}{\|x\circ b_1\|_1} \left(\|y\circ b_1\|_1 - \|x\circ b_1\|_1\right). 
\end{align*}
First, by \eqref{lem-xi1-0}, $b_1(k)\asymp \|\theta\|^{-1}$ for all $1\leq k\leq K$. It follows that $|b_1(k)|\leq C\|\theta\|^{-1}$ and $\|x\circ b_1\|_1\geq \|x\|_1\cdot C^{-1}\|\theta\|^{-1}\geq C^{-1}\|\theta\|^{-1}$. Hence,
\[
\frac{b_1(k)}{\|x\circ b_1\|_1}|x(k)-y(k)|\leq C|x(k)-y(k)|. 
\]
Second, by the triangle inequality, $|\|y\circ b_1\|_1 - \|x\circ b_1\|_1|\leq \|(y-x)\circ b_1\|_1$. Moreover, since $b_1(k)\asymp \|\theta\|^{-1}$ for all $k$, we have $\|(y-x)\circ b_1\|_1\leq C\|\theta\|^{-1}\|x-y\|_1$ and $\|x\circ b_1\|_1\geq C^{-1}\|\theta\|^{-1}$. It follows that
\[
\frac{y^*(k)}{\|x\circ b_1\|_1} \left|\|y\circ b_1\|_1 - \|x\circ b_1\|_1\right|\leq Cy^*(k)\cdot \|x-y\|_1. 
\]
Combining the above gives
\[
|x^*(k)-y^*(k)|\leq C|x(k)-y(k)| + Cy^*(k)\cdot\|x-y\|_1. 
\]
We sum over $k$ on both sides and note that $\sum_ky^*(k)=1$. It gives
\[
\|x^*-y^*\|_1\leq C\|x-y\|_1. 
\] 
This shows that $T_1$ is Lipschitz with respect to the $\ell^1$-norm. We then consider $T_1^{-1}$. Define $\tilde{b}_1\in\mathbb{R}^K$ by $\tilde{b}_1(k)=1/b_1(k)$, $1\leq k\leq K$. We can rewrite
\[
T_1^{-1}(x) = \frac{x\circ \tilde{b}_1}{\|x\circ \tilde{b}_1\|_1}. 
\]
 $T_1^{-1}$ has a similar form as $T_1$, where the vector $\tilde{b}_1$ satisfies that $\tilde{b}_1(k)\asymp \|\theta\|$ for all $k$. Hence, we can similarly prove that $T_1^{-1}$ is Lipschitz with respect to the $\ell^1$-norm. This proves \eqref{lemR-goal1}.
 
Next, we prove the claim about the connection between $\|r_i-r_j\|$ and $\|w_i-w_j\|$. Let ${\cal S}_0$ be the same as before, and let ${\cal S}^{ideal}={\cal S}^{ideal}(v_1,v_2,\ldots,v_K)\subset\mathbb{R}^{K-1}$ denote the Ideal Simplex. Let $B=[b_1,b_2,\ldots,b_K]$ be as in Lemma~\ref{lem:B}. Define a mapping:
\[
T_2: {\cal S}_0\to {\cal S}^{ideal}, \qquad \mbox{where}\quad 
\begin{pmatrix} 1\\ T_2(x) \end{pmatrix} = \underbrace{\begin{pmatrix}1 & \cdots & 1\\
v_1 & \cdots & v_K\end{pmatrix}}_{\equiv Q}x. 
\]
By Lemma~\ref{lem:ideal}, $r_i=T_2(w_i)$, for all $1\leq i\leq n$. To show the claim, it suffices to
show that $T_2$ and $T_2^{-1}$ are both Lipschitz with respect to the $\ell^2$-norm, whose Lipschitz constants are $\sqrt{K}$ and $1/\sqrt{K}$, respectively. In other words, we want to prove, for any $x,y\in {\cal S}_0$, 
\beq \label{lemR-goal2}
C^{-1}\sqrt{K}\|x-y\|\leq \|T_2(x)-T_2(y)\|\leq C\sqrt{K}\|x-y\|. 
\eeq 
We now show \eqref{lemR-goal2}. Note that $Qx=(1_K'x, T_2(x))'$. Since $1_K'x=1_K'y=1$, we have 
\[
\|T_2(x)-T_2(y)\|^2 = \|Qx-Qy\|^2=(x-y)'Q'Q(x-y). 
\]
It suffices to show that 
\beq \label{Q-eigenvalues}
\|Q\|\leq C\sqrt{K}, \qquad\mbox{and}\qquad \|Q^{-1}\|\leq C/\sqrt{K}. 
\eeq
By \eqref{DefineV}, we can re-write
\[
Q' = [\diag(b_1)]^{-1}B. 
\]
By \eqref{lem-xi1-0}, $b_1(k)\asymp \|\theta\|^{-1}$ for all $k$. By \eqref{BB'}, $BB'=K\|\theta\|^{-2}G^{-1}$; we note that by Assumption 2, $\|G\|\leq C$ and $\|G^{-1}\|\leq C$; it follows that $\|B\|\leq C\sqrt{K}\|\theta\|^{-1}$ and $\|B^{-1}\|\leq C\|\theta\|/\sqrt{K}$. Combining them gives \eqref{Q-eigenvalues}. Then, \eqref{lemR-goal2} follows.  

Last, we prove the claims about the Ideal Simplex (IS). Let $e_1,e_2,\ldots,e_K$ be the standard basis vectors of $\mathbb{R}^K$. It is seen that $v_k=T_2(e_k)$, $1\leq k\leq K$. By \eqref{lemR-goal2}, for $k\neq \ell$, 
\[
\|v_k-v_\ell\|\asymp \sqrt{K}\|e_k-e_\ell\|\asymp \sqrt{K}. 
\]
By definition of $Q$ and \eqref{Q-eigenvalues}, for all $1\leq k\leq K$, 
\[
\|v_k\|\leq \|Q\| = O(\sqrt{K}). 
\]
The above give the desired claims.  \qed

\section{Spectral Analysis of $A$ and Large-deviation Bounds for $\hat{R}$} \label{sec:app-RealCase}
We conduct spectral analysis for $A$.  
In Section~\ref{subsec:app-A}, we give the large deviation bounds for eigenvalues of $A$. In Sections~\ref{subsec:app-lem21proof}, we study the eigenvectors of $A$ and state a key technical lemma. In Section~\ref{subsec:app-thm21proof}, we prove Theorem~\ref{thm:r-entrywise-bound} in the paper, which is about the row-wise large deviation bound for $\hat{R}$. In Section~\ref{subsec:hatR-l2}, we give the $\ell^2$-norm large deviation bound for $\hat{R}$.  In Section~\ref{subsec:hatR-l2}, we give a useful property of the rotation matrix $H$.

\subsection{The eigenvalues of $A$} \label{subsec:app-A}  
Let $\hat{\lambda}_1, \hat{\lambda}_2,\ldots, \hat{\lambda}_K$ be the $K$ largest eigenvalues of $A$ (in magnitude), sorted descendingly in magnitude. 


\begin{lemma} \label{lem:Aeigenvals}
Under conditions of Theorem~\ref{thm:r-entrywise-bound}, with probability $1-o(n^{-3})$, $
\max_{1\leq k\leq K}|\hat{\lambda}_k -\lambda_k|\leq C \sqrt{\theta_{\max}\|\theta\|_1}$. 
\end{lemma}

\noindent{\it Proof of Lemma~\ref{lem:Aeigenvals}}:
By Weyl's inequality, $\max_{1\leq k\leq K}|\hat{\lambda}_k - \lambda_k|\leq \|A -\Omega\|$. 
To show the claim, it suffices to show that 
with probability $1-o(n^{-3})$,
\beq  \label{lem-adjacency-3}
\|A - \Omega\|\leq C\sqrt{\theta_{\max}\|\theta\|_1}. 
\eeq
The following inequality is useful:
\beq \label{err-order}
(\theta_{\max}\|\theta\|_1)/\log(n)\to\infty. 
\eeq
To see why \eqref{err-order} is true, we rewrite $err_n=(\theta_{\max}/\theta_{\min})\|\theta\|^{-2}\sqrt{\theta_{\max}\|\theta\|_1\log(n)}$. Since $\theta_{\max}\geq\theta_{\min}$ and $\theta_{\max}\|\theta\|_1\geq \|\theta\|^2$, we immediately have $err_n\geq \|\theta\|^{-1}\sqrt{\log(n)}$. Therefore, the assumption $err_n\to 0$ implies that $\|\theta\|^2/\log(n)\to\infty$. Then \eqref{err-order} is also true because $\theta_{\max}\|\theta\|_1\geq \|\theta\|^2$.

We now prove \eqref{lem-adjacency-3}. 
Write  
\[
A - \Omega  = W + \mathrm{diag}(\Omega), \qquad \mbox{where }\;  W \equiv A - E[A]. 
\]
Note that $\pi_i'P\pi_j=\sum_{k,\ell}\pi_i(k)\pi_j(\ell)P_{k\ell}\leq \|P\|_{\max}\|\pi_i\|_1\|\pi_j\|_1\leq C$. It follows that 
\[
\Omega(i,j)\leq C\theta(i)\theta(j). 
\]
Note that $\Omega(i,i)=\theta^2(i)(\pi_i'P\pi_i)\leq C\theta^2(i)$. As a result, 
\beq  \label{lem-adjacency-1}
\|\mathrm{diag}(\Omega)\|\leq C\theta^2_{\max}\leq C\sqrt{\theta_{\max}\|\theta\|_1},
\eeq
where the last inequality follows from \eqref{err-order} and $\theta_{\max}^2\leq C\ll\sqrt{\log(n)}$ . We then apply the non-asymptotic bounds for random matrices in \cite{bandeira2016sharp} to bound $\|W\|$. By Corollary 3.12 and Remark 3.13 of \cite{bandeira2016sharp}, for the $n\times n$ symmetric matrix $W$ whose upper triangle contains independent entries, for any $\epsilon>0$, there exists a universal constant $\tilde{c}_\epsilon>0$ such that for every $t\geq 0$, 
\beq \label{sharpRMTnorm}
\mathbb{P}\bigl( \|W\|> (1+\epsilon)2\sqrt{2}\tilde{\sigma} +t \bigr)\leq ne^{-t^2/(\tilde{c}\tilde{\sigma}_*^2)},
\eeq
where
\[
\tilde{\sigma} = \max_i\sqrt{\sum_j \mathbb{E}[W(i,j)^2]}, \qquad \tilde{\sigma}_*=\max_{ij}\|W(i,j)\|_\infty. 
\]
We fix $\epsilon=1/2$ in \eqref{sharpRMTnorm} and write $\tilde{c}=\tilde{c}_{\epsilon}$ for short. For $t=2\tilde{\sigma}_*\sqrt{\tilde{c}\log(n)}$, it follows from \eqref{sharpRMTnorm} that with probability $1-o(n^{-3})$,
\[
\|W\| \leq 3\sqrt{2}\max_i\sqrt{\sum_j \mathbb{E}[W(i,j)^2]} +C\tilde{\sigma}_*\sqrt{\log(n)}. 
\]
Note that $\tilde{\sigma}_*\leq 1$ and $\max_i \{\sum_j \mathbb{E}[W(i,j)^2]\}\leq \max_i \{ \sum_j\Omega(i,j) \} \leq C\max_i\{\sum_{j}\theta(i)\theta(j)\}\leq C\theta_{\max}\|\theta\|_1$. We plug them into the above inequality and apply \eqref{err-order}. It follows that, with probability $1-o(n^{-3})$, 
\beq  \label{lem-adjacency-2}
\|W\| \leq C\sqrt{\theta_{\max}\|\theta\|_1} + C\sqrt{\log(n)}\leq C\sqrt{\theta_{\max}\|\theta\|_1}.   
\eeq
Combining \eqref{lem-adjacency-1} and \eqref{lem-adjacency-2} gives \eqref{lem-adjacency-3}. \qed

\subsection{The eigenvectors of $A$} \label{subsec:app-lem21proof}

We state a main technical lemma about the eigenvectors of $A$. For $1\leq k\leq K$, let $\hat{\xi}_k$ be the eigenvector associated with $\hat{\lambda}_k$. Write $\hat{\Xi}_0=[\hat{\xi}_2,\hat{\xi}_3,\ldots,\hat{\xi}_K]\in\mathbb{R}^{n,K-1}$, and let $\hat{\Xi}_{0,i}'$ denote its $i$th row, $1\leq i\leq n$.  

\begin{lemma} \label{lem:Aeigenvecs}
Suppose the  conditions of Theorem~\ref{thm:r-entrywise-bound} hold.  With probability $1-o(n^{-3})$, there exist $\omega\in\{\pm 1\}$ and an orthogonal matrix $X\in\mathbb{R}^{K-1,K-1}$ (both $\omega$ and $X$ depend on $A$ and are stochastic) such that 
\begin{itemize}
\item[(a)] $\|\omega\hat{\xi}_1-\xi_1\|\leq C\|\theta\|^{-2}K\sqrt{\theta_{\max} \|\theta\|_1}$;
\item[(b)] $\|\hat{\Xi}_0X- \Xi_0\|_F \leq  C\beta_n^{-1}\|\theta\|^{-2}K^{3/2}\sqrt{\theta_{\max} \|\theta\|_1}$; 
\item[(c)] $\|\omega\hat{\xi}_1-\xi_1\|_\infty\leq C\|\theta\|^{-3}\theta^{3/2}_{\max}K\sqrt{\|\theta\|_1\log(n)}$; 
\item[(d)] $\max_{1\leq i\leq n} \|X'\hat{\Xi}_{0,i}- \Xi_{0,i}\|\leq C\beta_n^{-1}\|\theta\|^{-3}\theta^{3/2}_{\max}K^{3/2}\sqrt{\|\theta\|_1\log(n)}$. 
\end{itemize}
If $\beta_n=o(1)$, then the factor $K$ in the bounds for $\|\omega\hat{\xi}_1-\xi_1\|$ and $\|\omega\hat{\xi}_1-\xi_1\|_\infty$ can be removed. 
\end{lemma}

\noindent{\it Proof of Lemma~\ref{lem:Aeigenvecs}}: 
We first prove claims (a)-(b). The proof is based on the the classical sin-theta theorem \cite{sin-theta}, where below is a simpler version \cite[Theorem 10]{CMW13b}. 
\begin{lemma} \label{lem:sine-theta}
Let $M$ and $\hat{M}$ be two $n\times n$ symmetric matrices. For $1\leq k\leq n$, let $d_k$ be the $k$-th largest eigenvalue of $M$, $\eta_k$ and $\hat{\eta}_k$ be the eigenvector associated with the $k$-th largest eigenvalue of $M$ and $\hat{M}$, respectively. Suppose for some $\delta>0$ and $1\leq k_1\leq k_2\leq n$, we have $d_{k_1-1}>d_{k_1}+\delta$, $d_{k_2+1}< d_{k_2}-\delta$ and $\|\hat{G}-G\|\leq \delta/2$. Write $U=[\eta_{k_1},\cdots,\eta_{k_2}]$ and $\hat{U}=[\hat{\eta}_{k_1},\cdots,\hat{\eta}_{k_2}]$. Then, $\|\hat{U}\hat{U}' - UU'\|\leq 2\delta^{-1}\|\hat{G}-G\|$. 
\end{lemma}

We divide all eigenvalues of $\Omega$ into four groups: (i) $\lambda_1$, (ii) positive eigenvalues among $\lambda_2,\ldots,\lambda_K$, (iii) zero eigenvalues, and (iv) negative eigenvalues among $\lambda_2,\ldots,\lambda_K$. Define $\Xi_{01}$ and $\Xi_{02}$ as the submatrices of $\Xi_0$ by restricting to columns corresponding to eigenvalues in groups (ii) and (iv), respectively. By dividing the empirical eigenvalues and eigenvectors in a similar way, we can define $\hat{\Xi}_{01}$ and $\hat{\Xi}_{02}$. Now, $\xi_1$, $\Xi_{01}$ and $\Xi_{02}$ contain the eigenvectors associated with eigenvalues in groups (i), (ii) and (iv), respectively. By Lemma~\ref{lem:eigenOmega}, the gap between eigenvalues in group (i) and those in other groups is $\lambda_1-|\lambda_2|\geq C^{-1}\lambda_1\geq C^{-1}K^{-1}\|\theta\|^2$, and the eigen-gap between any two remaining groups is $\geq C\beta_nK^{-1}\|\theta\|^2$. It follows from Lemma~\ref{lem:sine-theta} that
\beq \label{adjacency-new1}
\|\hat{\xi}_1\hat{\xi}_1'- \xi_1\xi_1'\| =O\Bigl(\frac{K\|A-\Omega\|}{\|\theta\|^2}\Bigr), \quad \max_{t\in\{1,2\}}\{\|\hat{\Xi}_{0t}\hat{\Xi}_{0t}'- \Xi_{0t}\Xi_{0t}'\| \}= O\Bigl(\frac{K\|A-\Omega\|}{\beta_n\|\theta\|^2}\Bigr). 
\eeq
By elementary linear algebra, $(\hat{\xi}_1\hat{\xi}_1' - \xi_1\xi_1')$ has two nonzero eigenvalues $\pm[1-(\hat{\xi}_1'\xi_1)^2]^{1/2}$, where $|1-(\hat{\xi}_1'\xi_1)^2|\geq \min_{\pm}|1\pm \hat{\xi}_1'\xi_1|= (\min_{\pm}\|\hat{\xi}_1\pm \xi_1\|^2)/2$. It follows that 
\beq \label{adjacency-new2}
\min_{\pm}\|\hat{\xi}_1\pm \xi_1\|\leq \sqrt{2}\|\hat{\xi}_1\hat{\xi}_1'- \xi_1\xi_1'\|. 
\eeq
Moreover, by \cite[Lemma 2.4]{IFPCA}, there always is an orthogonal matrix $X_1$ such that $\| \hat{\Xi}_{01} - \Xi_{01}X_1\|_F \leq  \|\hat{\Xi}_{01}\hat{\Xi}'_{01} - \Xi_{01}\Xi'_{01}\|_F$. Since the rank of $(\hat{\Xi}_{01}\hat{\Xi}'_{01} - \Xi_{01}\Xi'_{01})$ is at most $2K$, we then have 
\[
\| \hat{\Xi}_{01} - \Xi_{01}X_1\|_F \leq \sqrt{2K}\| \hat{\Xi}_{01} - \Xi_{01}X_1\|.
\]
Similarly, there exists an orthogonal matrix $X_2$ such that $\| \hat{\Xi}_{02} - \Xi_{02}X_2\|_F \leq \sqrt{2K}\| \hat{\Xi}_{02} - \Xi_{02}X_2\|$. As a result, for the orthogonal matrix $X=\diag(X_1, X_2)$, 
\begin{align} \label{adjacency-new3}
\|\hat{\Xi}_0X - \Xi_0\|_F \leq 2\sqrt{K} \max_{t\in\{1,2\}}\{\|\hat{\Xi}_{0t}\hat{\Xi}_{0t}'- \Xi_{0t}\Xi_{0t}'\| \}.
\end{align}
Plugging \eqref{adjacency-new2}-\eqref{adjacency-new3} into \eqref{adjacency-new1} gives that with probability $1-o(n^{-3})$,
\begin{align*}
\min_{\pm}\|\hat{\xi}_1\pm \xi_1\|&=O\biggl( \frac{K\|A-\Omega\|}{\|\theta\|^2}\biggr)=O\biggl( \frac{K\sqrt{\theta_{\max}\|\theta\|_1}}{\|\theta\|^2}\biggr),\cr
\|\hat{\Xi}_0X - \Xi_0\|_F  &=O\biggl( \frac{K\sqrt{K}\|A-\Omega\|}{\beta_n\|\theta\|^2}\biggr)=O\biggl(\frac{\sqrt{K^3\theta_{\max}\|\theta\|_1}}{\beta_n\|\theta\|^2}\biggr), 
\end{align*}
where we have used \eqref{lem-adjacency-3}. This proves the first two items. 

We then prove claims (c)-(d). We borrow the techniques and some results from \cite{abbe2017entrywise}. The following lemma is adapted from \cite[Theorem 2.1]{abbe2017entrywise} and is proved below. A direct use of \cite[Theorem 2.1]{abbe2017entrywise} will lead to sub-optimal dependence on $\beta_n$ in the resulting bound, so we have to modify that theorem accordingly. 
\begin{lemma} \label{lem:AbbeFan17}
Let $M\in\mathbb{R}^{n,n}$ be a symmetric random matrix. Write $M^* = \mathbb{E}M$ and $K_0=\mathrm{rank}(M^*)$. For each $1\leq k\leq K_0$, let $d^*_k$ and $d_k$ be the $k$-th largest nonzero eigenvalue of $M^*$ and $M$, respectively, and let $\eta^*_k$ and $\eta_k$ be the corresponding eigenvector, respectively. Let $s$ and $r$ be two integers such that $1\leq r\leq K_0$ and $0\leq s\leq K_0-r$. Write $D=\diag(d_{s+1},d_{s+2},\ldots,d_{s+r})$, $D^*=\diag(d^*_{s+1},d^*_{s+2},\ldots,d^*_{s+r})$, 
\[
U=[\eta_{s+1}, \eta_{s+2},\ldots,\eta_{s+r}],\qquad \mbox{and} \qquad U^*=[\eta^*_{s+1}, \eta^*_{s+2},\ldots,\eta^*_{s+r}].
\]
Define $\Delta^*= \min\{d^*_s-d^*_{s+1}, d^*_{s+r}-d^*_{s+r-1}, \min_{1\leq j\leq r}|d^*_{s+j}|\}$ and define $\kappa=(\max_{1\leq j\leq r}|d^*_{s+j}|)/\Delta^*$. Below, the notation $\|\cdot\|_{2\to\infty}$ represents the maximum row-wise $\ell^2$-norm of a matrix, and $M^*_{m,\cdot}$ is the $m$-th row of $M^*$. Suppose for a number $\gamma>0$, the following assumptions are satisfied:
\begin{itemize}
\item A1 (Incoherence): $\max_{1\leq m\leq n}\|M^*_{m,\cdot}\|\leq \gamma\Delta^*$.
\item A2 (Independence): For any $1\leq m\leq n$, the entries of the $m$-th row and column of $M$ are independent with the other entries. 
\item A3 (Spectral norm concentration): For a number $\delta_0\in (0,1)$, $\mathbb{P}(\|M-M^*\|\leq \gamma\Delta^*)\geq 1-\delta_0$.  
\item A4 (Row concentration): There is a number $\delta_1\in (0,1)$ and a continuous non-decreasing function $\varphi(\cdot)$ with $\varphi(0)=0$ and $\varphi(x)/x$ being non-increasing in $\mathbb{R}^+$ such that, for any $1\leq m\leq n$ and non-stochastic matrix $Y\in\mathbb{R}^{n,r}$, 
\[
\mathbb{P}\left(\|(M-M^*)_{m,\cdot}Y\|_2 \leq \Delta^*\|Y\|_{2\to\infty}\varphi\Bigl(\frac{\|Y\|_F}{\sqrt{n}\|Y\|_{2\to\infty}}\Bigr)  \right)\geq 1-\delta_1/n. 
\]   
\end{itemize}
Let $I_0=(\{1,\ldots, s-1\}\cup \{s+r+1,\ldots,K_0\})\cap \{j: |d^*_j|>\max_{1\leq i\leq r}|d^*_{s+r}|\}$ and $\Delta_0^*=\min\{\min_{j\in I_0}|d_j^*-d^*_{s}|, \min_{j\in I_0}|d_j^*-d^*_{s+r}|\}$. Define $\widetilde{U}^*=[\eta_1,\ldots,\eta_{K_0}]$ and 
\[
\widetilde{\kappa}=\begin{cases}
\max_{j\in I_0} (|d_j^*|/\Delta_0^*), & \mbox{if $I_0\neq \emptyset$},\\
0 & \mbox{otherwise}. 
\end{cases}
\]
Then, with probability $1-\delta_0-2\delta_1$, for an orthogonal matrix $O\in\mathbb{R}^{r,r}$, 
\beq \label{newentrywise}
\|UO - MU^*(D^*)^{-1}\|_{2\to\infty}\leq  C\bigl[\kappa(\kappa+\varphi(1))(\gamma+\varphi(\gamma)) + \widetilde{\kappa}\gamma\bigr]\cdot \|\widetilde{U}^*\|_{2\to\infty}. 
\eeq
\end{lemma}

\noindent
{\it Proof of Lemma~\ref{lem:AbbeFan17}:} Fix $1\leq m\leq n$. Let $M^{(m)}$ be the matrix by setting the $m$-th row and the $m$-th column of $M$ to be zero. Let $\eta_1^{(m)},\eta_2^{(m)},\ldots,\eta_n^{(m)}$ be the eigenvectors of $M^{(m)}$. Write $U^{(m)}=[\eta_{s+1}^{(m)},\ldots,\eta_{s+r}^{(m)}]$. Let $H=U'U^*$, $H^{(m)}=(U^{(m)})'U^*$ and $V^{(m)}=U^{(m)}H^{(m)}-U^*$. We aim to prove 
\begin{align} \label{lemAbbe-4}
\|M_{m\cdot}V^{(m)}\| \leq &  6(\kappa + \widetilde{\kappa})\gamma \Delta^*\|\widetilde{U}^*\|_{2\to\infty}\cr
 &+ \Delta^*\varphi(\gamma)\bigl(4\kappa \|UH\|_{2\to\infty} + 6\|U^*\|_{2\to\infty}\bigr).
\end{align}
Once \eqref{lemAbbe-4} is obtained, the proof is almost identical to the proof of (B.26) in \cite{abbe2017entrywise}, except that we plug in \eqref{lemAbbe-4} instead of (B.32) in \cite{abbe2017entrywise}. This is straightforward, so we omit it. 

What remains is to prove \eqref{lemAbbe-4}. Without loss of generality, we only consider the case where $I_0\neq\emptyset$. In the proof of \cite[Lemma 5]{abbe2017entrywise}, it is shown that
\begin{align*}
\|M_{m\cdot}V^{(m)}\| & \leq \|M^*_mV^{(m)}\| + \|(M-M^*)_{m\cdot}V^{(m)}\|, \cr
\|(M-M^*)_{m\cdot}V^{(m)}\| &\leq \Delta^*\varphi(\gamma)\bigl(4\kappa \|UH\|_{2\to\infty} + 6\|U^*\|_{2\to\infty}\bigr).
\end{align*}
Combining them gives
\beq \label{lemAbbe-1}
\|M_{m\cdot}V^{(m)}\|\leq \|M^*_{m\cdot}V^{(m)}\| + \Delta^*\varphi(\gamma)\bigl(4\kappa \|UH\|_{2\to\infty} + 6\|U^*\|_{2\to\infty}\bigr).
\eeq
We further bound the first term in \eqref{lemAbbe-1}. Recall that $I_0$ is the index set of eigenvalues that are not contained in $D^*$ and have an absolute value larger than $\|D^*\|$. 
Let $\widetilde{M}^*=\sum_{j\in I_0}d_j^*\eta_j^*(\eta_j^*)'$. 
\begin{align*} 
\|M^*_{m\cdot}V^{(m)}\|& \leq \|\widetilde{M}_{m\cdot}^*V^{(m)}\| + \|(M^*_{m\cdot}-\widetilde{M}_{m\cdot}^*)V^{(m)}\|\cr
&\leq \|\widetilde{M}_{m\cdot}^*V^{(m)}\| + \|M^*-\widetilde{M}^*\|_{2\to\infty}\|V^{(m)}\|\cr
&\leq   \|\widetilde{M}_{m\cdot}^*V^{(m)}\| + 6\gamma \|M^*-\widetilde{M}^*\|_{2\to\infty},
\end{align*}
where the last line uses $\|V^{(m)}\|\leq 6\gamma$, by (B.12) of \cite{abbe2017entrywise}. Note that
$M^*-\widetilde{M}^*=\sum_{j\notin I_0}d_j^*\eta_j^*(\eta_j^*)'$. By definition of $I_0$, for any $j\notin I_0$, $|d_j^*|\leq \max_{1\leq i\leq r}|d_{s+r}^*|\leq \kappa\Delta^*$. It follows that
\[
\|M^*-\widetilde{M}^*\|_{2\to\infty}\leq \bigl(\max_{j\notin I_0}|d_j^*|\bigr)\|\widetilde{U}^*\|_{2\to\infty}\leq \kappa \Delta^*\|\widetilde{U}^*\|_{2\to\infty}. 
\]
Combining the above gives
\beq \label{lemAbbe-2}
\|M^*_{m\cdot}V^{(m)}\|\leq \|\widetilde{M}_{m\cdot}^*V^{(m)}\| + 6\kappa \gamma \Delta^*\|\widetilde{U}^*\|_{2\to\infty}. 
\eeq 
Write $D_0^*=\mathrm{diag}(d_j^*)_{j\in I_0}$, $U_0^*=[\eta^*_j]_{j\in I_0}$, $U_0=[\eta_j]_{j\in I_0}$, $U_0^{(m)}=[\eta^{(m)}_j]_{j\in I_0}$, and $H_0^{(m)}=(U_0^{(m)})'U_0^*$. We similarly have $\|U_0^{(m)}H_0^{(m)}-U^*_0\|\leq 6\gamma_0$, where $\gamma_0$ is defined in the same way as $\gamma$ but is with respect to the eigen-gap $\Delta^*_0$. It is not hard to see that $\gamma_0=\gamma\Delta^*/\Delta^*_0$. Hence, 
\[
\|U_0^{(m)}H_0^{(m)}-U^*_0\|\leq 6\gamma\Delta^*/\Delta^*_0. 
\]
By mutual orthogonality of eigenvectors, $(U_0^{(m)})'U^{(m)}=0$ and $(U^*_0)'U^*=0$. It follows that
\begin{align*}
\|\widetilde{M}_{m\cdot}^*V^{(m)}\| &= \|e_m'[U_0^*\Lambda^*_0(U_0^*)'][U^{(m)}H^{(m)}-U^*]\|\cr
&= \|e_m'[U_0^*\Lambda^*_0(U_0^*)']U^{(m)}H^{(m)}\|\cr
&\leq \|e_m'[U_0^*\Lambda^*_0(U_0^*)']U^{(m)}\|\cr
&= \|e_m'U_0^*\Lambda^*_0(U_0^* - U_0^{(m)}H_0^{(m)})'U^{(m)}\|\cr
&\leq \|e_m'U_0^*\Lambda^*_0(U_0^* - U_0^{(m)}H_0^{(m)})'\|\cr
&\leq \|\widetilde{U}^*\|_{2\to\infty}\cdot\|\Lambda_0^*\|\cdot \|U_0^* - U_0^{(m)}H_0^{(m)}\|\cr
&\leq 6(\|\Lambda_0\|^*/\Delta^*_0)\cdot \gamma\Delta^* \|\widetilde{U}^*\|_{2\to\infty}. 
\end{align*} 
We plug it into \eqref{lemAbbe-2} and note that $\widetilde{\kappa}=\|\Lambda_0\|^*/\Delta^*_0$. It gives
\beq \label{lemAbbe-3}
\|M^*_{m\cdot}V^{(m)}\|\leq 6(\kappa + \widetilde{\kappa})\gamma \Delta^*\|\widetilde{U}^*\|_{2\to\infty}. 
\eeq
Combining \eqref{lemAbbe-1} and \eqref{lemAbbe-3} gives \eqref{lemAbbe-4}. \qed  

\bigskip

We now come back to the proof of Lemma~\ref{lem:Aeigenvecs}. We have divided nonzero eigenvalues of $\Omega$ into four groups: (i) $\lambda_1$, (ii) positive eigenvalues in $\lambda_2,\ldots,\lambda_K$, (iii) zero eigenvalues, and (iv) negative eigenvalues in $\lambda_2,\ldots,\lambda_K$. 
We shall apply Lemma~\ref{lem:AbbeFan17} to each of the four groups. To save space, we only consider applying it to group (ii). The proof for other groups is similar and omitted. 

Now, $M=A$ and $M^*=\Omega=\mathrm{diag}(\Omega)+(A-\mathbb{E}A)$. We check conditions A1-A4. By Lemma~\ref{lem:eigenOmega}, $\Delta^*\geq C\beta_nK^{-1}\|\theta\|^2$ and $\kappa\leq C$. For an appropriately large constant $\tilde{C}>0$, we take  
\[
\gamma = \tilde{C}\beta_n^{-1}\|\theta\|^{-2}K\sqrt{\theta_{\max}\|\theta\|_1}. 
\]
Consider A1. Since $\Omega(i,j)\leq C\theta(i)\theta(j)$, we have $\max_{1\leq i\leq n}\|\Omega_{i,\cdot}\|\leq C\theta_{\max}\|\theta\|$. From the universal inequality $\|\theta\|\leq \sqrt{\theta_{\max}\|\theta\|_1}$ and the assumption $\theta_{\max}=O(1)$, this term is  $O(\sqrt{\theta_{\max}\|\theta\|_1})$, which is bounded by $\gamma\Delta^*$ when $\tilde{C}$ is appropriately large. Hence, A1 is satisfied. A2 is satisfied because the upper triangle of $A$ contains independent variables. By \eqref{lem-adjacency-3}, A3 is satisfied for $\delta_0=o(n^{-3})$. We then verify A4. Since $\|\mathrm{diag}(\Omega)\|\leq C$, 
\beq \label{entrywise-0}
\|\mathrm{diag}(\Omega)_{i,\cdot}Y\|_2\leq C\|Y\|_{2\to\infty}, \qquad 1\leq i\leq n. 
\eeq
Fix $1\leq i\leq n$ and $1\leq k\leq r$. Let $y_k\in\mathbb{R}^n$ be the $k$-th column of $Y$. 
Using the Bernstein's inequality, for any $t\geq 0$, 
\beq \label{entrywise-1}
\mathbb{P}\bigl(|y_k'(A-\mathbb{E}A)_{i,\cdot}| > t\bigr)\leq 2\exp\left( - \frac{t^2/2}{\sum_{j=1}^n\Omega(i,j)y^2_k(j) + t\|y_k\|_\infty/3}\right). 
\eeq
Note that $\sum_j\Omega(i,j)y_k^2(j)\leq C \|y_k\|_\infty^2\theta_{\max}\|\theta\|_1$. Moreover, $\theta_{\max}\|\theta\|_1\gg \log(n)$ by \eqref{err-order}. We take $t=C\|y_k\|_\infty \sqrt{\theta_{\max}\|\theta\|_1\log(n)}$ for a large enough constant $C>0$. It follows that with probability $1-o(n^{-4})$,
\[
|y_k'(A-\mathbb{E}A)_{i,\cdot}| \leq \|y_k\|_\infty\cdot C\sqrt{\theta_{\max}\|\theta\|_1\log(n)}. 
\]
Combining it with the probability union bound and \eqref{entrywise-0}, with probability $1-o(n^{-3})$,
\begin{align} \label{getPhi-1}
\|(A-\Omega)_{i,\cdot}Y\|_2 &\leq C\sqrt{\theta_{\max}\|\theta\|_1\log(n)}\cdot \|Y\|_{2\to\infty}\cr
&\leq  \Delta^*\|Y\|_{2\to\infty}\cdot \frac{C\sqrt{\theta_{\max}\|\theta\|_1\log(n)}}{K^{-1}\beta_n\|\theta\|^2}. 
\end{align}
Moreover, in \eqref{entrywise-1}, if we use an alternative bound $\sum_j\Omega(i,j)y_k^2(j)\leq \|y_k\|^2\theta_{\max}^2$, we obtain a different bound as follows: With probability $1-o(n^{-4})$,
\[
|y_k'(A-\mathbb{E}A)_{i,\cdot}| \leq C\max\bigl\{\|y_k\|\theta_{\max}\sqrt{\log(n)},\; \|y_k\|_\infty\log(n)\bigr\}. 
\] 
Due to the probability union bound and \eqref{entrywise-0}, with probability $1-o(n^{-3})$,
\begin{align} \label{getPhi-2}
\|&(A-\Omega)_{i,\cdot}Y\|_2 \leq C\max\bigl\{\|Y\|_F\theta_{\max}\sqrt{\log(n)},\; \|Y\|_{2\to\infty}\log(n)\bigr\}\cr
&\leq \Delta^*\|Y\|_{2\to\infty}\, \max\left\{ \frac{\theta_{\max}\sqrt{n\log(n)}}{K^{-1}\beta_n\|\theta\|^2}\frac{\|Y\|_F}{\sqrt{n}\|Y\|_{2\to\infty}},\; \frac{\log(n)}{K^{-1}\beta_n\|\theta\|^2}\right\}. 
\end{align}
Let $t_1=C(K^{-1}\beta_n\|\theta\|^2)^{-1}\sqrt{\theta_{\max}\|\theta\|_1\log(n)}$, $t_2=C(K^{-1}\beta_n\|\theta\|^2)^{-1}\theta_{\max}\sqrt{n\log(n)}$, and $t_3=C(K^{-1}\beta_n\|\theta\|^2)^{-1}\log(n)$. Define the function
\[
\widetilde{\varphi}(x) = \min\{t_1, \;\max\{t_2x, t_3\}\}. 
\]
Then, \eqref{getPhi-1}-\eqref{getPhi-2} together imply that with probability $1-o(n^{-3})$,
\beq \label{getPhi-3}
\|(A-\mathbb{E}A)_{i,\cdot}Y\|_2\leq \Delta^*\|Y\|_{2\to\infty} \widetilde{\varphi}\Bigl(\frac{\|Y\|_F}{\sqrt{n}\|Y\|_{2\to\infty}}\Bigr).
\eeq
We look at the function $\widetilde{\varphi}(x)$. Note that $(\sqrt{n}\|Y\|_{2\to\infty})^{-1}\|Y\|_F$ takes values in the interval $[n^{-1/2},1]$. By \eqref{err-order}, $t_1\gg t_3$. Moreover, since $\|\theta\|_1\leq n\theta_{\max}$, when $x=1$, $t_2x\geq Ct_1$. Last, when $x=n^{-1/2}$, $t_2x\ll t_3$. Combining the above, we conclude that in $[n^{-1/2}, \infty)$, the function $\widetilde{\varphi}(x)$ first stays flat at $t_3$, then linearly increases to $t_1$ and then stays flat at $t_1$. Hence, we construct a function $\varphi(x)$, which linearly increases from $0$ to $t_3$ for $x\in [0, n^{-1/2}]$, then linear increases from $t_3$ to $t_1$ for $x\in [n^{-1/2}, t_2/t_1]$, and then stays constant as $t_1$ for $x\in [t_2/t_1,\infty)$. It is seen that  $\varphi(0)=0$, $\varphi(x)/x$ is non-increasing, and $\widetilde{\varphi}(x)\leq \varphi(x)\leq t_1$ in the interval $[n^{-1/2}, 1]$. By \eqref{getPhi-3} and that $\widetilde{\varphi}(x)\leq \varphi(x)$, A4 is satisfied with $\delta_1=o(n^{-3})$. Furthermore, since $\varphi(x)\leq t_1$, 
\[
\varphi(\gamma)\leq \frac{C\sqrt{\theta_{\max}\|\theta\|_1\log(n)}}{K^{-1}\beta_n\|\theta\|^2}. 
\]
So far, we have shown that A1-A4 hold. 

We now apply Lemma~\ref{lem:AbbeFan17}. As mentioned, we only study the eigenvectors in group (ii), which correspond to positive eigenvalues among $\lambda_2,\ldots,\lambda_K$. Let $\Lambda_1$ be the diagonal matrix consisting of these eigenvalues and let $\Xi_{01}$ be the matrix formed by associated eigenvectors. Define their empirical counterparts, $\hat{\Lambda}_1$ and $\hat{\Xi}_{01}$, in the same way. In Lemma~\ref{lem:AbbeFan17}, we take $U=\hat{\Xi}_{01}$, $U^*=\Xi_{01}$, and $\widetilde{U}^*=\Xi$. Since $\lambda_2,\ldots,\lambda_K$ are at the same order, $\kappa\leq C$. Also, $\widetilde{\kappa}\leq \lambda_1/(\lambda_1-|\lambda_2|)\leq C$ by our assumption. 
It follows from \eqref{newentrywise} that there exists an orthogonal matrix $O$ such that 
\[
 \|\hat{\Xi}_{01}O - A\Xi_{01}\Lambda_1^{-1}\|_{2\to\infty}
\leq \frac{C\sqrt{\theta_{\max}\|\theta\|_1\log(n)}}{K^{-1}\beta_n\|\theta\|^2}\|\Xi\|_{2\to\infty}. 
\]
By Lemma~\ref{lem:xi1}, $\|\Xi\|_{2\to\infty}=O(\sqrt{K}\|\theta\|^{-1}\theta_{\max})$. Plugging it into the above inequality, we find that
\beq \label{entrywise-2}
\|\hat{\Xi}_{01}O - A\Xi_{01}\Lambda_1^{-1}\|_{2\to\infty}\leq \frac{C\theta_{\max}^{3/2}K^{3/2}\sqrt{\|\theta\|_1\log(n)}}{\beta_n\|\theta\|^3}. 
\eeq
By definition of eigen-decomposition, $\Omega\Xi_{01}=\Xi_{01}\Lambda_1$. It follows that
\[
A\Xi_{01}\Lambda_1^{-1}=\Omega\Xi_{01}\Lambda_1^{-1} + (A-\Omega)\Xi_{01}\Lambda_1^{-1}=\Xi_{01} + (A-\Omega)\Xi_{01}\Lambda_1^{-1}. 
\]
Plugging it into \eqref{entrywise-2} yields 
\beq \label{entrywise-3}
\|\hat{\Xi}_{01}O - \Xi_{01}\|_{2\to\infty}\leq \frac{C\theta_{\max}^{3/2}K^{3/2}\sqrt{\|\theta\|_1\log(n)}}{\beta_n\|\theta\|^3} + \|(A-\Omega)\Xi_{01}\Lambda_1^{-1}\|_{2\to\infty}. 
\eeq
To bound the second term on the right hand side, we apply the first line of \eqref{getPhi-1} by letting $Y=\Xi_{01}$. It turns out that with probability $1-o(n^{-3})$, 
\begin{align} \label{entrywise-4}
\|(A-\Omega)\Xi_{01}\Lambda_1^{-1}\|_{2\to\infty} & \leq (\max_{1\leq i\leq n}\| (A-\Omega)_{i,\cdot}\Xi_{01}\|_2)\cdot \|\Lambda_1^{-1}\|\cr
&\leq C\sqrt{\theta_{\max}\|\theta\|_1\log(n)}\cdot \|\Xi_{01}\|_{2\to\infty}\cdot\|\Lambda_1^{-1}\|\cr
&\leq C\sqrt{\theta_{\max}\|\theta\|_1\log(n)}\cdot \sqrt{K}\|\theta\|^{-1}\theta_{\max}\cdot K\beta_n^{-1}\|\theta\|^{-2}, 
\end{align}
where in the last inequality, the bound of $\|\Lambda_1^{-1}\|$ is from Lemma~\ref{lem:eigenOmega} and the bound of $\|\Xi_{01}\|_{2\to\infty}$ is from Lemma~\ref{lem:xi1}. Combining \eqref{entrywise-3}-\eqref{entrywise-4} gives
\[
\|\hat{\Xi}_{01}O - \Xi_{01}\|_{2\to\infty}\leq \frac{C\theta_{\max}^{3/2}K^{3/2}\sqrt{\|\theta\|_1\log(n)}}{\beta_n\|\theta\|^3}. 
\] 
Note that the left hand side only involves eigenvectors in group (ii). We can prove similar results for the other three groups of eigenvectors. For group (i), $\Delta^*\geq CK^{-1}\|\theta\|^{-1}$ and $\|\widetilde{U}^*\|_{2\to\infty}\leq C\|\theta\|^{-1}\theta_{\max}$, and the resulting bound is
\[
\|\omega\hat{\xi}_1-\xi_1\|_\infty \leq \frac{C\theta_{\max}^{3/2}K\sqrt{\|\theta\|_1\log(n)}}{\|\theta\|^3}.
\]

Furthermore, if $\beta_n=o(1)$, by Lemma~\ref{lem:eigenOmega}, $\lambda_1-|\lambda_2|\geq C^{-1}\lambda_1\geq C^{-1}K\|\theta\|^2$. Compared with the case of $\beta_n\geq c$, the $\Delta^*$ of group (i) is larger by a factor of $K$, so all the bounds concerning $\hat{\xi}_1$ are reduced by a factor of $K$. \qed

\subsection{Proof of Theorem~\ref{thm:r-entrywise-bound}}  \label{subsec:app-thm21proof}
Without loss of generality, we assume $T=\infty$, so that no thresholding is applied in obtaining $\hat{R}$. Note that $\max_i\|r_i\|\leq \max_k\|v_k\|\leq C\sqrt{K}$ by Lemma~\ref{lem:R}. For any threshold $\sqrt{K}\ll T<\infty$, the threshold always reduces errors. Therefore, the error bounds for the case of no thresholding immediately imply the error bounds for the case of thresholding.

The second claim is straightforward. We only show the first claim. By Lemma~\ref{lem:xi1}, we can choose the sign of $\xi_1$ such that it is a strictly positive vector. By definition of $err_n$, we can re-write
\[
err_n = \frac{\|\theta\|}{\theta_{\min}}\cdot \frac{\theta^{3/2}_{\max}\sqrt{\|\theta\|_1\log(n)}}{\|\theta\|^3}. 
\]
Then, the statements (c)-(d) of Lemma~\ref{lem:Aeigenvecs} can be re-expressed as
\beq \label{proof-thm21-1}
\|\omega\hat{\xi}-\xi\|_\infty=O\Bigl( \frac{\theta_{\min}}{\|\theta\|}Kerr_n\Bigr),\quad \max_{1\leq i\leq n} \|X'\hat{\Xi}_{i,0}-\Xi_{i,0}\| =O \Bigl(\frac{\theta_{\min}}{\|\theta\|}K^{3/2}\beta_n^{-1}err_n\Bigr). 
\eeq
We now show the claim. Let $(\omega, X)$ be the same as in Lemma~\ref{lem:Aeigenvecs}, and define $H = \omega X'\in\mathbb{R}^{K-1, K-1}$. 
Fix $i$. By definition of $(r_i,\hat{r}_i)$ and $H$, 
\[
r_i = \frac{1}{\xi_1(i)}\Xi_{i,0}, \qquad H\hat{r}_i = \omega X'\hat{r}_i = \frac{1}{\omega \hat{\xi}_1(i)}X'\hat{\Xi}_{i,0}. 
\]
It follows that
\begin{align*} 
H\hat{r}_i - r_i &=\frac{1}{\omega\hat{\xi}_1(i)}(X'\hat{\Xi}_{i,0}-\Xi_{i,0}) + \Bigl[ \frac{1}{\omega\hat{\xi}_1(i)}-\frac{1}{\xi_1(i)} \Bigr]\Xi_{i,0}\cr
&  = \frac{1}{\omega\hat{\xi}_1(i)}(X'\hat{\Xi}_{i,0}-\Xi_{i,0}) - \frac{\omega\hat{\xi}_1(i)-\xi_1(i)}{\omega\hat{\xi}_1(i)} r_i. 
\end{align*}
First, by Lemma~\ref{lem:xi1}, $\xi_1(i)\geq C\theta_{\min}/\|\theta\|$; also, by \eqref{proof-thm21-1}, $|\omega\hat{\xi}_1(i)-\xi_1(i)|\ll \theta_{\min}/\|\theta\|$. We thus have  
$\omega \hat{\xi}_1(i)\geq \xi_1(i)/2\geq C\theta_{\min}/\|\theta\|$. Second, using the first bullet point of Lemma~\ref{lem:R}, we have $\|r_i\|\leq \max_{k}\|v_k\|\leq C\sqrt{K}$. Plugging these results into the above equation gives
\beq \label{proof-thm21-2}
\|H\hat{r}_i - r_i \| \leq \frac{C\|\theta\|}{\theta_{\min}}\bigl( \|X'\hat{\Xi}_{i,0}-\Xi_{i,0}\| + \sqrt{K}|\omega\hat{\xi}_1(i)-\xi_1(i)|\bigr).
\eeq
The claim follows by plugging \eqref{proof-thm21-1} into \eqref{proof-thm21-2}. \qed

\subsection{The $\ell^2$-norm deviation bound for $\hat{R}$} \label{subsec:hatR-l2}

Theorem~\ref{thm:r-entrywise-bound} is about the row-wise large deviation bound for $\hat{R}$. For completeness of theory, we also present the $\ell^2$-norm deviation bound for $\hat{R}$. This result will be useful in the proofs of Theorems~\ref{thm:SVS1}-\ref{thm:SVS2} about faster rates of Mixed-SCORE. 
Recall the following definition: 
\[
err_n^* =  [(\theta_{\max}^{1/2}\bar{\theta}^{3/2})/(\theta_{\min} \bar{\theta}_*)]   \cdot    
(n \bar{\theta}^2)^{-1/2}.  
\]

\begin{lemma} \label{lem:hatR-l2}
Under conditions of Theorem~\ref{thm:r-entrywise-bound}, with probability $1-o(n^{-3})$, 
\[
n^{-1}\sum_{i=1}^n \|H\hat{r}_i - r_i \|^2\leq CK^3\beta_n^{-2}(err_n^*)^2.
\] 
\end{lemma}

\noindent{\it Proof of Lemma~\ref{lem:hatR-l2}}: As explained in the proof of Theorem~\ref{thm:r-entrywise-bound}, we only need to prove the claim for the special case of $T=\infty$ in obtaining $\hat{R}$ (i.e., no thresholding is applied). By definition of $err_n^*$, we can re-write it as
\[
err^*_n = \frac{\|\theta\|}{\theta_{\min}\sqrt{n}}\cdot \frac{\sqrt{\theta_{\max}\|\theta\|_1}}{\|\theta\|^2}. 
\]
Then, the first two bullet points of Lemma~\ref{lem:Aeigenvecs} can be re-expressed as
\[
\|\omega\hat{\xi}-\xi\|=O\biggl( \frac{\theta_{\min}\sqrt{n}}{\|\theta\|}Kerr^*_n\biggr),\quad \|\hat{\Xi}_{0}X-\Xi_{0}\|_F =O \biggl(\frac{\theta_{\min}\sqrt{n}}{\|\theta\|} K^{3/2}\beta_n^{-1}err^*_n\biggr). 
\]
Combining it with \eqref{proof-thm21-2} gives
\[
n^{-1}\sum_{i=1}^n \|H\hat{r}_i - r_i \|^2 \leq \frac{C\|\theta\|^2}{n\theta^2_{\min}}\bigl(\|\hat{\Xi}_0X - \hat{\Xi}_0\|_F^2 + K\|\omega\hat{\xi}_1-\xi_1\|^2\bigr)\leq CK^3\beta_n^{-2}(err_n^*)^2. 
\]
This proves the claim.  \qed

\subsection{A property of the rotation matrix $H$}

\begin{lemma} \label{lem:H}
Let $H$ be the orthogonal matrix in Theorem~\ref{thm:r-entrywise-bound}. 
With probability $1-o(n^{-3})$, $\|H\mathrm{diag}(\hat{\lambda}_2,\ldots,\hat{\lambda}_K)-\mathrm{diag}(\hat{\lambda}_2,\ldots,\hat{\lambda}_K)H\|\leq C\sqrt{\theta_{\max}\|\theta\|_1}$.
\end{lemma}

\noindent{\it Proof of Lemma~\ref{lem:H}}: 
Write for short $\hat{\Lambda}_0=\mathrm{diag}(\hat{\lambda}_2,\ldots,\hat{\lambda}_K)$. Let $\hat{\Xi}_0$, $\hat{\Xi}_0$, $\omega$ and $X$ be the same as in Lemma~\ref{lem:Aeigenvecs}. In the proof of Theorem~\ref{thm:r-entrywise-bound}, we have seen that 
\[
H=\omega X', \qquad\mbox{where}\quad \omega\in\{\pm 1\}. 
\]
It follows that  
\begin{align} \label{proof-lemH-1}
\|H\hat{\Lambda}_0 - \hat{\Lambda}_0 H\| & = \|(H\hat{\Lambda}_0 - \hat{\Lambda}_0 H)'\| = \|X\hat{\Lambda}_0 - \hat{\Lambda}_0 X\| \cr
&= \| (\hat{\Xi}_0'\Xi_0)\hat{\Lambda}_0 - \hat{\Lambda}_0 (\hat{\Xi}_0'\Xi_0)  + (H- \hat{\Xi}_0'\Xi_0)\hat{\Lambda}_0 -\hat{\Lambda}_0 (H- \hat{\Xi}_0'\Xi_0) \|\cr
&\leq \| (\hat{\Xi}_0'\Xi_0)\hat{\Lambda}_0 - \hat{\Lambda}_0 (\hat{\Xi}_0'\Xi_0)\| + 2\| \hat{\Xi}_0'\Xi_0 - X \|\cdot \|\hat{\Lambda}_0\|. 
\end{align}
We shall apply \cite[Lemma 2]{abbe2017entrywise}: in our setting, their notations $H$ and $\mathrm{sgn}(H)$ correspond to our notations of $\hat{\Xi}_0'\Xi_0$ and $X$.  By their Lemma 2, 
\beq \label{proof-lemH-2}
\| \hat{\Xi}_0'\Xi_0 - X \|^{1/2}\leq C\|A-\Omega\|/\Delta^*, \quad \|(\hat{\Xi}_0'\Xi_0)\hat{\Lambda}_0 - \hat{\Lambda}_0 (\hat{\Xi}_0'\Xi_0)\|\leq 2\|A-\Omega\|,   
\eeq
where $\Delta^*$ is the eigen-gap quantity defined in the proof of Lemma~\ref{lem:Aeigenvecs} and satisfies $\Delta^*\geq C\beta_nK^{-1}\|\theta\|^2$. Additionally, by Lemma~\ref{lem:eigenOmega} and Lemma~\ref{lem:Aeigenvals}, $\|\hat{\Lambda}_0\| \lesssim \|\Lambda_0\|\leq C\beta_nK^{-1}\|\theta\|^2\leq C\Delta^*$, with probability $1-o(n^{-3})$. Combining these with \eqref{proof-lemH-1}-\eqref{proof-lemH-2}, we have: with probability $1-o(n^{-3})$, 
\begin{align*}
\|H\hat{\Lambda}_0 - \hat{\Lambda}_0 H\| 
&  \leq  \|(\hat{\Xi}_0'\Xi_0)\hat{\Lambda}_0 - \hat{\Lambda}_0 (\hat{\Xi}_0'\Xi_0)\| + 2\| \hat{\Xi}_0'\Xi_0 - X \|\cdot \|\hat{\Lambda}_0\|\cr
&\leq 2\|A-\Omega\| + C(\|A-\Omega\|/\Delta^*)^{2}\cdot C\Delta^* \cr
&\leq C \|A-\Omega\|\cr
&\leq C\sqrt{\theta_{\max}\|\theta\|_1},
\end{align*}
where the third line is because $\|A-\Omega\|\ll \Delta^*$ and the last line is from \eqref{lem-adjacency-3}. \qed

\section{Vertex Hunting} \label{sec:app-VH}


Mixed-SCORE as a generic algorithm, where the VH step is a plug-in step. 
To analyze the errors of Mixed-SCORE, we must first understand the errors of different VH approaches. 


\begin{definition}[Efficiency and strong efficiency of Vertex Hunting] \label{def:effVH} 
A Vertex Hunting algorithm is said to be efficient if it satisfies 
$\max_{1\leq k\leq K}\|H\hat{v}_k - v_k\|\leq C \max_{1\leq i\leq n}\|H\hat{r}_i - r_i\|$,  
and it is said to be strongly efficient if  
$\max_{1\leq k\leq K}\|H\hat{v}_k - v_k\|\leq C \bigl(n^{-1}\sum_{i=1}^n\|H\hat{r}_i - r_i\|^2\bigr)^{1/2}$,  
where $H$ is the same orthogonal matrix as in Theorem~\ref{thm:r-entrywise-bound}. 
\end{definition} 
Consider all  $4$ VH approaches: SVS, SVS*, CVS, and SP in Table~\ref{tab:CompSVS}. 
We show 
\begin{itemize} 
\itemsep0em 
\item All approaches are efficient under some regularity conditions. 
\item SVS and SVS* are also strongly efficient in some settings (however, CVS and SP are generally not strongly efficient; this is because SVS and SVS* use a denoise stage while CVS and SP do not). 
\end{itemize}  


\subsection{Efficiency of SP and CVS}

The next lemma gives the efficiency of CVS and SP.    
\begin{lemma}[Efficiency of CVS and SP] \label{lem:SP+CVS}
Suppose conditions of Theorem~\ref{thm:main-rev} hold. Suppose we apply either CVS or SP algorithm  to the $n$ rows of $\hat{R}$. With probability $1-o(n^{-3})$, the estimated $\hat{v}_1,\ldots,\hat{v}_K$ satisfy that $
\max_{1\leq k\leq K}\|H\hat{v}_k - v_k\|\leq C \max_{1\leq i\leq n}\|H\hat{r}_i - r_i\|$. 
Therefore, both the CVS and  SP algorithms are  efficient. 
\end{lemma}    

\noindent{\it Proof of Lemma~\ref{lem:SP+CVS}}: Without loss of generality, we only consider the case that $H$ equals to the identity matrix. When $H$ is not the identity matrix, noticing that $\max_{1\leq k\leq K}\|H\hat{v}_k - v_k\|= \max_{1\leq k\leq K}\|\hat{v}_k - H'v_k\|$, we only need to plug $H'v_1,\ldots,H'v_K$ into the proof below. 

We first prove the efficiency of the CVS algorithm. Write $\hat{h}=\max_{1\leq i\leq n}\|\hat{r}_i-r_i\|$. We aim to show 
\beq \label{VHproof-1}
\min_{1\leq \ell\leq K}\|v_k- \hat{v}_\ell\|\leq C_0 \hat{h}, \qquad \mbox{for all }1\leq k\leq K. 
\eeq
It means for each true vertex $v_k$, there is at least one of $\{\hat{v}_1,\hat{v}_2,\ldots,\hat{v}_K\}$ that is within a distance of $C_0\hat{h}$ to $v_k$. At the same time, since $\hat{h}=o(\sqrt{K})$ and the distance between any two vertices is $\geq C\sqrt{K}$ (see Lemma~\ref{lem:R}), each $\hat{v}_\ell$ cannot be simultaneously within a distance $C_0\hat{h}$ to two vertices. The above imply that there is a one-to-one correspondence between true and estimated vertices such that for each true vertex the corresponding estimated vertex is within a distance $C_0\hat{h}$ to it. The claim then follows.   

It remains to show \eqref{VHproof-1}. Fix $1\leq k\leq K$. Recall that $w_i$ is the unique weight vector such that $r_i=\sum_{s=1}^K w_i(s)v_s$, $1\leq i\leq n$. For a constant $C_1>0$ to be decided, let
\[
{\cal V}_{0k}=\bigl\{1\leq i\leq n: w_i(k)\geq 1- C_1K^{-1/2}\hat{h}\bigr\}. 
\]
Let $\hat{i}_s$ be such that $\hat{v}_s = \hat{r}_{\hat{i}_s}$, $1\leq s\leq K$.
We shall first prove that 
\beq \label{VHproof-2}
\{\hat{i}_1,\hat{i}_2,\ldots,\hat{i}_K\}\cap {\cal V}_{0k}\neq \emptyset. 
\eeq
This means at least one of the estimated vertices has to come from the point set $\{\hat{r}_i: i\in {\cal V}_{0k}\}$. We shall next prove that 
\beq \label{VHproof-3}
\max_{i\in {\cal V}_{0k}}\|\hat{r}_i - v_k\|\leq C_0 \hat{h}.  
\eeq
Then, the estimated vertex which comes from $\{\hat{r}_i: i\in {\cal V}_{0k}\}$ is guaranteed to be within a distance $C_0\hat{h}$ to the true $v_k$, i.e., \eqref{VHproof-1} holds. 

It remains to show \eqref{VHproof-2}-\eqref{VHproof-3}. 
First, consider \eqref{VHproof-2}. In the proof of Lemma~\ref{lem:R}, we introduce a one-to-one linear mapping $T_2$ from the standard simplex ${\cal S}_0$ to the Ideal Simplex ${\cal S}^{ideal}$ such that $T_2(w_i)=r_i$ for all $1\leq i\leq n$. We have shown that both $T_2$ and $T_2^{-1}$ are Lipschitz with the Lipschitz constants at the order of $\sqrt{K}$ and $1/\sqrt{K}$, respectively.  As a result, there is a constant $C_2>1$ such that, for any $w, \tilde{w}\in {\cal S}_0$,  
\beq \label{norm-preserve}
C_2^{-1}\sqrt{K}\|w-\tilde{w}\|\leq \|T_2(w)-T_2(\tilde{w})\|\leq C_2\sqrt{K}\|w-\tilde{w}\|.
\eeq
Below, we first use \eqref{norm-preserve} to show the distance from $v_k$ to the convex hull of $\{r_i: i\notin {\cal V}_{0k}\}$ is sufficiently large, and then prove \eqref{VHproof-2} by contradiction. We take $C_1= 5C_2$. Take an arbitrary point $x^*$ from the convex hull ${\cal H}\{r_i: i\notin {\cal V}_{0k}\}$. 
Since $T_2$ is a linear mapping,  $y^*=T_2^{-1}(x^*)$ is a convex combination of $\{w_i: i\notin {\cal V}_{0k}\}$. By definition, for each $i\notin {\cal V}_{0k}$, $0\leq w_i(k)\leq 1- C_1K^{-1/2}\hat{h}$. As a result, $y^*(k)$, as a convex combination of $\{w_i(k): i\notin {\cal V}_{0k}\}$, also satisfies that $0\leq y^*(k)\leq 1-C_1K^{-1/2}\hat{h}$. This implies 
\[
\|T_2^{-1}(x^*)-e_k\|= \|y^*- e_k\|\geq C_1K^{-1/2}\hat{h}, \qquad \mbox{for any }x^*\in {\cal H}\{r_i: i\notin {\cal V}_{0k}\}. 
\]
Combining it with \eqref{norm-preserve}, we have 
\[
\|x^* - v_k\|=\|T_2(y^*) - T_2(e_k)\|\geq C_2^{-1}\sqrt{K}\cdot C_1K^{-1/2}\hat{h}\geq 5\hat{h}.
\]
Since $x^*$ is taken arbitrarily from the convex hull ${\cal H}\{r_i: i\notin {\cal V}_{0k}\}$, we have
\beq \label{VHproof-4}
d\bigl(v_k, {\cal H}\{r_i: i\notin {\cal V}_{0k}\}\bigr)\geq 5\hat{h}. 
\eeq
Come back to the proof of \eqref{VHproof-2}. When this claim is not true, the estimated simplex $\widehat{\cal S}$ is contained in the convex hull of $\{\hat{r}_i: i\notin {\cal V}_{0k}\}$. It follows that 
\begin{align*}
d(v_k, \widehat{\cal S}) & \geq d\bigl(v_k, {\cal H}\{\hat{r}_i: i\notin {\cal V}_{0k}\}\bigr)\cr
&\geq d\bigl(v_k, {\cal H}\{r_i: i\notin {\cal V}_{0k}\}\bigr) - \hat{h}\cr
&\geq 4\hat{h}. 
\end{align*} 
Let $j_k$ be a pure node of community $k$. Then, $\|\hat{r}_{j_k}-v_k\|=\|\hat{r}_{j_k}-r_{j_k}\|\leq \hat{h}$. It follows that 
\beq \label{VHproof-5}
\max_{1\leq i\leq n}d(\hat{r}_i, \widehat{\cal S}) \geq d(\hat{r}_{j_k}, \widehat{\cal S})\geq d(v_k, \widehat{\cal S}) - \hat{h} \geq 3\hat{h}. 
\eeq
At the same time, consider the simplex $\widehat{\cal S}^*$ formed by $\hat{r}_{j_1},\hat{r}_{j_2},\ldots,\hat{r}_{j_K}$, where $j_s$ is a pure node of community $s$, for $1\leq s\leq K$. Note that $r_{i_1},r_{i_2},\ldots,r_{i_K}$ form the Ideal Simplex ${\cal S}^*$ and $\max_{1\leq i\leq n}d(r_i, {\cal S}^*)=0$.  It follows that 
\beq \label{VHproof-6}
\max_{1\leq i\leq n}d(\hat{r}_i, \widehat{\cal S}^*)\leq \max_{1\leq i\leq n}d(r_i, {\cal S}^*) + 2\hat{h}\leq 2\hat{h}. 
\eeq
Note that $\widehat{\cal S}$ is the solution of the combinatory search step. It has to satisfy
\[
\max_{1\leq i\leq n}d(\hat{r}_i, \widehat{\cal S})\leq \max_{1\leq i\leq n}d(\hat{r}_i, \widehat{\cal S}^*).
\]
This yields a contradiction to \eqref{VHproof-5}-\eqref{VHproof-6}. Hence, \eqref{VHproof-2} must be true. 

Next, consider \eqref{VHproof-3}. It is easy to see that 
\begin{align*}
\max_{i\in {\cal V}_{0k}}\|\hat{r}_i - v_k\| & \leq \max_{i\in {\cal V}_{0k}}\|r_i - v_k\|+ \hat{h}\cr
& = \max_{i\in {\cal V}_{0k}}\|T_2(w_i) - T_2(e_k)\|+ \hat{h}\cr
&\leq C_2\sqrt{K}\max_{i\in {\cal V}_{0k}}\|w_i - e_k\| + \hat{h}, 
\end{align*}
where we have used \eqref{norm-preserve} in the last line. For any $i\in {\cal V}_{0k}$, $\|w_i - e_k\|^2=[1-w_i(k)]^2+\sum_{\ell\neq k}w^2_i(\ell)\leq [1-w_i(k)]^2 + [\sum_{\ell\neq k}w_i(\ell)]^2\leq 2(C_1K^{-1/2}\hat{h})^2=50C_2^2K^{-1}\hat{h}^2$. It follows that
\[
\max_{i\in {\cal V}_{0k}}\|\hat{r}_i - v_k\|\leq (5\sqrt{2}C^2_2 + 1)\hat{h}. 
\]
Hence, \eqref{VHproof-3} is true by choosing $C_0=5\sqrt{2}C_2^2+1$.

We then prove the efficiency of the SP algorithm. 
For space limit, the exact description of the SP algorithm is not given in the main paper. We include it here:  
\begin{itemize}
\item Initialize $Y_i=(1, \hat{r}'_i)'\in\mathbb{R}^K$, for $1\leq i\leq n$.
\item At iteration $k=1,2,\ldots,K$: Find $i_k=\mathrm{argmax}_{1\leq i\leq n}\|Y_i\|$ and let $u_k=Y_{i_k}/\|Y_{i_k}\|$. Set the $k$-th estimated vertex as $\hat{v}_k=\hat{r}_{i_k}$. Update $Y_i$ to $(1-u_ku_k')Y_i$, for $1\leq i\leq n$. 
\end{itemize}
This algorithm has been analyzed in various literature. We only need to adapt the existing results. The next lemma is from \cite[Theorem 3]{gillis2013fast}. 
\begin{lemma} \label{lem:Gillis}
Fix $m\geq r$ and $n\geq r$. 
Consider a matrix $Y=SM+Z$, where $S\in\mathbb{R}^{m\times r}$ has a full column rank, $M\in\mathbb{R}^{r\times n}$ is a nonnegative matrix such that the sum of each column is at most $1$, and $Z=[Z_1, \ldots,Z_n]\in\mathbb{R}^{m\times n}$. Suppose $M$ has a submatrix equal to $I_r$. Write $\epsilon=\max_{1\leq i\leq n}\|Z_i\|$. Suppose $\epsilon=O(\frac{\sigma_{\min}(S)}{\sqrt{r}\kappa^2(S)})$, where $\sigma_{\min}(S)$ and $\kappa(S)$ are the minimum singular value and condition number of $S$, respectively. If we apply the SP algorithm to columns of $Y$, then it outputs an index set ${\cal K}\subset\{1,2,\ldots,n\}$ such that $|{\cal K}|=r$ and $\max_{1\leq k\leq r}\min_{j\in {\cal K}}\|S_k -Y_j\| = O(\epsilon\kappa^2(S))$, where $S_k$ is the $k$-th column of $S$. 
\end{lemma}

Given ${\cal K}$, the estimated vertices by SP are $\{Y_j\}_{j\in {\cal K}}$. Hence, the above lemma says the maximum $\ell^2$-error on estimating vertices is $
O(\epsilon \kappa^2(S))=O\bigl( \kappa^2(S)\max_{1\leq i\leq n}\|Z_i\|\bigr)$. 

In our setting, we apply SP to $Y_i=(1,\hat{r}_i')'$, $1\leq i\leq n$. We shall re-write the data in the same form as in Lemma~\ref{lem:Gillis}. Recall that $H$ is the orthogonal matrix in Theorem~\ref{thm:r-entrywise-bound} and $v_1,\ldots,v_K$ are vertices of the Ideal Simplex. By definition,
\[
\begin{pmatrix}
1 & \cdots & 1\\
H^{-1}v_1 & \cdots & H^{-1}v_K
\end{pmatrix} w_i =\begin{pmatrix}1\\ H^{-1}r_i\end{pmatrix}.
\] 
Let $\tilde{v}_k=(1, (H^{-1}v_k)')'$, $\tilde{r}_i=(1, (H^{-1}r_i)')'$, $z_i=(0, (\hat{r}_i-H^{-1}r_i)')'$, $1\leq k\leq K$, $1\leq i\leq n$. It is seen that
\[
(1, \hat{r}_i')'\equiv Y_i = [\tilde{v}_1,\ldots,\tilde{v}_K]w_i + z_i. 
\]
Write $Y=[Y_1,\ldots,Y_n]\in\mathbb{R}^{K\times n}$, $\tilde{V}=[\tilde{v}_1,\ldots,\tilde{v}_K]\in\mathbb{R}^{K\times K}$, $W=[w_1,\ldots,w_n]\in\mathbb{R}^{K\times n}$, and $Z=[z_1,\ldots,z_n]\in\mathbb{R}^{K\times n}$. The above can be re-written as
\beq \label{proof-SP-1}
Y = \tilde{V}W + Z. 
\eeq
This reduces to the form in Lemma~\ref{lem:Gillis} with $m=K$. 
To apply Lemma~\ref{lem:Gillis}, we note that $\tilde{V}$ can be re-written as
\[
\tilde{V}=\mathrm{diag}(1, H^{-1})\cdot Q, \qquad \mbox{where}\quad Q = \begin{pmatrix}
1 & \cdots & 1\\
v_1 & \cdots & v_K
\end{pmatrix}. 
\]
Since $\mathrm{diag}(1, H^{-1})$ is an orthogonal matrix, the singular values of $\tilde{V}$ are the same as the singular values of $Q$. Moreover, by \eqref{Q-eigenvalues}, all the singular values of $Q$ are at the order of $\sqrt{K}$. It follows that
\beq \label{proof-SP-2}
\sigma_{\min}(\tilde{V})\asymp \sqrt{K}, \qquad \kappa(\tilde{V})\asymp 1. 
\eeq
In particular, $\tilde{V}$ has a full rank, and $\frac{\sigma_{\min}(\tilde{V})}{\sqrt{K}\kappa^2(\tilde{V})}\asymp 1$. By Lemma~\ref{lem:Gillis}, the maximum $\ell^2$-error on estimating vertices is $O(\max_{1\leq i\leq n}\|Z_i\|)=O(\max_{1\leq i\leq n}\|\hat{r}_i-H^{-1}r_i\|)= O(\max_{1\leq i\leq n}\|H\hat{r}_i-r_i\|)$. The claim follows immediately. \qed

\subsection{Strong efficiency of SVS and SVS$^*$}

SVS and SVS$^*$ both have a  denoise stage,  
where we use $k$-means to reduce the $n$ rows of $\hat{R}$ into $L$ ``cluster centers", with 
an $L$ that is (usually a few times) larger than $K$. We have seen that the denoise stage makes 
SVS and SVS$^*$ more accurate numerically (see Figure~\ref{fig:VHcompare}). We now give a theoretical justification, where we show that SVS and SVS$^*$ are
strongly efficient (see Definition~\ref{def:effVH}). Without loss of generality, we focus on SVS. The analysis of SVS$^*$ is very similar, which is discussed in the remark in the end. 

 
First, consider Setting 1. Let ${\cal S}_0 = {\cal S}_0(e_1, e_2, \ldots, e_K)$ be the standard simplex in $\mathbb{R}^K$, 
where the vertices $e_1, e_2, \ldots, e_K$ are the standard Euclidean basis vectors of $\mathbb{R}^{K}$.  Fix a density $g$ defined over ${\cal S}_0$ and let ${\cal R} = \{\pi \in {\cal S}_0: g(\pi) > 0\}$ be the support of $g$. We suppose there is a constant $c_0 > 0$ such that 
\begin{equation} \label{Baymodel1} 
\mbox{${\cal R}$ is an open subset of ${\cal S}_0$, and $\mathrm{distance}(e_k, {\cal R}) \geq c_0$, $1\leq k\leq K$}. 
\end{equation} 
Let $\delta_{v}(\pi)$ denote the point mass at  $\pi = v$. Let $\eps_1,\ldots,\eps_K>0$ be constants such that $\sum_{k = 1}^K \eps_k < 1$.
We invoke a random design model where $\pi_i$'s are $iid$ drawn from a mixture
\begin{equation} \label{Baymodel2} 
f(\pi) = \sum_{k=1}^K \eps_k \cdot \delta_{e_k}(\pi) + \Bigl(1 - \sum_{k = 1}^K \eps_k\Bigr) \cdot  g(\pi). 
\end{equation}
\begin{lemma}[Efficiency of SVS, Setting 1] \label{lem:VH1}
Suppose conditions of Theorem~\ref{thm:main-rev} hold. Additionally, suppose $K$ is fixed and rows of $\Pi$ are iid generated from \eqref{Baymodel1}-\eqref{Baymodel2}. We apply the SVS algorithm to rows of $\hat{R}$ with an $L$ that does not change with $n$. Then, there exists $L_0=L_0(g, \epsilon_1,\ldots,\epsilon_K)$ such that, as long as $L\geq L_0$, with probability $1-o(n^{-3})$, the estimated $\hat{v}_1,\ldots,\hat{v}_K$ satisfy $\max_{1\leq k\leq K}\|H\hat{v}_k - v_k\|\leq C \max_{1\leq i\leq n}\|H\hat{r}_i - r_i\|$. 
As a result, the SVS algorithm is efficient. 
\end{lemma}
Lemma \ref{lem:VH1} is proved in Section~\ref{subsubsec:proof-lem-VH1}. Its proof utilizes the Borel-Lebesgue covering theorem to characterize the local centers produced in the denoise stage. 

{\bf Remark}. A noteworthy implication of Lemma~\ref{lem:VH1} is that the performance of SVS is robust to the choice of $L$: an overshooting of $L$ only has negligible effects (so as long as computation is not a serious issue, we can choose a larger $L$ in SVS).  
This is intuitively explained as follows. As $L$ increases, more local centers emerge, and we have two representative scenarios. In the first scenario,  new ``local centers" emerge  in the interior of the Ideal Simplex, while ``local centers" that fall close to one of the vertices of Ideal Simplex remain unaffected.  In this case, as ``local centers" that fall in the interior of the Ideal Simplex  won't be selected in the second stage of SVS, the estimated vertices remain roughly the same as $L$ increases. In the second scenario, 
near a vertex of the Ideal Simplex, the number of ``local centers" increases as $L$ increases. 
However, all these ``local centers" remain close to the vertex, and in its second stage, SVS selects one of these 
``local centers" as the estimated vertex. In this case, the estimates of vertices also remain 
roughly the same as $L$ increases. The above heuristic explanation is made rigorous in the proof of Lemma \ref{lem:VH1}. 
 
Next, consider Setting 2.  Let ${\cal N}_k=\{1\leq i\leq n: \pi_i(k)=1\}$ be the set of pure nodes of community $k$,  $1\leq k\leq K$,  
and let ${\cal M}=\{1\leq i\leq n: \max_{1\leq k\leq K}\pi_i(k)<1\}$ be the set of all mixed nodes.  
We assume there are constants $c_1, c_2\in (0,1)$ such that  
\beq \label{cond-pure}
\min_{1\leq k\leq K} |{\cal N}_k|\geq c_1n,  \qquad  \min_{1\leq k\leq K}\sum_{i \in {\cal N}_k}  \theta^2(i)  \geq c_2\|\theta\|^2. 
\eeq
Furthermore, for a fixed integer $L_0 \geq 1$, we assume there is a partition of ${\cal M}$,  ${\cal M}={\cal M}_1\cup \cdots\cup{\cal M}_{L_0}$,  a set of PMF's $\gamma_1,\cdots, \gamma_{L_0}$, and constants   $c_3, c_4>0$  such that  ($e_k$:  $k$-th standard basis vector  of $\mathbb{R}^{K}$)  
\beq \label{cond1-pi}
\Big\{ \min_{1\leq j\neq \ell\leq L_0} \|\gamma_j-\gamma_\ell\|, \; \min_{1\leq \ell\leq L_0, 1\leq k\leq K } \|\gamma_\ell -e_k\| \Big\}\geq c_3, 
\eeq 
and for each $1 \leq \ell \leq L_0$ (note: $err_n$ is the same as that in \eqref{def:err}),  
\beq \label{cond2-pi}
|{\cal M}_\ell| \geq c_4 |{\cal M}|\geq n\beta_n^{-2}err_n^2, 
 \qquad \max_{i\in{\cal M}_\ell}\|\pi_i - \gamma_\ell\| \leq 1/\log(n).   
\eeq 
In this setting, we assume that the true $\pi_i$'s form several {\it loose clusters}, where the $\pi_i$'s in the same cluster are within a distance of $O(\frac{1}{\log(n)})$ from each other. We note that $\frac{1}{\log(n)}$ is much larger than the order of noise, $ \max_{1\leq i\leq n}\|H\hat{r}_i-r_i\|$ (see Theorem~\ref{thm:r-entrywise-bound}). Hence, the assumed clustering structure is ``loose".
\begin{lemma}[Strong efficiency of SVS, Setting 2] \label{lem:VH2}
Suppose conditions of Theorem~\ref{thm:main-rev} hold. Additionally, suppose $K$ is fixed and $(\Theta,\Pi)$ satisfy  \eqref{cond-pure}-\eqref{cond2-pi}. For any integer $L\geq 1$, denote by $\epsilon_L(\hat{R})$ the sum of squared residuals of applying $k$-means to rows of $\hat{R}$ to get $L$ clusters. We apply the SVS algorithm to rows of $\hat{R}$, with a data-drive choice of $L$:
\begin{equation} \label{chooseLtheory} 
\hat{L}_n(A) = \min\{L\geq K+1:  \epsilon_{L}(\hat{R})<\epsilon_{L-1}(\hat{R})/\log(\log(n)) \}. 
\end{equation}  
With probability $1-o(n^{-3})$, the estimated $\hat{v}_1,\ldots,\hat{v}_K$ satisfy
\begin{equation} \label{chooseLtheoryadd} 
\max_{1\leq k\leq K}\|H\hat{v}_k - v_k\|\leq C\Bigl(n^{-1}\sum_{i=1}^n \|H\hat{r}_i - r_i\|^2\Bigr)^{1/2}.  
\end{equation} 
As a result, the SVS algorithm is strongly efficient. 
\end{lemma}

Lemma~\ref{lem:VH2} is proved in Section~\ref{subsubsec:proof-lem-VH2}. The proof requires unconventional analysis of $k$-means. The challenge comes from that the clusters of $\pi_i$'s are loose. Using the conventional analysis of $k$-means, the VH error is governed by the largest within-cluster variance, which can be as large as $O(\frac{1}{\log(n)})$ for loose clusters (see \eqref{cond2-pi}). The key of the proof is to show that the loose clusters in the interior have negligible effects on the estimated vertices.

{\bf Remark}. Lemmas~\ref{lem:VH1}-\ref{lem:VH2} can be easily extended to SVS$^*$. Let $\hat{h}=\max_{i}\|H\hat{r}_i-r_i\|$. In the proofs of these lemmas, we have shown the following properties of the k-means cluster centers: With high probability, (a) all k-means centers are within a distance of $O(\hat{h})$ to the Ideal Simplex, and (b) for each vertex $v_k$, there is at least one k-means center that is within a distance of $O(\hat{h})$ to $v_k$. SVS$^*$ applies SP to these k-means centers. Therefore, we can apply Lemma~\ref{lem:SP+CVS} pretending that the k-means centers are the data points. This gives the desired claims for SVS$^*$.

\subsubsection{Proof of Lemma~\ref{lem:VH1}} \label{subsubsec:proof-lem-VH1}
Lemma~\ref{lem:VH1} follows directly from the next lemma:
\begin{lemma} \label{lem:VH1detail}
Suppose the conditions of Lemma~\ref{lem:VH1} hold. We apply the SVS algorithm to $\{\hat{r}_i\}_{i=1}^n$ with $L$ being a properly large constant. Write $\hat{h}=\max_{1\leq i\leq n}\|H\hat{r}_i-r_i\|$. The following statements are true. 
\begin{itemize}
\item In the local clustering sub-step, all the local centers output by $k$-means are within a distance of $C\hat{h}$ to the Ideal Simplex. Moreover, for each true vertex $v_k$, there is at least one local center that is within a distance of $C\hat{h}$ to it, $1\leq k\leq K$. 
\item The combinatorial search sub-step selects exactly one local center among those within a distance of $C\hat{h}$ to a true $v_k$, $1\leq k\leq K$. As a result, up to a permutation of estimated vertices, $\max_{1\leq k\leq K}\|H\hat{v}_k - v_k\|\leq C\hat{h}$.
\end{itemize}  
\end{lemma} 
\noindent

\bigskip
\noindent
{\it Proof of Lemma~\ref{lem:VH1detail}}: As explained in the proof of Lemma~\ref{lem:SP+CVS}, we can assume $H=I_{K-1}$ without loss of generality. 

We first argue that, once the first bullet point is proved, the second bullet point follows directly. Let $\hat{m}_1, \hat{m}_2,\ldots,\hat{m}_L$ be the local centers by $k$-means. The combinatorial search step of SVS is an application of CVS on these local centers, and we hope to apply Lemma~\ref{lem:SP+CVS}. 
Note that when the first bullet point of the claim is true, we have:
\begin{itemize}
\item $d(\hat{m}_j, {\cal S}^{ideal})\leq C\hat{h}$, $1\leq j\leq L$. 
\item For each $1\leq k\leq K$, there exists $j_k$ such that $\|\hat{m}_{j_k}-v_k\|\leq C\hat{h}$. 
\end{itemize}
By Lemma~\ref{lem:R}, the distance between two different $v_k$ and $v_\ell$ is lower bounded by a constant times $\sqrt{K}$, while $\hat{h}=o(\sqrt{K})$. As a result, any $\hat{m}_j$ cannot be simultaneously within a distance of $C\hat{h}$ to two vertices, which implies that $j_1,j_2,\ldots,j_K$ are distinct. Define
\[
m_j = \begin{cases}
\mathrm{argmin}_{x\in {\cal S}^{ideal}}\|x - \hat{m}_j\|, & j\notin\{j_1,j_2,\ldots,j_K\},\\
v_k, & j=j_k, 1\leq k\leq K. 
\end{cases}
\]
We then have
\begin{itemize}
\item The points $m_1,m_2,\ldots,m_L$ are in the Ideal Simplex ${\cal S}^{ideal}$.
\item $\|\hat{m}_j-m_j\|\leq C\hat{h}$, $1\leq j\leq L$. 
\item For each $1\leq k\leq K$, there is at least one $m_j$ located at the vertex $v_k$. 
\end{itemize}
If we view $\hat{m}_1,\hat{m}_2,\ldots,\hat{m}_L$ as the data points 
and view $m_{j_1},\ldots,m_{j_K}$ as the ``pure nodes",  we can apply Lemma~\ref{lem:SP+CVS} to get $
\max_{1\leq k\leq K}\|\hat{v}_k - v_k\|\leq C\max_{1\leq j\leq L}\|\hat{m}_j-m_j\|\leq C\hat{h}$.

Therefore, it suffices to prove the first bullet point of the claim. For any $L\geq 1$, let $RSS(L)$ be the objective achieved by applying $k$-means to mixed $r_i$'s assuming $\leq L$ clusters:
\[
RSS(L)=\min_{\text{$L$ cluster centers}}\sum_{\text{mixed nodes }i}\|r_i - (\mbox{closest-cluster-center})\|^2. 
\]
In preparation, we study $RSS(L)$ as a function of $L$. 
 
We provide an upper bound of $RSS(L)$ by constructing a feasible solution to the $k$-means problem. In the proof of Lemma~\ref{lem:R}, we see that there is a one-to-one mapping $T=T_2\circ T_1$ from the standard simplex ${\cal S}_0$ to the Ideal Simplex ${\cal S}^{ideal}$ such that $r_i=T(\pi_i)$ and that (note: we have used that $K$ is a constant)
\beq \label{Lipschitz}
C^{-1}\|x-y\|\leq \|T(x)-T(y)\|\leq C\|x-y\|, \qquad \mbox{for any $x,y\in {\cal S}_0$}. 
\eeq
For an integer $s=\lfloor L^{\frac{1}{K-1}}-1\rfloor$, we consider the following choice of centers: 
\[
\Bigl\{T(x): \mbox{$x\in {\cal S}_0$, entries of $x$ take value on } \bigl\{0, \frac{1}{s}, ..., \frac{s-1}{s}, 1\bigr\}\Bigr\}. 
\]
The total number of centers is bounded by $(s+1)^{K-1}\leq L$. We then assign each $r_i$ to the nearest center. The $\ell^\infty$-distance from each $\pi_i$ to the nearest $x$ above is at most $1/s$, so the Euclidean distance is at most $\sqrt{K}/s$; combining it with  \eqref{Lipschitz}, the Euclidean distance from $r_i=T(\pi_i)$ to the nearest $T(x)$ above is at most $C\sqrt{K}/s$. 
It follows that 
\[
RSS(L) \leq  n(C\sqrt{K}/s)^2.  
\]
The choice of $s$ guarantees that $s> L^{\frac{1}{K-1}}-2$. 
As a result, for a constant $\tilde{c}$ that does not depend on $L$, 
\beq \label{RSS}
RSS(L) \leq n\cdot \tilde{c} L^{- \frac{2}{K-1}}.  
\eeq

We are now ready to prove the first bullet point. Note that each $\hat{r}_i$ is within a distance $C\hat{h}$ to the corresponding $r_i$ and that all the $r_i$'s are in the Ideal Simplex. Hence, all data points $\{\hat{r}_i\}_{i=1}^n$ are within a distance $C\hat{h}$ to the Ideal Simplex. It is easy to see that all local centers output by $k$-means must also be within a distance $C\hat{h}$ to the Ideal Simplex. What remains is to show that there is at least one local center within a distance of $C\hat{h}$ to each true vertex $v_k$. Fix $v_k$. Our strategy is as follows: for a constant $\ell_0$ to be decided, 
\begin{itemize}
\item[(a)] We first show that there exists at least one local center within a distance $\ell_0$ to $v_k$. 
\item[(b)] We then show that, for each local center within a distance $\ell_0$ to $v_k$, the associated data cluster consists of only pure $\hat{r}_i$ from community $k$. 
\end{itemize}
Then, by the nature of $k$-means, such a local center equals to the average of all the $\hat{r}_i$ assigned to this cluster. Since each $\hat{r}_i$ corresponds to a pure node of community $k$, it is within a distance $C\hat{h}$ to $v_k$. As a result, the local center must also be within a distance $C\hat{h}$ to $v_k$. This gives the first bullet point. 

What remains is to prove (a) and (b). Fix $v_k$. Consider (a). Suppose there are no local centers within a distance $\ell_0$ to $v_k$. Then, each pure $r_i$ from community $k$ has a distance $> \ell_0$ to the nearest local center; hence, the distance from $\hat{r}_i$ to the nearest local center is at least $\ell_0-C\hat{h}\geq \ell_0/2$. At the same time, by the generating process of $\pi_i$'s, with probability $1-o(n^{-3})$, the number of pure nodes of community $k$ is at least $n\eps_k/2$. These pure nodes contribute a sum-of-squares of
\[
\geq (n\eps_k/2)\cdot (\ell_0/2)^2= n(\ell_0^2\eps_k/8). 
\] 
Additionally, the mixed $\hat{r}_i$'s are assigned to at most $L$ clusters. Since $\|\hat{r}_i-x\|^2\geq \|r_i-x\|^2/2-O(\hat{h}^2)$ for any point $x$, we immediately know that the sum-of-squares contributed by mixed $\hat{r}_i$'s is
\[
\geq \frac{1}{2}RSS(L) - O(n\hat{h}^2). 
\]
Combining the above, the objective attained by $k$-means is 
\beq \label{kmeans-1}
\geq \frac{1}{2}RSS(L) + n(\ell_0^2\eps_k/9)
\eeq
At the same time, we construct an alternative solution by letting $(L-K)$ of the local centers be those associated with $RSS(L-K)$, letting the remaining $K$ centers be $v_1,v_2,\ldots,v_K$, and assigning each $\hat{r}_i$ to the center closest to the corresponding $r_i$. Since $\|\hat{r}_i-x\|^2\leq 2\|r_i-x\|^2+O(\hat{h}^2)$, the sum of squares attained by this solution is 
\beq \label{kmeans-2}
\leq 2 RSS(L-K) + O(n\hat{h}^2). 
\eeq 
A contradiction is obtained as long as
\begin{align*}
2RSS(L-K) - \frac{1}{2} RSS(L+K) & < n(\ell_0^2\eps_k/9) - O(n\hat{h}^2)\cr
&<  n(\ell_0^2/10). 
\end{align*}
According to \eqref{RSS}, the above is true if we choose $L>(20\tilde{c}/\ell_0^2)^{\frac{K-1}{2}}$. This proves (a). 

Consider (b). Fix $k$. Let $\hat{m}^*$ be a local center such that $\|\hat{m}^* - v_k\|\leq \ell_0$. By the assumption \eqref{Baymodel1}, for any $\pi_i\neq e_k$, its distance to $e_k$ ($e_k$ is the $k$-th standard basis of $\mathbb{R}^K$) is at least $c_0$. Combining it with \eqref{Lipschitz}, for any node $i$ that is not a pure node of community $k$, the distance from $r_i$ to $v_k$ is at least $C^{-1}c_0$. As a result, for any such node, 
\[
\|\hat{r}_i - \hat{m}^*\|\geq C^{-1}c_0 - \ell_0- C\hat{h}. 
\]
By taking $\ell_0=C^{-1}c_0/4.1$, for any node $i$ not pure of community $k$, 
\beq \label{VH2proof-temp}
\mbox{the distance from $\hat{r}_i$ to the center $\hat{m}^*$ is at least } 3\ell_0.
\eeq
We shall also show that, for any node $i$ not pure of community $k$,  
\beq \label{VH2proof-temp2}
\mbox{the distance from $\hat{r}_i$ to the nearest center is at most } 2.5\ell_0. 
\eeq
By \eqref{VH2proof-temp}-\eqref{VH2proof-temp2}, these nodes cannot be assigned to $\hat{m}^*$. Therefore, the cluster associated with $\hat{m}^*$ consists of only those $\hat{r}_i$ such that $i$ is a pure node of community $k$. This proves (b). 

What remains is to prove \eqref{VH2proof-temp2}. If $i$ is a pure node of a different community $\ell$, then by (a) above, the distance from $r_i=v_{\ell}$ to the nearest center is $\ell_0+C\hat{h} < 2.5\ell_0$. Hence, we only need to consider $i$ that is a mixed node.   
Since $\max_{i}\|\hat{r}_i-r_i\|\leq C\hat{h} \ll 0.5\ell_0$, it suffices to show that
\beq \label{VH2proof-key}
\mbox{the distance from a mixed $r_i$ to the nearest center is at most } 2\ell_0. 
\eeq
Let ${\cal S}_0={\cal S}_0(e_1,\ldots,e_K)\in\mathbb{R}^K$ be the standard $(K-1)$-simplex, and denote by ${\cal B}(x; c)$ an open ball in ${\cal S}_0$ centered at $x$ with a radius $c$; we notice that here an ``open ball" means the intersection of ${\cal S}_0$ and an open ball in $\mathbb{R}^K$. Let $\bar{\cal R}$ be the closure of ${\cal R}$, where ${\cal R}$ is the support of $f(\cdot)$. We consider the open cover of $\bar{\cal R}$: 
\[
\bigl\{ {\cal B}(x, C^{-1}\ell_0): x\in {\cal R} \bigr\}. 
\]
Since $\bar{\cal R}$ is closed and bounded, it is a compact set. 
According to the Borel-Lebesgue covering theorem, the above open cover has a finite sub-cover:
\[
\bigl\{ {\cal B}(x_1, C^{-1}\ell_0), {\cal B}(x_2, C^{-1}\ell_0),  \ldots, {\cal B}(x_p, C^{-1}\ell_0) \bigr\}, \qquad \mbox{where } x_1,\ldots,x_p\in {\cal R}.  
\] 
This means each $\pi_i\neq e_k$ is contained in one ${\cal B}(x_j, C^{-1}\ell_0)$. Recalling that $T$ is the mapping in \eqref{Lipschitz}, define
\[
{\cal B}^*_j =T\bigl( {\cal B}(x_j, C^{-1}\ell_0)\bigr), \qquad 1\leq j\leq p.  
\]
Then, $r_i=T(\pi_i)$ is contained in ${\cal B}^*_j$. Moreover, for any $y,\tilde{y}\in {\cal B}^*_j$, $\|y-\tilde{y}\|\leq C\max_{x,\tilde{x}\in {\cal B}(x_j, C^{-1}\ell_0)}\leq 2\ell_0$. Therefore, if we can show that  
\beq  \label{VH2proof-key0}
\mbox{each ${\cal B}^*_j$ contains at least one local center}, 1\leq j\leq p,  
\eeq
then the distance from $r_i$ to this local center is bounded by $2\ell_0$. This gives
\eqref{VH2proof-key}, and in turn gives \eqref{VH2proof-temp2}.  

What remains is to prove \eqref{VH2proof-key0}. Note that ${\cal R}$ is an open set. 
By definition of open sets, for each of $x_1,x_2,\ldots,x_p$, there is a $\tau_j>0$ such that the closed ball $\bar{\cal B}(x_j, \tau_j)$ is contained in ${\cal R}$. We define the closed balls 
\[
{\cal BB}_j\equiv \bar{\cal B}\bigl(x_j, \min\{\tau_j, C^{-1}\ell_0/2\} \bigr), \qquad 1\leq j\leq p.  
\]
Let $\omega_j=\int f(\pi)1\{\pi\in {\cal BB}_j\}d\pi = (1-\sum_{k=1}^K\eps_k)\int g(\pi)1\{\pi\in {\cal BB}_j\}d\pi$, $1\leq j\leq p$. Note that each of these closed balls is contained in the support of $g$ with a nonzero radius and that $g$ as a probability density is measurable. We immediately know that $\omega_j>0$. From the assumption \eqref{Baymodel2} and elementary large-deviation inequalities (e.g., the Hoeffding's inequality), we know that with probability $1-o(n^{-3})$, for $1\leq j\leq p$, 
\beq  \label{VH2proof-temp3}
\mbox{the number of $\pi_i$'s contained in ${\cal BB}_j$ is at least $n\omega_j/2$}. 
\eeq
With \eqref{VH2proof-temp3}, we now prove \eqref{VH2proof-key0} by contradiction. Suppose \eqref{VH2proof-key0} does not hold, i.e., there exists ${\cal B}_j^*$ such that
\[
{\cal B}_j^*\cap \{\hat{m}_1,\hat{m}_2,\ldots,\hat{m}_L\}=\emptyset,
\]
where $\hat{m}_1,\hat{m}_2,\ldots,\hat{m}_L$ are the local centers output by $k$-means. 
By definition of ${\cal B}_j^*$ and the fact that $T$ is a one-to-one mapping, we have
\[
{\cal B}(x_j, C^{-1}\ell_0)\cap \bigl\{T^{-1}(\hat{m}_1), T^{-1}(\hat{m}_2), \ldots, T^{-1}(\hat{m}_L)\bigr\}=\emptyset. 
\]
Note that ${\cal BB}_j$ is a ball also centered at $x_j$ but with a radius no larger than half of the radius of ${\cal B}(x_j, C^{-1}\ell_0)$. As a result, for any $x\in {\cal BB}_j$, its distance to the nearest one of $T^{-1}(\hat{m}_1),\cdots,T^{-1}(\hat{m}_L)$ is at least $C^{-1}\ell_0/2$; combining it with \eqref{Lipschitz}, the distance from $T(x)$ to the nearest one of $\hat{m}_1,\hat{m}_2,\ldots,\hat{m}_L$ is at least $C^{-2}\ell_0/2$. It follows that 
\[
\mbox{for any $\pi_i\in {\cal BB}_j$}, \min_{1\leq s\leq L}\|r_i - \hat{m}_s\|\geq C^{-2}\ell_0/2.  
\]
Note that $\max_i\|\hat{r}_i-r_i\|\leq C\hat{h} = o(1)$. We further conclude that 
\beq \label{VH2proof-temp4}
\begin{array}{c}
\mbox{for any $\pi_i\in {\cal BB}_j$, the distance from $\hat{r}_i$}\\
\mbox{to the nearest local center is $\geq C^{-2}\ell_0/3$}. 
\end{array}  
\eeq
Combining \eqref{VH2proof-temp3}-\eqref{VH2proof-temp4}, the sum-of-squares attained by $k$-means is
\[
\geq (C^{-2}\ell_0/3)^2\cdot (n\omega_j/2)\geq n(\omega_{\min}C^{-4}\ell^2_0/18),  
\]
where $\omega_{\min}=\min\{\omega_1,\ldots,\omega_p\}$. At the same time, the objective attained by $k$-means should be
\[
\leq RSS(L) + n(C\hat{h}^2). 
\] 
A contradiction is obtained as long as 
\beq\label{VH2proof-temp5}
RSS(L) < n(\omega_{\min}C^{-4}\ell^2_0/18) -  n(C\hat{h}^2). 
\eeq
Comparing it with \eqref{RSS}, as long as $L>(\frac{19C^4\tilde{c}}{\ell_0^2\omega_{\min}})^{\frac{K-1}{2}}$, the inequality \eqref{VH2proof-temp5} will be true. We then have a contradiction, which implies that \eqref{VH2proof-key0} must hold. The proof is now complete. \qed

\subsubsection{Proof of Lemma~\ref{lem:VH2}} \label{subsubsec:proof-lem-VH2}
Lemma~\ref{lem:VH2} follows directly from the next lemma:
\begin{lemma}  \label{lem:VH2detail}
Suppose the conditions of Lemma~\ref{lem:VH2} hold.  We apply the SVS algorithm to $\{\hat{r}_i\}_{i=1}^n$ with $L = \hat{L}_n(A)$, where $\hat{L}_n(A)$ is defined in \eqref{chooseLtheory}. Let $\hat{h}^*=\sqrt{n^{-1}\sum_{i=1}^n\|H\hat{r}_i-r_i\|^2}$ and $\hat{h}=\max_{1\leq i\leq n}\|H\hat{r}_i-r_i\|$. With probability $1 - o(n^{-3})$, the following statements are true.
\begin{itemize} 
\item $\hat{L}_n(A)=L_0+K$.  
\item The local clustering sub-step identifies $(L_0 + K)$ local centers, where there is a unique  $(K-1)$-simplex such that  $K$ of these centers (denoted by $\hat{v}_1, \hat{v}_2, \ldots, \hat{v}_K$) are its vertices, and all other centers are within a distance of $C\hat{h}$ to this simplex. These $K$ local centers will be identified by the combinatorial search sub-step. 
\item The above $K$ local centers satisfy $\hat{v}_k=|{\cal N}_k|^{-1}\sum_{i\in {\cal N}_k}\hat{r}_i$, $1\leq k\leq K$.  
As a result, up to a permutation of estimated vertices, $\max_{1\leq k\leq K}\|H\hat{v}_k - v_k\|\leq C\hat{h}^*$.
\end{itemize} 
\end{lemma} 

\bigskip
\noindent
{\it Proof of Lemma~\ref{lem:VH2detail}}: As explained in the proof of Lemma~\ref{lem:SP+CVS}, we can assume $H=I_{K-1}$ without loss of generality. 
By Theorem~\ref{thm:r-entrywise-bound} and Lemma~\ref{lem:hatR-l2}, with probability $1-o(n^{-3})$, 
\beq \label{VH3proof-0}
\hat{h}\equiv\max_{1\leq i\leq n}\|\hat{r}_i-r_i\|\leq \frac{Cerr_n}{\beta_n}, \quad n(h^*)^2\equiv\sum_{i=1}^n \|\hat{r}_i - r_i\|^2\leq \frac{Cn(err_n^*)^2}{\beta_n^2},
\eeq
where we have absorbed the factors of $K$ into the constants. 
We also note that $err_n^*\leq err_n/\sqrt{\log(n)}$. Below, we restrict to the event of \eqref{VH3proof-0}.

First, we study $\hat{L}_n(A)$. Recall that $\gamma_1,\gamma_2,\ldots,\gamma_{L_0}$ are as in \eqref{cond1-pi}. Let $T$ be the mapping as in \eqref{Lipschitz}; note that $T(\pi_i)=r_i$ for $1\leq i\leq n$. Introduce
\[
m_j = T(\gamma_j), \qquad 1\leq j\leq L_0. 
\]
By \eqref{Lipschitz}, the assumptions \eqref{cond1-pi}-\eqref{cond2-pi} imply that the distance between any two of $\{v_1,v_2,\ldots,v_K,m_1,m_2,\ldots,m_{L_0}\}$ is at least $c$, and $\max_{i\in {\cal M}_j}\|r_i-m_j\|\leq C_1/\log(n)$, where $c>0$ and $C_1>0$ are constants. In particular,
\[
\alpha_n^2\leq \frac{C|{\cal M}|}{n\log(n)}, \qquad \mbox{where}\quad \alpha^2_n\equiv  n^{-1}\sum_{j=1}^{L_0}\sum_{i\in{\cal M}_j}\|r_i - m_j\|^2. 
\] 
We now study $\eps_L(\hat{R})$. When $L=L_0+K$, by choosing this choice of centers $\{v_1,\ldots,v_K,m_1,\ldots,m_{L_0}\}$, it is easy to see that
\beq \label{chooseLproof-1}
\eps_{L_0+K}(\hat{R})\leq n\alpha_n^2 + C\sum_{i=1}^n\|\hat{r}_i-r_i\|^2\leq \frac{C|{\cal M}|}{\log(n)}, 
\eeq
where the last inequality is due to \eqref{VH3proof-0} and the assumption that $|{\cal M}|\geq n \beta_n^{-2}err_n^2\geq n\beta_n^{-2}(err_n^*)^2\log(n)$. 
When $K\leq L<L_0+K$, suppose there are $L_1$ of $\{v_1,v_2,\ldots,v_K\}$ and $L_2$ of $\{m_1,m_2,\ldots,m_{L_0}\}$ such that no local centers are within a distance of $c/3$ of them. Since the distance between any two of $\{v_1,v_2,\ldots,v_K,m_1,m_2,\ldots,m_{L_0}\}$ is at least $c$, we have that $(L_1+L_2)$ is at least $(L_0+K)-L$. For any such $v_k$ and $i\in {\cal N}_k$ or such $m_j$ and $i\in {\cal M}_j$, the distance from $\hat{r}_i$ to the nearest local center is at least $c/3-\hat{h}\geq c/4$. It follows that 
\beq \label{chooseLproof-2}
\eps_{L}(\hat{R}) \geq (c/4)^2\cdot(L_1\min_k|{\cal N}_k| + L_2\min_j |{\cal M}_j|)\geq C|{\cal M}|, 
\eeq
where the last inequality is due to $\min_k|{\cal N}_k|\geq c_1n$ and $\min_j|{\cal M}_j|\geq c_4|{\cal M}|$. 
At the same time, by choosing the centers to be $\{v_1,v_2,\ldots,v_K\}$ and $(L-K)$ of $\{m_1,m_2,\ldots,m_{L_0}\}$, 
\beq \label{chooseLproof-3}
\eps_L(\hat{R})\leq C(L_0+K-L)|{\cal M}| + C\sum_{i=1}^n\|\hat{r}_i-r_i\|^2\leq C|{\cal M}|.  
\eeq
By \eqref{chooseLproof-1}-\eqref{chooseLproof-3}, 
\[
\eps_{L}(\hat{R})/\eps_{L-1}(\hat{R})
\begin{cases}
\leq C/\log(n), & L=L_0+K,\cr
\geq C, & K+1\leq L\leq L_0+K. 
\end{cases}
\]
Hence, the definition of $\hat{L}_n(A)$ in \eqref{chooseLtheory} yields $\hat{L}_n(A)=L_0+K$. This proves the first bullet point.  

Next, we consider the second bullet point. Suppose for $L_1$ of $\{v_1,v_2,\ldots,v_K\}$ and $L_2$ of $\{m_1,m_2,\ldots,m_{L_0}\}$, there are no local centers are within a distance of $c/4$ of them. When $L_1+L_2\geq 1$, using similar arguments as those for proving \eqref{chooseLproof-2}, we can see that the associated sum-of-squares is lower bounded by $C|{\cal M}|$. However, in \eqref{chooseLproof-1}, we have seen that the sum-of-squares attained by $k$-means is at most $C|{\cal M}|/\log(n)$. Hence, the above situation is impossible, i.e., for each of $\{v_1,v_2,\ldots,v_K,m_1,\ldots,m_{L_0}\}$, there is at least one local center within a distance $c/4$ to it. Since that the distance between any two of $\{v_1,v_2,\ldots,v_K,m_1,\ldots,m_{L_0}\}$ is at least $c$, these $(L_0+K)$ local centers must be distinct. Noting that there are at most $\hat{L}_n(A)=L_0+K$ cluster centers in total, we find that 
\beq \label{VH3proof-1}
\begin{array}{c}
\mbox{there is exactly one local center within a distance $c/4$}\\
\mbox{ to each of $\{v_1,v_2,\ldots,v_K,m_1,m_2\ldots,m_{L_0}\}$}.
\end{array} 
\eeq
Denote by $\hat{m}^*_{(k)}$ the local center nearest to $v_k$ and by $\hat{m}_{(j)}$ the local center nearest to $m_j$, $1\leq k\leq K$, $1\leq j\leq L_0$. For any $i\in {\cal N}_k$, the distance from $\hat{r}_i$ to $\hat{m}^*_{(k)}$ is at most $c/4+ O(\hat{h})\leq c/3$, but its distance to any other local center is at least $c-c/4-O(\hat{h})\geq 2c/3$; hence, $\hat{r}_i$ can only be assigned to the cluster associated with $\hat{m}^*_{(k)}$. Similarly, for any $i\in {\cal M}_j$, the distance from $\hat{r}_i$ to $\hat{m}_{(j)}$ is at most $c/4+O(\frac{1}{\log(n)})+O(\hat{h})\leq c/3$, but the distance to any other local center is at least $c-c/4-O(\frac{1}{\log(n)})-O(\hat{h})\geq 2c/3$; so $\hat{r}_i$ must be assigned to $\hat{m}_{(j)}$. We have proved that
\beq \label{VH3proof-2}
\left\lbrace\begin{array}{l}
\mbox{the cluster associated with $\hat{m}^*_{(k)}$ is }\{\hat{r}_i: i\in {\cal N}_k\}, 1\leq k\leq K,\cr
\mbox{the cluster associated with $\hat{m}_{(j)}$ is }\{\hat{r}_i: i\in {\cal M}_j\}, 1\leq j\leq L_0. 
\end{array}\right.
\eeq
Then, it is easy to see that 
\begin{itemize}
\item All the local centers are within a distance $\hat{h}$ to the Ideal Simplex.
\item Each $\hat{m}^*_{(k)}$ is within a distance $C\hat{h}$ to $v_k$, $1\leq k\leq K$.
\item Each $\hat{m}_{(j)}$ is within a distance $C/\log(n)$ to $m_j$, $1\leq j\leq L_0$.
\end{itemize}
We now show that $\hat{m}^*_{(1)},\hat{m}^*_{(2)},\dots,\hat{m}^*_{(K)}$ will be selected by the combinatorial search. The proof is similar to that of Lemma~\ref{lem:SP+CVS} but is simpler. Suppose one $\hat{m}^*_{(k)}$ is not selected by the combinatorial search. By \eqref{VH3proof-2}, the other local centers are contained in the convex hull ${\cal H}\{\hat{r}_i: i\notin {\cal N}_k\}$. Hence, the estimated simplex $\hat{\cal S}\subset {\cal H}\{\hat{r}_i: i\notin {\cal N}_k\}$.
We notice that the distance from $e_k$ to the convex hull of all $\pi_i\neq e_k$ is lower bounded by a constant, as a result of the assumptions \eqref{cond1-pi}-\eqref{cond2-pi}. Using \eqref{Lipschitz}, we know that the distance from $v_k$ to the convex hull ${\cal H}\{r_i: i\notin {\cal N}_k\}$ is also lower bounded by a constant. Then, 
\begin{align*}
d(\hat{m}^{*}_{(k)}, \hat{\cal S}) &\geq d\bigl(\hat{m}^{*}_{(k)},\; {\cal H}\{\hat{r}_i: i\notin {\cal N}_k\})\cr
&\geq d(v_k, {\cal H}\{r_i: i\notin {\cal N}_k\}) -O(\hat{h})\cr
&\geq C. 
\end{align*} 
At the same time, if we pick the $K$ local centers $\hat{m}^*_{(1)},\hat{m}^*_{(2)},\dots,\hat{m}^*_{(K)}$,
\[
\max_{1\leq j\leq L_0}d\bigl(\hat{m}_j, {\cal S}(\hat{m}^*_{(1)},\hat{m}^*_{(2)},\dots,\hat{m}^*_{(K)})\bigr)\leq C\hat{h}. 
\]
This yields a contradiction since $\hat{h}=o(1)$. As a result, all of $\hat{m}^*_{(1)},\hat{m}^*_{(2)},\dots,\hat{m}^*_{(K)}$ will be selected by the combinatorial search. 

Last, we prove the third bullet point. So far, we have seen that $\hat{v}_k=\hat{m}^*_{(k)}$ (up to a label permutation). By \eqref{VH3proof-2} and the nature of $k$-means solutions,
\[
\hat{v}_k = |{\cal N}_k|^{-1}\sum_{i\in {\cal N}_k}\hat{r}_i, \qquad 1\leq k\leq K. 
\]
We note that $0\leq \sum_{i\in {\cal N}_k}\|\hat{r}_i-\hat{v}_k\|^2=\sum_{i\in {\cal N}_k}\{\|\hat{r}_i-v_k\|^2-2(\hat{v}_k-v_k)'(\hat{r}_i-v_k)+\|\hat{v}_k-v_k\|^2)\}=\sum_{i\in {\cal N}_k}\|\hat{r}_i-v_k\|^2-|{\cal N}_k|\|\hat{v}_k-v_k\|^2$. As a result, 
\[
\|\hat{v}_k - v_k\|^2 \leq \frac{1}{|{\cal N}_k|}\sum_{i\in {\cal N}_k}\|\hat{r}_i-v_k\|^2 \leq \frac{1}{|{\cal N}_k|}\sum_{i=1}^n\|\hat{r}_i-r_i\|^2, \quad 1\leq k\leq K. 
\]
Since $|{\cal N}_k|\geq c_1n$, it follows that
\beq \label{VHtighter}
\max_{1\leq k\leq K}\|\hat{v}_k - v_k\| \leq C\sqrt{n^{-1}\sum_{i=1}^n\|\hat{r}_i-r_i\|^2} \leq C\hat{h}^*. 
\eeq
This proves the third bullet point. \qed

\section{Rates of Convergence of Mixed-SCORE} \label{sec:app-Rates}

We prove the main results about Mixed-SCORE, including Theorems~\ref{thm:main-rev}-\ref{thm:SVS2}. 
\subsection{Proofs of Theorem~\ref{thm:main-rev}} \label{sec:mainproof}
Let $H$ be the orthogonal matrix as in Theorem~\ref{thm:r-entrywise-bound}. 
We aim to show that, with probability $1-o(n^{-3})$, for all $1\leq i\leq n$, 
\beq \label{decomp}
\|\hat{\pi}_i-\pi_i\|_1 \leq C\|H\hat{r}_i-r_i\|+C\max_{1\leq k\leq K}\|H\hat{v}_k-v_k\|+CKerr_n. 
\eeq
Once \eqref{decomp} is true, by efficiency of the VH algorithm (see Definition~\ref{def:effVH}) and the bound in Theorem~\ref{thm:r-entrywise-bound}, we immediately have that, with probability $1-o(n^{-3})$, 
\beq \label{mainproof-target}
\max_{1\leq i\leq n}\|\hat{\pi}_i-\pi_i\|_1\leq CK^{3/2}\beta_n^{-1}err_n. 
\eeq
Note that $\|\hat{\pi}_i-\pi_i\|^2\leq \|\hat{\pi}_i-\pi_i\|_\infty\|\hat{\pi}_i-\pi_i\|_1\leq \|\hat{\pi}_i-\pi_i\|_1^2$. It follows that $\frac{1}{n}\sum_{i=1}^n\|\hat{\pi}_i-\pi_i\|^2\leq \max_{1\leq i\leq n}\|\hat{\pi}_i-\pi_i\|^2\leq \max_{1\leq i\leq n}\|\hat{\pi}_i-\pi_i\|_1^2\leq CK^3\beta_n^{-2}err_n^2$, with probability $1-o(n^{-3})$. Moreover, $\sum_{i=1}^n\|\hat{\pi}_i-\pi_i\|^2\leq 2$ always holds. Combining these arguments gives
\[
\mathbb{E}\Bigl[\frac{1}{n}\sum_{i=1}^n\|\hat{\pi}_i-\pi_i^2\|\Bigr]\leq CK^3\beta_n^{-2}err_n^2+o(n^{-3}). 
\]
This proves the first claim. The second claim follows directly by noting that $err^2_n\leq (n\bar{\theta}^2)^{-1}$ if $\theta_{\max}\leq C\theta_{\min}$.

Below, we show \eqref{decomp}. In the Membership Reconstruction (MR) step, we compute $\hat{w}_i$ and $\hat{b}_1$, then use them to construct 
\beq \label{pihat-star}
\hat{\pi}_i^*(k) = \max\{0, \hat{w}_i(k)/\hat{b}_1(k)\}, \qquad 1\leq k\leq K, 
\eeq
and then estimates $\pi_i$ by $\hat{\pi}_i = \hat{\pi}_i^*/ \|\hat{\pi}_i^*\|_1$. We shall study $\hat{w}_i$ and $\hat{b}_1$ separately and then combine their error bounds to get \eqref{decomp}.

First, we study $\hat{w}_i$. By definition,
\beq \label{mainproof-add}
\underbrace{\begin{pmatrix}
1 & \cdots & 1\\
v_1 & \cdots & v_K
\end{pmatrix}}_{\equiv Q} w_i =\begin{pmatrix}1\\ r_i\end{pmatrix}, \qquad  
\underbrace{\begin{pmatrix}
1 & \cdots & 1\\
H\hat{v}_1 & \cdots & H\hat{v}_K
\end{pmatrix}}_{\equiv \hat{Q}} \hat{w}_i = \begin{pmatrix}1\\H\hat{r}_i\end{pmatrix}. 
\eeq 
We thus write
\begin{eqnarray*}
\hat{w}_i - w_i &=& \hat{Q}^{-1}\begin{pmatrix}1\\H\hat{r}_i\end{pmatrix} - Q^{-1}\begin{pmatrix}1\\r_i\end{pmatrix}\cr
&=& \hat{Q}^{-1}\biggl[\begin{pmatrix}1\\H\hat{r}_i\end{pmatrix}- \begin{pmatrix}1\\r_i\end{pmatrix}\biggr] - (Q^{-1}-\hat{Q}^{-1})\begin{pmatrix}1\\r_i\end{pmatrix}\cr
&=& \hat{Q}^{-1}\begin{pmatrix}0\\H\hat{r}_i-r_i\end{pmatrix} - \hat{Q}^{-1}(\hat{Q}-Q)Q^{-1}\begin{pmatrix}1\\r_i\end{pmatrix}\cr
&=& \hat{Q}^{-1}\begin{pmatrix}0\\H\hat{r}_i-r_i\end{pmatrix} - \hat{Q}^{-1}(\hat{Q}-Q)w_i. 
\end{eqnarray*}
It follows that 
\beq \label{ThmMainrev-1}
\|\hat{w}_i-w_i\| \leq \|\hat{Q}^{-1}\|\cdot\bigl( \|H\hat{r}_i-r_i\|+\|(\hat{Q}-Q)w_i\|\bigr). 
\eeq
This matrix $Q$ is studied in the proof of Lemma~\ref{lem:R}, where we prove $\|Q^{-1}\|=O(1/\sqrt{K})$; see \eqref{Q-eigenvalues}. This means the minimum singular value of $Q$ is lower bounded by $C\sqrt{K}$. Moreover, $\|\hat{Q}-Q\|\leq \|\hat{Q}-Q\|_F\leq \sqrt{K}\max_{1\leq k\leq K}\|H\hat{v}_k-v_k\|=o(\sqrt{K})$. As a result, the minimum singular value of $\hat{Q}$ is also lower bounded by $C\sqrt{K}$. It leads to
\[
\|\hat{Q}^{-1}\|\leq C/\sqrt{K}. 
\]
We note that $(\hat{Q}-Q)w_i\in\mathbb{R}^K$ is a vector whose first entry is $0$ and whose remaining entries are equal to $\sum_{k=2}^K w_i(k)(\hat{v}_k-v_k)\in\mathbb{R}^{K-1}$. Since $w_i$ contains the coefficients of writing $r_i$ as a convex combination of $v_1,\ldots,v_K$, we have $\|w_i\|_1=1$. Therefore, 
\[
\|(\hat{Q}-Q)w_i\| =\Bigl\|\sum_{k=1}^K w_i(k)(H\hat{v}_k-v_k)\Bigr\|\leq \sum_{k=1}^Kw_i(k)\|H\hat{v}_k-v_k\|\leq \max_{1\leq k\leq K}\|H\hat{v}_k-v_k\|. 
\]
Plugging in the above results into \eqref{ThmMainrev-1} gives
\beq \label{ThmMainrev-2}
\|\hat{w}_i-w_i\|\leq CK^{-1/2}\bigl(\|H\hat{r}_i - r_i\| + \max_{1\leq k\leq K}\|H\hat{v}_k-v_k\|\bigr).
\eeq

Next, we study $\hat{b}_1$. 
Recall that 
\[
\hat{b}_1(k)=[ \hat{\lambda}_1 + \hat{v}_k'\mathrm{diag}(\hat{\lambda}_2,\cdots,\hat{\lambda}_K)\hat{v}_k]^{-1/2}.
\]
By Lemma~\ref{lem:ideal}, $b_1(k)$ has the same form except that $(\hat{\lambda}_k, \hat{v}_k)$ are replaced with their population counterparts. Letting $\Lambda_0=\mathrm{diag}(\lambda_2,\cdots,\lambda_K)$ and $\hat{\Lambda}_0 = \mathrm{diag}(\hat{\lambda}_2,\cdots,\hat{\lambda}_K)$, we write
\[
\frac{1}{\hat{b}^2_1(k)}= \hat{\lambda}_1 + \hat{v}_k'\hat{\Lambda}_0\hat{v}_k, \qquad \frac{1}{b^2_1(k)}= \lambda_1 + v_k'\Lambda_0v_k. 
\]
By direct calculations,
\begin{align*}
|&\frac{1}{\hat{b}_1^2(k)} - \frac{1}{b^2_1(k)}| \leq |\hat{\lambda}_1-\lambda_1| + |\hat{v}_k'\hat{\Lambda}_0\hat{v}_k-v_k'\Lambda_0v_k| \cr
&= |\hat{\lambda}_1-\lambda_1| + |\hat{v}_k'H'H\hat{\Lambda}_0\hat{v}_k-v_k'\Lambda_0v_k| \cr
&\leq |\hat{\lambda}_1-\lambda_1| + |\hat{v}_k'H'\hat{\Lambda}_0H\hat{v}_k-v_k'\hat{\Lambda}_0v_k| + |\hat{v}_k'H'(H\hat{\Lambda}_0-\hat{\Lambda}_0H)\hat{v}_k| + |v_k'(\hat{\Lambda}_0-\Lambda_0)v_k|\cr
&\leq  |\hat{\lambda}_1-\lambda_1| + |\hat{v}_k'H'\hat{\Lambda}_0H\hat{v}_k-v_k'\hat{\Lambda}_0v_k| + \|\hat{v}_k\|^2\|H\hat{\Lambda}_0- \hat{\Lambda}_0H \|+\|v_k\|^2\|\hat{\Lambda}_0-\Lambda_0\|\cr
&\leq (1+\|v_k\|^2)|\max_{\ell}|\hat{\lambda}_\ell-\lambda_\ell| + |\hat{v}_k'H'\hat{\Lambda}_0H\hat{v}_k-v_k'\hat{\Lambda}_0v_k| + \|\hat{v}_k\|^2\|H\hat{\Lambda}_0- \hat{\Lambda}_0H \|.
\end{align*}
First, by Lemma~\ref{lem:Aeigenvals}, $\max_{\ell}|\hat{\lambda}_\ell-\lambda_\ell|\leq C\sqrt{\theta_{\max}\|\theta\|_1}$. Second, 
by Lemma~\ref{lem:H}, $\|H\hat{\Lambda}_0-\hat{\Lambda}_0H\|\leq C\sqrt{\theta_{\max}\|\theta\|_1}$. Third, by Lemma~\ref{lem:R}, $\|v_k\|\leq C\sqrt{K}$; since $\max_{\ell}\|\hat{v}_\ell-v_\ell\|=o(\sqrt{K})$, it follows that $\|\hat{v}_k\|\leq C\sqrt{K}$. Combining the above gives
\beq \label{ThmMainrev-3}
|\frac{1}{\hat{b}_1^2(k)} - \frac{1}{b^2_1(k)}|\leq |\hat{v}_k'H'\hat{\Lambda}_0H\hat{v}_k-v_k'\hat{\Lambda}_0v_k|+CK\sqrt{\theta_{\max}\|\theta\|_1}. 
\eeq
Since $\hat{v}_k'H'\hat{\Lambda}_0H\hat{v}_k=v_k'\hat{\Lambda}_0v_k+2v_k'\hat{\Lambda}_0(H\hat{v}_k-v_k)+(H\hat{v}_k-v_k)'\hat{\Lambda}_0(H\hat{v}_k-v_k)$, we have
\[
|\hat{v}_k'H'\hat{\Lambda}_0H\hat{v}_k-v_k'\hat{\Lambda}_0v_k| \leq  2\|v_k\|\|\hat{\Lambda}_0\|\|H\hat{v}_k-v_k\|+\|\hat{\Lambda}_0\|\|H\hat{v}_k-v_k\|^2. 
\]
By Lemma~\ref{lem:eigenOmega} and Lemma~\ref{lem:Aeigenvals}, $\|\Lambda_0\|\leq C\beta_nK^{-1}\|\theta\|^2$ and $\|\hat{\Lambda}_0-\Lambda_0\|\leq C\sqrt{\theta_{\max}\|\theta\|_1}=o(K\beta_n^{-1}\|\theta\|^2)$. It follows that $\|\hat{\Lambda}_0\|\leq C\beta_nK^{-1}\|\theta\|^2$. Also, as we have argued before, $\|v_k\|\leq C\sqrt{K}$ and $\|H\hat{v}_k-v_k\|=o(\sqrt{K})$. Plugging these results into the above inequality gives
\[
|\hat{v}_k'H'\hat{\Lambda}_0H\hat{v}_k-v_k'\hat{\Lambda}_0v_k| \leq CK^{-1/2}\beta_n\|\theta\|^2 \|H\hat{v}_k-v_k\|. 
\]
We then plug it into \eqref{ThmMainrev-3} to get
\beq \label{ThmMainrev-4}
|\frac{1}{\hat{b}_1^2(k)} - \frac{1}{b^2_1(k)}|\leq CK^{-1/2}\beta_n\|\theta\|^2 \|H\hat{v}_k-v_k\|+CK\sqrt{\theta_{\max}\|\theta\|_1}. 
\eeq
In the proof of Lemma~\ref{lem:xi1}, we have shown $b_1(k)\asymp \|\theta\|^{-1}$; see \eqref{lem-xi1-0}. Then, $\frac{1}{b_1^2(k)}\asymp\|\theta\|^2$. Combining it with \eqref{ThmMainrev-4}, we have $\frac{1}{\hat{b}_1^2(k)}= \frac{1}{b_1^2(k)}[1+o(1)]\asymp\|\theta\|^2$. It follows that
\begin{align}  \label{ThmMainrev-5}
|\frac{1}{\hat{b}_1(k)} - &\frac{1}{b_1(k)}| = |\frac{1}{\hat{b}_1(k)} + \frac{1}{b_1(k)}|^{-1}\cdot |\frac{1}{\hat{b}_1^2(k)} - \frac{1}{b^2_1(k)}|\cr
&\leq C\|\theta\|^{-1}\cdot |\frac{1}{\hat{b}_1^2(k)} - \frac{1}{b^2_1(k)}|\cr
&\leq CK^{-1/2}\beta_n\|\theta\| \|H\hat{v}_k-v_k\|+C\|\theta\|^{-1}K\sqrt{\theta_{\max}\|\theta\|_1}\cr 
&\leq CK^{-1/2}\beta_n\|\theta\| \|H\hat{v}_k-v_k\| + CK\|\theta\|err_n,
\end{align}
where the last line is because $err_n=(\theta_{\max}/\theta_{\min})\cdot\|\theta\|^{-2}\sqrt{\theta_{\max}\|\theta\|_1\log(n)}\gg \|\theta\|^{-2}\sqrt{\theta_{\max}\|\theta\|_1}$.

Last, we combine the results for $(\hat{w}_i,\hat{b}_1)$ to prove \eqref{decomp}. Recall that $\hat{\pi}_i^*$ is as defined in \eqref{pihat-star}. Introduce its non-stochastic counterpart $\pi_i^*$ by
\beq \label{pi-star}
\pi_i^*(k)=w_i(k)/b_1(k), \qquad 1\leq k\leq K. 
\eeq
Since $\pi^*_i(k)\geq 0$, in \eqref{pihat-star}, the operation of truncating at zero can only make it closer to $\pi^*_i(k)$. It follows that  
\begin{align} \label{ThmMainrev-add1}
|\hat{\pi}^*_i(k)-\pi_i^*(k)| 
&\leq |\hat{w}_i(k)/\hat{b}_1(k)-\pi_i^*(k)|\cr
&= |\hat{w}_i(k)/\hat{b}_1(k)-w_i(k)/b_1(k)|\cr
&\leq \frac{1}{\hat{b}_1(k)}|\hat{w}_i(k)-w_i(k)| + w_i(k)|\frac{1}{\hat{b}_1(k)}-\frac{1}{b_1(k)}|. 
\end{align}
We sum over $k$ on both sides and note that $\hat{b}_1(k)\asymp \|\theta\|^{-1}$ (see the paragraph above \eqref{ThmMainrev-5}) and $\|w_i\|_1=1$. It yields
\begin{align} \label{ThmMainrev-6}
\|\hat{\pi}_i^*&-\pi_i^*\|_1 \leq C\|\theta\|\|\hat{w}_i-w_i\|_1 + |\frac{1}{b_1(k)}-\frac{1}{\hat{b}_1(k)}|\cr
&\leq C\|\theta\|\sqrt{K}\|\hat{w}_i-w_i\| + \max_{1\leq k\leq K}|\frac{1}{b_1(k)}-\frac{1}{\hat{b}_1(k)}|\cr
&\leq C\|\theta\|\bigl(\|H\hat{r}_i-r_i\|+ \max_{1\leq k\leq K}\|H\hat{v}_k-v_k\|+Kerr_n\bigr),
\end{align}
where in the second line we have used Cauchy-Schwarz inequality and in the last line we have plugged in \eqref{ThmMainrev-2} and \eqref{ThmMainrev-5}. By definition,    
$\hat{\pi}_i=\hat{\pi}_i^*/\|\hat{\pi}_i^*\|_1$.  
By the triangular inequality, 
\begin{align} \label{ThmMainrev-add2}
|\hat{\pi}_i(k)-\pi_i(k)| & \leq \frac{1}{\|\pi_i^*\|_1} |\hat{\pi}_i^*(k)-\pi_i^*(k)| + \hat{\pi}_i^*(k) |\frac{1}{\|\hat{\pi}_i^*\|_1}- \frac{1}{\|\pi_i^*\|_1}|\cr
&=  \frac{1}{\|\pi_i^*\|_1} |\hat{\pi}_i^*(k)-\pi_i^*(k)| + \frac{\hat{\pi}_i(k)}{ \|\pi_i^*\|_1} |\|\hat{\pi}_i^*\|_1- \|\pi_i^*\|_1|\cr
&\leq \frac{1}{\|\pi_i^*\|_1} \bigl(|\hat{\pi}_i^*(k)-\pi_i^*(k)| +\hat{\pi}_i(k)\|\hat{\pi}_i^*-\pi_i^*\|_1\bigr),
\end{align}
where the last inequality is because $|\|\hat{\pi}_i^*\|_1 - \|\pi_i^*\|_1|\leq \|\hat{\pi}_i^* - \pi^*_i\|_1$. We sum over $k$ on both sides and note that $\sum_k\hat{\pi}_i(k)=1$ by definition. It follows that
\[
\|\hat{\pi}_i-\pi_i\|_1\leq \frac{1}{\|\pi_i^*\|_1}\cdot 2\|\hat{\pi}_i^*-\pi_i^*\|_1. 
\]
By \eqref{pi-star}, $\|\pi^*\|_1\geq \|w_i\|_1\cdot \min_{k}\frac{1}{b_1(k)}$. In the paragraph above \eqref{ThmMainrev-5}, we  have seen that $b_1(k)\asymp\|\theta\|^{-1}$. This suggests that $\|\pi_i^*\|_1\geq C\|\theta\|$. As a result,
\begin{align}  \label{ThmMainrev-final}
\|\hat{\pi}_i&-\pi_i\|_1\leq C\|\theta\|^{-1}\cdot \|\hat{\pi}_i^*-\pi_i^*\|_1\cr
&\leq C\bigl(\|H\hat{r}_i-r_i\|+ \max_{1\leq k\leq K}\|H\hat{v}_k-v_k\|+Kerr_n\bigr). 
\end{align}
This gives \eqref{decomp}. The proof is now complete. \qed

\subsection{Proof of Theorem~\ref{thm:other-params}}

First, consider $\hat{P}-P$. Let $Q$ and $\hat{Q}$ be the same as in \eqref{mainproof-add}.  Then, 
\[
P=\diag(b_1) Q'\Lambda Q \diag(b_1), \qquad \hat{P} = \diag(\hat{b}_1)\hat{Q}'\hat{\Lambda}\hat{Q} \diag(\hat{b}_1). 
\]
It follows that
\begin{align} \label{thm-otherparam-1}
\|\hat{P}-P\| & \leq  \|\hat{Q}\diag(\hat{b}_1)\|^2 \|\hat{\Lambda}-\Lambda\| +  \|\hat{Q}\diag(\hat{b}_1) - Q\diag(b_1)\|\|\Lambda\|\|\hat{Q}\diag(\hat{b}_1)\|\cr
&\qquad + \|Q\diag(b_1)\|\|\Lambda\|\|\hat{Q}\diag(\hat{b}_1)-Q\diag(b)_1\|. 
\end{align}
Recall that we have the following facts (they hold with probability $1-o(n^{-3})$): 
\begin{itemize}
\item $\|\Lambda\|\leq C\|\theta\|^{-1}$ (by Lemma~\ref{lem:eigenOmega}); $\|\hat{\Lambda}-\Lambda\|\leq C\sqrt{\theta_{\max}\|\theta\|}\ll \|\theta\|^2err_n$ (by Lemma~\ref{lem:Aeigenvals} and the definition of $err_n$). 
\item $\|Q\|\leq C\sqrt{K}$ (by Lemma~\ref{lem:R}); $\|\hat{Q}-\hat{Q}\|\leq C\sqrt{K}\max_{1\leq k\leq K}\|H\hat{v}_k-v_k\|\leq CK^2\beta_n^{-1}err_n$ (by Theorem~\ref{thm:r-entrywise-bound} and the definitions of $Q$ and $\hat{Q}$). 
\item $C^{-1}\|\theta\|^{-1}\leq b_1(k)\leq C\|\theta\|^{-1}$, for $1\leq k\leq K$ (by \eqref{lem-xi1-0} in the proof of Lemma~\ref{lem:xi1});  $|\frac{1}{\hat{b}_1(k)}-\frac{1}{b_1(k)}|\leq C K^{-1/2}\beta_n\|\theta\|\|H\hat{v}_k-v_k\|+CK\|\theta\|err_n\leq CK\|\theta\|err_n$ (by \eqref{ThmMainrev-5} in the proof of Theorem~\ref{thm:main-rev}).  
\end{itemize}
From the third bullet point, $|\hat{b}_1(k)-b_1(k)|\leq C\|\theta\|^{-2}|\frac{1}{\hat{b}_1(k)}-\frac{1}{b_1(k)}|\leq CK\|\theta\|^{-1}err_n$. From the second bullet point, $\|\hat{Q}-Q\|\leq CK^2\beta_n^{-1}err_n$, and $\|\hat{Q}\|\leq 2\|Q\|\leq C\sqrt{K}$. As a result,  
\begin{align} \label{thm-otherparam-2}
\|\hat{Q}\diag(\hat{b}_1)-Q\diag(b)_1\| & \leq \|\hat{Q}\|\|\diag(\hat{b}_1)-\diag(b_1)\| + \|\hat{Q}-Q\|\|\diag(b_1)\|\cr
&\leq C\sqrt{K}\cdot K\|\theta\|^{-1}err_n + CK^2\beta_n^{-1}err_n\cdot \|\theta\|^{-1}\cr
&\leq C(K^{3/2}+K^2\beta_n^{-1})\|\theta\|^{-1}err_n. 
\end{align}
It further implies $\|\hat{Q}\diag(\hat{b}_1)\|\leq 2\|Q\diag(b)_1\|\leq C\sqrt{K}\|\theta\|^{-1}$. We then plug these results into \eqref{thm-otherparam-1} and use the first bullet point above. It gives
\begin{align} \label{thm-otherparam-3}
\|\hat{P}-P\| & \leq \|\hat{Q}\diag(\hat{b}_1)\|^2 \|\hat{\Lambda}-\Lambda\| + 3\|\hat{Q}\diag(\hat{b}_1) - Q\diag(b_1)\|\|\Lambda\|\|\hat{Q}\diag(\hat{b}_1)\| \cr
& \leq C(\sqrt{K}\|\theta\|^{-1})^2\cdot \|\theta\|^2err_n + C(K^{3/2}+K^2\beta_n^{-1})\|\theta\|^{-1}err_n\cdot \|\theta\|^2\cdot \sqrt{K}\|\theta\|^{-1}\cr
&\leq C(K^2+K^{3/2}\beta_n^{-1})err_n. 
\end{align}
This proves the first claim. 

Second, consider $\|\hat{\Theta}-\Theta\|_F^2$, which by definition is equal to $\sum_{i=1}^n|\hat{\theta}(i)-\theta(i)|^2$. Recall that $\theta(i)=\xi_1(i)/(\pi_i'b_1)$ and $\hat{\theta}(i)=\hat{\xi}_1(i)/(\hat{\pi}_i'\hat{b}_1)$. It follows that
\begin{align*}
|\hat{\theta}(i)-\theta(i)| & \leq \frac{1}{|\pi_i'b_1|}|\hat{\xi}_1(i)-\xi_1(i)| + |\hat{\xi}_1(i)|\Bigl|\frac{1}{\hat{\pi}_i'\hat{b}_1}-\frac{1}{\pi_i'b_1}\Bigr|\cr
&\leq \frac{1}{|\pi_i'b_1|} |\hat{\xi}_1(i)-\xi_1(i)| + |\hat{\xi}_1(i)|\cdot \frac{|\hat{\pi}_i'\hat{b}_1 - \pi_i'b_1|}{|\hat{\pi}_i'\hat{b}_1||\pi_i'b_1|}\cr
&\leq \frac{|\hat{\xi}_1(i)-\xi_1(i)|}{|\pi_i'b_1|}  + \frac{|\hat{\xi}_1(i)|}{|\hat{\pi}_i'\hat{b}_1||\pi_i'b_1|} \bigl(\|\hat{\pi}_i-\pi_i\|_1\|b_1\|_\infty + \|\hat{\pi}_i\|_1\|\hat{b}_1-b_1\|_\infty\bigr). 
\end{align*}
Note that $\|\hat{\pi}_i\|_1=1$, $b_1(k)\asymp \|\theta\|^{-1}$, and $\|\hat{b}_1-b_1\|_\infty\leq CK\|\theta\|^{-1}err_n = o(\|\theta\|^{-1})$. 
It further implies $\pi_i'b_1\asymp \hat{\pi}_i'\hat{b}_1\asymp \|\theta\|^{-1}$. We plug these results into the above inequality to get
\[
|\hat{\theta}(i)-\theta(i)| \leq C\|\theta\| |\hat{\xi}_1(i)-\xi_1(i)| + C\|\theta\||\hat{\xi}_1(i)|\|\hat{\pi}_i-\pi_i\|_1  +  CK\|\theta\|err_n|\hat{\xi}_1(i)|. 
\]
We take the sum of squares of $i=1,2,\ldots,n$ on both sides and note that $\|\hat{\xi}\|=1$. Moreover, by Lemma~\ref{lem:Aeigenvecs}, $\|\hat{\xi}_1-\xi_1\|\leq C\|\theta\|^{-2}K\sqrt{\theta_{\max}\|\theta\|_1}\ll Kerr_n$. It follows that 
\begin{align} \label{thm-otherparam-4}
\|\hat{\Theta}-\Theta\|^2_F & \leq C\|\theta\|^2\|\hat{\xi}_1-\xi_1\|^2 + C\|\theta\|^2\Bigl(\max_{1\leq i\leq n}\|\hat{\pi}_i-\pi_i\|_1^2\Bigr) + CK^2\|\theta\|^2err_n^2\cr
&\leq C\|\theta\|^2\bigl( K^2err_n^2 + K^3\beta_n^{-2}err_n^2 + CK^2err_n^2\bigr)\cr
&\leq \|\theta\|^2\cdot CK^3\beta_n^{-2}err_n^2. 
\end{align}
This proves the second claim. \qed

\subsection{Proofs of Theorems~\ref{thm:CVS+SP}, \ref{thm:SVS1} and \ref{thm:SVS2}}
Theorem~\ref{thm:CVS+SP} is a direct consequence of Theorem~\ref{thm:main-rev} and Lemma~\ref{lem:SP+CVS}. 
For Theorem~\ref{thm:SVS1} and Theorem~\ref{thm:SVS2}, their first claims about the VH step follow from Lemma~\ref{lem:VH1} and Lemma~\ref{lem:VH2}, respectively. 
We now show their second claims, where we aim to obtain a faster rate for $\frac{1}{n}\sum_{i=1}^n\|\hat{\pi}_i-\pi_i\|^2$ when the VH step is strongly efficient.

In \eqref{ThmMainrev-add2}, we have shown that for every $1\leq k\leq K$, 
\[
|\hat{\pi}_i(k)-\pi_i(k)| \leq \frac{1}{\|\pi_i^*\|_1} \bigl(|\hat{\pi}_i^*(k)-\pi_i^*(k)| +\hat{\pi}_i(k)\|\hat{\pi}_i^*-\pi_i^*\|_1\bigr). 
\]
Taking the sum of squares over $k$ on both sides and using the universal inequality $(a+b)^2\leq 2a^2+2b^2$, we have
\[
\|\hat{\pi}_i-\pi_i\|^2\leq \frac{2}{\|\pi_i^*\|_1^2}\bigl(\|\hat{\pi}^*_i-\pi^*_i\|^2 + \|\hat{\pi}_i\|^2\cdot \|\hat{\pi}_i^*-\pi_i^*\|_1^2\bigr). 
\]
In the paragraph above \eqref{ThmMainrev-final}, we have shown that $\|\pi_i^*\|_1\geq C\|\theta\|$. Additionally, $\|\hat{\pi}_i\|^2\leq \|\hat{\pi}_i\|_1\|\hat{\pi}_i\|_\infty\leq 1$. It follows that 
\beq \label{proof-L2err-1}
\|\hat{\pi}_i-\pi_i\|^2\leq \frac{C}{\|\theta\|^2}\bigl(\|\hat{\pi}^*_i-\pi^*_i\|^2 + \|\hat{\pi}_i^*-\pi_i^*\|_1^2\bigr). 
\eeq

In light of \eqref{proof-L2err-1}, we first derive upper bounds for $\|\hat{\pi}^*_i-\pi^*_i\|$ and $\|\hat{\pi}^*_i-\pi^*_i\|_1$, respectively. By \eqref{ThmMainrev-add1} and  \eqref{ThmMainrev-5}, 
\begin{align*} \label{proof-L2err-2}
|\hat{\pi}^*_i(k)-\pi_i^*(k)| & \leq \frac{1}{\hat{b}_1(k)}|\hat{w}_i(k)-w_i(k)| + w_i(k)|\frac{1}{\hat{b}_1(k)}-\frac{1}{b_1(k)}|,\cr
|\frac{1}{\hat{b}_1(k)} - \frac{1}{b_1(k)}| &\leq CK^{-1/2}\beta_n\|\theta\| \|H\hat{v}_k-v_k\|+C\|\theta\|^{-1}K\sqrt{\theta_{\max}\|\theta\|_1}. 
\end{align*}
Also, $\hat{b}_1(k)\asymp b_1(k)\asymp \|\theta\|^{-1}$ (see the paragraph above \eqref{ThmMainrev-5}). It follows that
\[
|\hat{\pi}^*_i(k)-\pi_i^*(k)|\leq C\|\theta\|\, |\hat{w}_i(k)-w_i(k)| + Cw_i(k)\biggl(\frac{\beta_n\|\theta\|  \|H\hat{v}_k-v_k\|}{\sqrt{K}}+\frac{K\sqrt{\theta_{\max}\|\theta\|_1}}{\|\theta\|}\biggr) . 
\]
Note that 
\[
err_n^*=[\|\theta\|/(\theta_{\min}\sqrt{n})]\cdot\|\theta\|^{-2}\sqrt{\theta_{\max}\|\theta\|_1}\geq \|\theta\|^{-2}\sqrt{\theta_{\max}\|\theta\|_1}. 
\]
We further have 
\beq \label{proof-L2err-2}
|\hat{\pi}^*_i(k)-\pi_i^*(k)|\leq C\|\theta\||\hat{w}_i(k)-w_i(k)| + Cw_i(k) \|\theta\|\Bigl(K^{-1/2}\beta_n \|H\hat{v}_k-v_k\|+Kerr_n^* \Bigr).
\eeq
It follows that
\begin{align*}
\|\hat{\pi}^*_i-\pi_i^*\|^2 & \leq C\|\theta\|^2\Bigl[ \|\hat{w}_i-w_i\|^2 + \|w_i\|^2 \Bigl( K^{-1}\beta^2_n \max_{1\leq k\leq K}\|H\hat{v}_k-v_k\|^2+K^2 (err_n^*)^2 \Bigr)\Bigr],\cr
\|\hat{\pi}^*_i-\pi^*_i\|_1 & \leq C\|\theta\|\Bigl[ \|\hat{w}_i-w_i\|_1 + \|w_i\|_1 \Bigl( K^{-1/2}\beta_n \max_{1\leq k\leq K}\|H\hat{v}_k-v_k\| +K err_n^* \Bigr)\Bigr]. 
\end{align*}
Note that $\|w_i\|_1=1$, $\|w_i\|^2\leq\|w_i\|_1\|w_i\|_\infty\leq 1$, and $\|\hat{w}_i-w_i\|_1\leq \sqrt{K}\|\hat{w}_i-w_i\|$. Additionally, by  \eqref{ThmMainrev-2}, 
\[
\|\hat{w}_i-w_i\|  \leq CK^{-1/2}\bigl(\|H\hat{r}_i - r_i\| + \max_{1\leq k\leq K}\|H\hat{v}_k-v_k\|\bigr).
\]
Combining the above gives
\begin{align}  \label{proof-L2err-3}
\|\hat{\pi}^*_i-\pi_i^*\|^2 & \leq C\|\theta\|^2\Bigl( K^{-1}\|H\hat{r}_i - r_i\|^2 + K^{-1} \max_{1\leq k\leq K}\|H\hat{v}_k-v_k\|^2+K^2 (err_n^*)^2 \Bigr),\cr
\|\hat{\pi}^*_i-\pi^*_i\|_1 & \leq C\|\theta\|\Bigl( \|H\hat{r}_i - r_i\| +  \max_{1\leq k\leq K}\|H\hat{v}_k-v_k\| +K err_n^* \Bigr). 
\end{align}

Next, we plug \eqref{proof-L2err-3} into \eqref{proof-L2err-1} to get 
\[
\|\hat{\pi}_i-\pi_i\|^2\leq C\|H\hat{r}_i-r_i\|^2+C\Bigl(\max_{1\leq k\leq K}\|H\hat{v}_k-v_k\|\Bigr)^2+CK^2(err^*_n)^2. 
\]
Summing over $i$ on both sides gives 
\[
n^{-1}\sum_{i=1}^n\|\hat{\pi}_i-\pi_i\|^2 \leq Cn^{-1}\sum_{i=1}^n\|H\hat{r}_i-r_i\|^2 + C\Bigl(\max_{1\leq k\leq K}\|H\hat{v}_k-v_k\|\Bigr)^2+CK^2(err^*_n)^2. 
\]
By strong efficiency of the VH step, $\max_{1\leq k\leq K}\|H\hat{v}_k-v_k\|\leq \sqrt{n^{-1}\sum_{i=1}^n\|H\hat{r}_i-r_i\|^2}$ (see Definition~\ref{def:effVH}). 
It follows that
\[
n^{-1}\sum_{i=1}^n\|\hat{\pi}_i-\pi_i\|^2 \leq Cn^{-1}\sum_{i=1}^n\|H\hat{r}_i-r_i\|^2+CK^2(err^*_n)^2. 
\]
Using Lemma~\ref{lem:hatR-l2}, $n^{-1}\sum_{i=1}^n \|H\hat{r}_i - r_i \|^2\leq CK^3\beta_n^{-2}(err_n^*)^2$. Therefore, 
\[
n^{-1}\sum_{i=1}^n\|\hat{\pi}_i-\pi_i\|^2\leq CK^3\beta_n^{-2}(err_n^*)^2+CK^2(err_n^*)^2\leq CK^3\beta_n^{-2}(err_n^*)^2. 
\]
Additionally, $err_n^*=[\|\theta\|/(\sqrt{n}\theta_{\max})]\cdot err_n/\sqrt{\log(n)}\leq err_n/\sqrt{\log(n)}$. We thus have 
\beq \label{proof-L2err-final}
n^{-1}\sum_{i=1}^n\|\hat{\pi}_i-\pi_i\|_1^2\leq CK^3\beta_n^{-1}(err_n^*)^2\leq \frac{C K^3\beta_n^{-1}err_n^2}{\log(n)}.  
\eeq
This proves the claim.

\section{More Simulation Results}

We present additional simulation results. They are not included in the main article due to space limit. For most experiments below, we set $n=500$ and $K=3$. For $0\leq n_0\leq 160$, let each community have $n_0$ number of pure nodes. Fixing $x\in (0,1/2)$, let the mixed nodes have four different memberships $(x, x, 1-2x)$, $(x, 1-2x, x)$, $(1-2x, x, x)$ and $(1/3, 1/3, 1/3)$, each with $(500-3n_0)/4$ number of nodes. Fixing $\rho\in (0,1)$, the matrix $P$ has diagonals $1$ and off-diagonals $\rho$. Fixing $z\geq 1$, we generate the degree parameters such that $1/\theta(i)\overset{iid}{\sim} U(1,z)$, where $U(1,z)$ denotes the uniform distribution on $[1,z]$. The tuning parameter $L$ is selected as in \eqref{chooseL}. For each setting, we report $n^{-1}\sum_{i=1}^n\|\hat{\pi}_i-\pi_i\|^2$ averaged over $100$ repetitions.

\begin{figure}[htb] 
\centering
\includegraphics[width=.4\textwidth, height=.3\textwidth, trim = 0mm 5mm 0mm 7mm, clip=true]{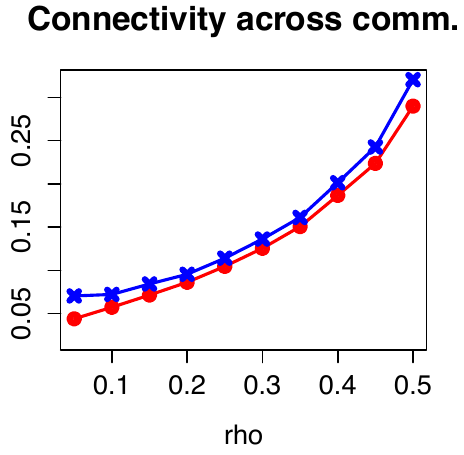}
\caption{Estimation errors of Mixed-SCORE and OCCAM ($y$-axis: $n^{-1}\sum_{i=1}^n\|\hat{\pi}_i - \pi_i\|^2$). 
} \label{fig:comparison-add}
\end{figure}

{\bf Experiment 5: Connectivity across communities}.   
Fix $(x, n_0, z)=(0.4, 80, 5)$ and let $\rho$ range in $\{0.05, 0.1, 0.15, \cdots, 0.5\}$. The larger $\rho$, the more edges across different communities. The results are presented in Figure~\ref{fig:comparison-add}. We see that the performance of Mixed-SCORE improves as $\rho$ decreases. One possible reason is that, for $\rho$ large, it is relatively more difficult to identify the vertices of the Ideal Simplex. Furthermore, Mixed-SCORE is better than OCCAM in all settings. 

\spacingset{1}
\spacingset{1.45}

 {\bf Experiment 6: Mixed memberships taking continuous values}. In this experiment, we generate the mixed memberships from a continuous distribution. Set $(n,K)=(500,3)$ and let $P$ have diagonals $1$ and off-diagonals $0.3$. Each community has $n_0=25$ pure nodes. The $\pi_i$ of remaining nodes are $iid$ drawn as follows: We generate $\pi_i(1)$ and $\pi_i(2)$ independently from $U(1/6, 1/2)$ and set $\pi_i(3)=1-\pi_i(1)-\pi_i(2)$. The degree parameters $\theta(i)$ are $iid$ drawn from $\alpha_n\cdot U(1,2)$, where $\alpha_n>0$ controls the sparsity of the network. Let $\alpha_n$ range in $\{0.02, 0.04, 0.06, \cdots, 0.20\}$. The results are presented in Table~\ref{tb:AB}. This setting does not satisfy the regularity conditions \eqref{cond1-pi}-\eqref{cond2-pi} on $\pi_i$'s, however, Mixed-SCORE still has a good performance and outperforms OCCAM. It suggests that the regularity conditions on $\pi_i$'s are only for theoretical convenience, and our method indeed works for broader settings.

\begin{table}[htb!]  
\centering
\caption{Estimation errors in Experiment 6, where $\pi_i$'s take continuous values.} \label{tb:AB}
\begin{tabular}{l |  c  | c  |  c  | c|c | c  | c|c|c| l  }
\hline
$\alpha_n$ &  0.02 & 0.04 & 0.06 & 0.08 & 0.10 & 0.12 & 0.14 & 0.16 & 0.18 & 0.20\\
\hline
Mixed-SCORE & .38 & .35  & .36 &  .32 & .30 & .28 & .23 & .18 & .15 & .12  \\
OCCAM & .44 & .42 & .41 & .41 & .38 & .36 & .32  & .28  & .26 & .23  \\
\hline
\end{tabular}
\end{table}

\begin{figure}[htb]
\centering
\includegraphics[width=.325\textwidth, height=.26\textwidth, trim = 8mm 14mm 10mm 8mm, clip=true]{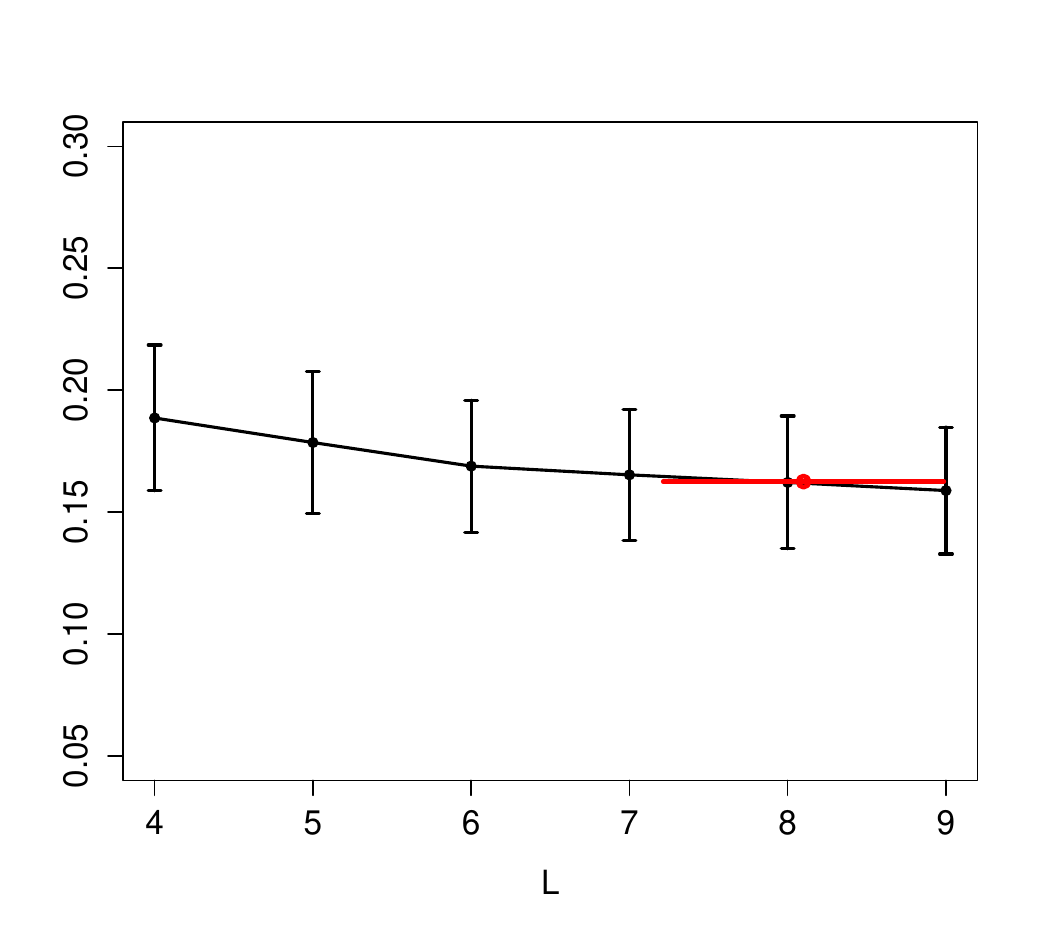}
\includegraphics[width=.325\textwidth, height=.26\textwidth, trim = 8mm 14mm 10mm 8mm, clip=true]{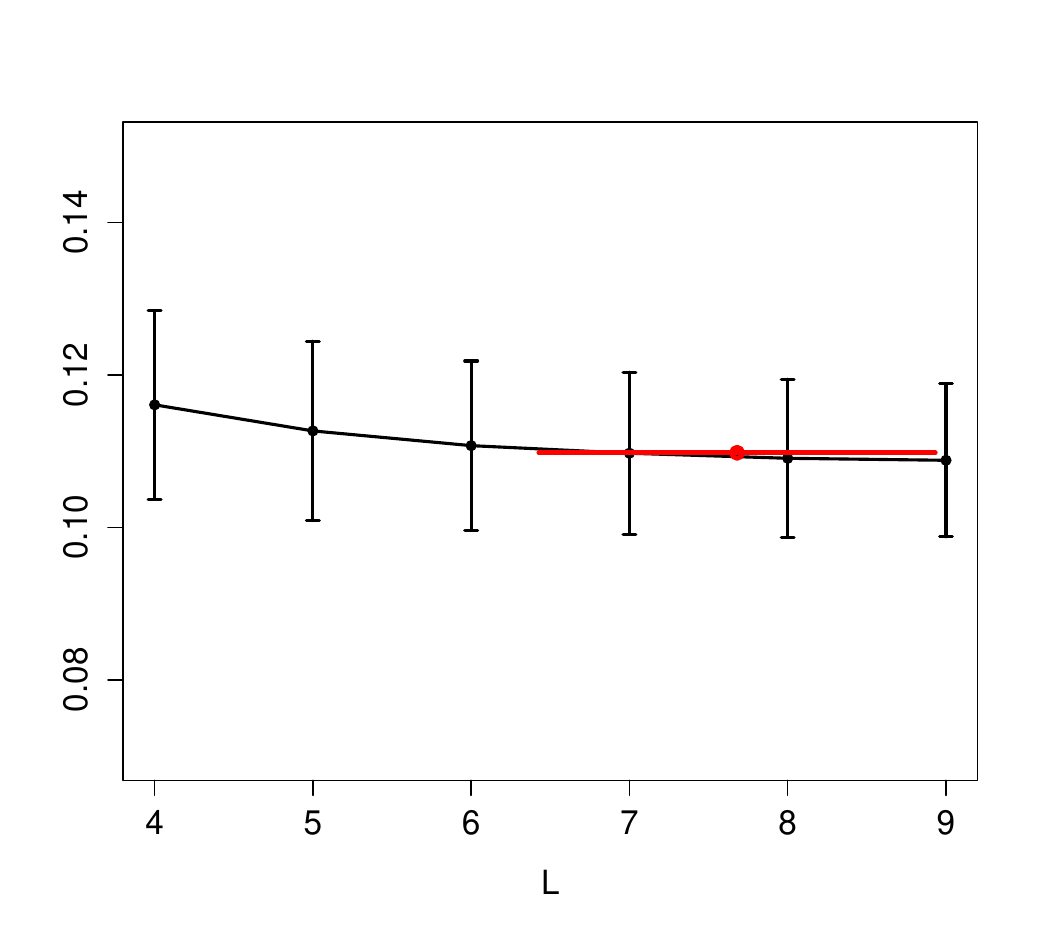}
\includegraphics[width=.325\textwidth, height=.26\textwidth, trim = 8mm 14mm 10mm 8mm, clip=true]{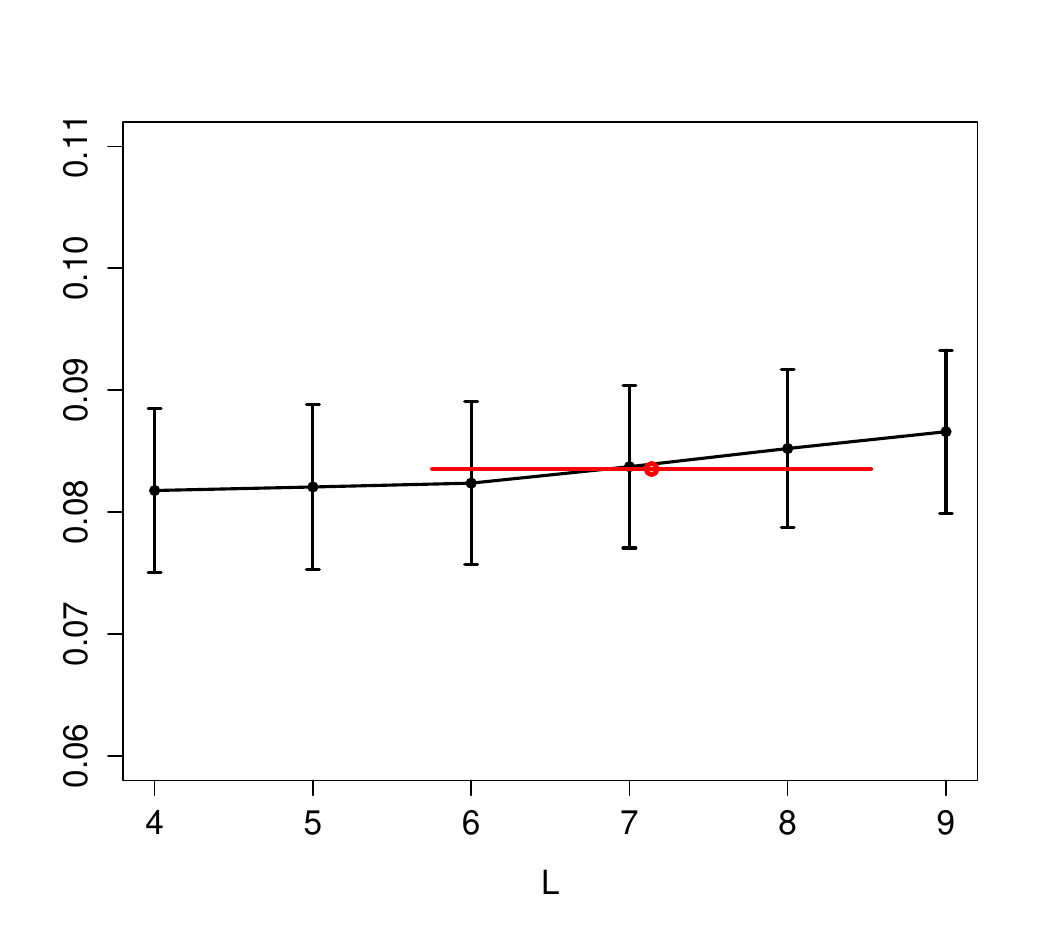}
\caption{Performance of Mixed-SCORE as the tuning parameter $L$ varies ($y$-axis: estimation errors; $\hat{L}_n^*(A)$ is plotted in red; both mean and standard deviation are displayed). From left to right, there are $60,80,100$ pure nodes in each community, respectively. }\label{fig:choice-k}
\end{figure}

\begin{figure}[tb]
\centering
\includegraphics[width=.325\textwidth, height=.22\textwidth, trim = 30mm 30mm 10mm 26mm, clip=true]{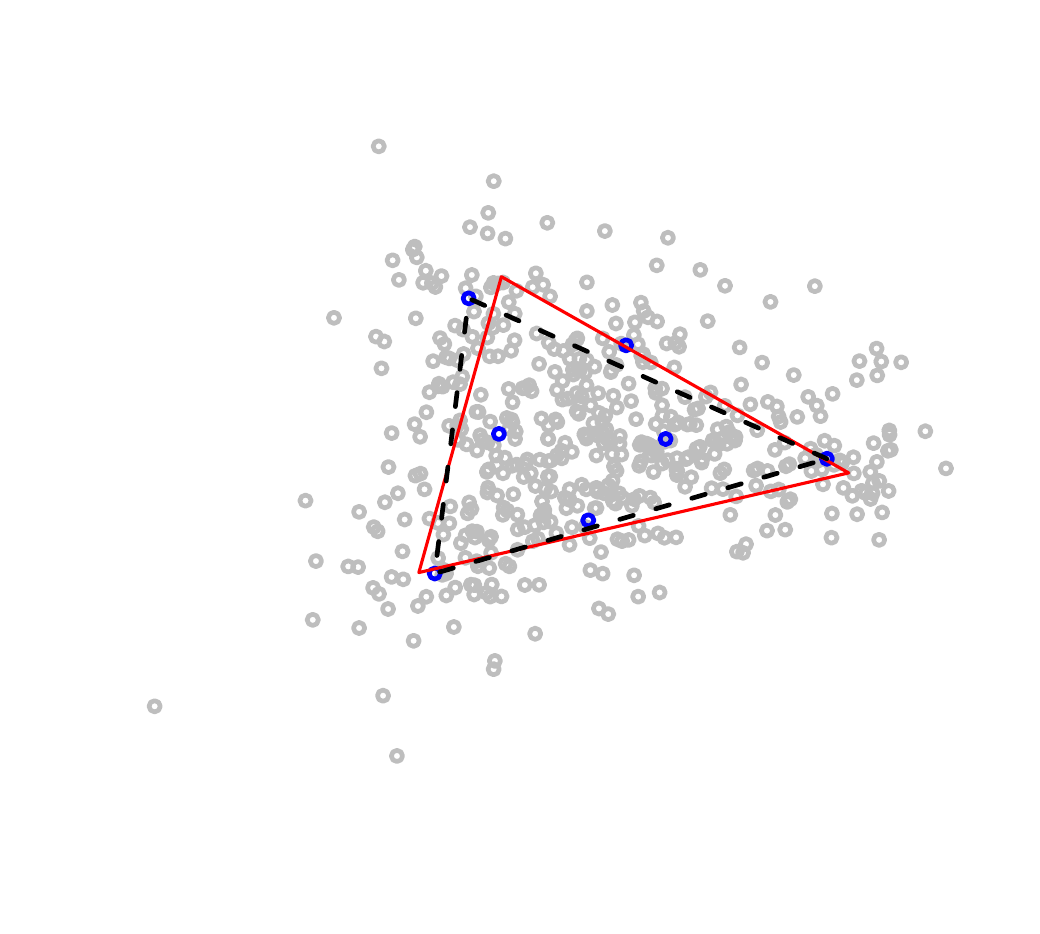}
\includegraphics[width=.325\textwidth, height=.22\textwidth, trim = 30mm 30mm 10mm 26mm, clip=true]{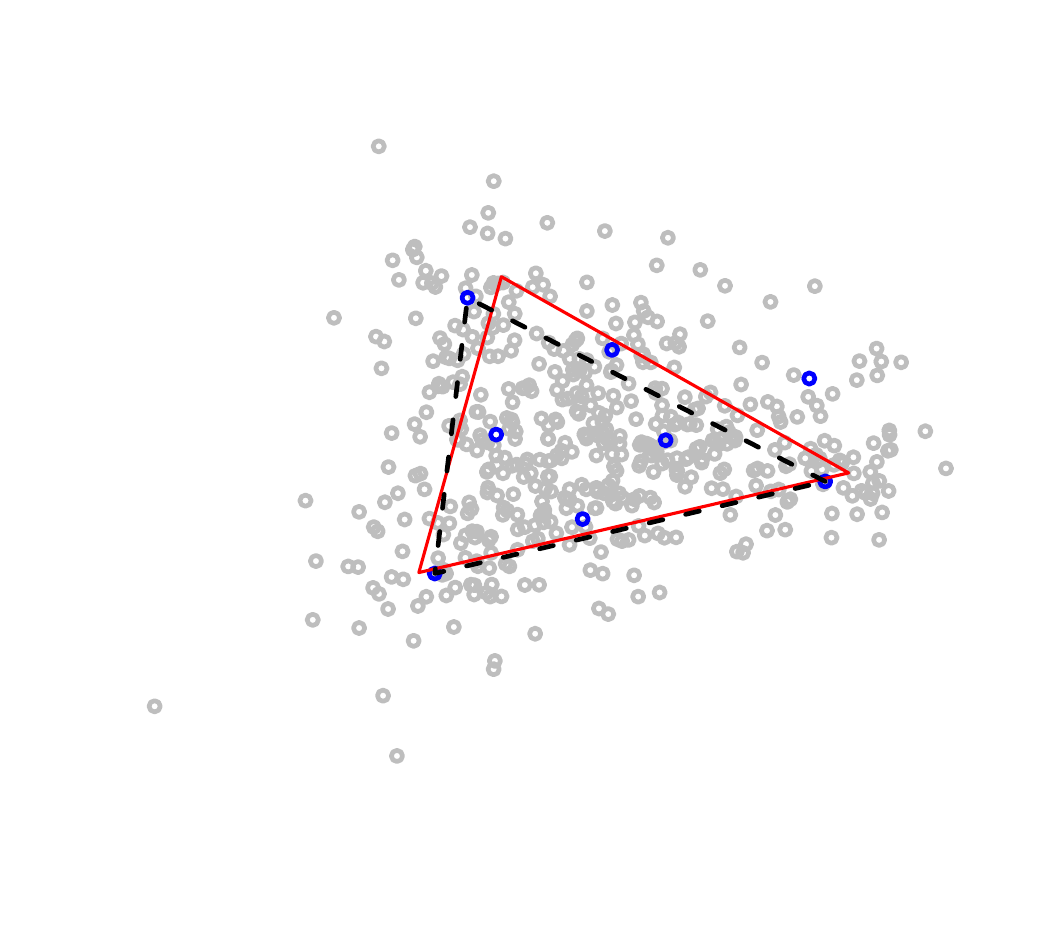}
\includegraphics[width=.325\textwidth, height=.22\textwidth, trim = 30mm 30mm 10mm 26mm, clip=true]{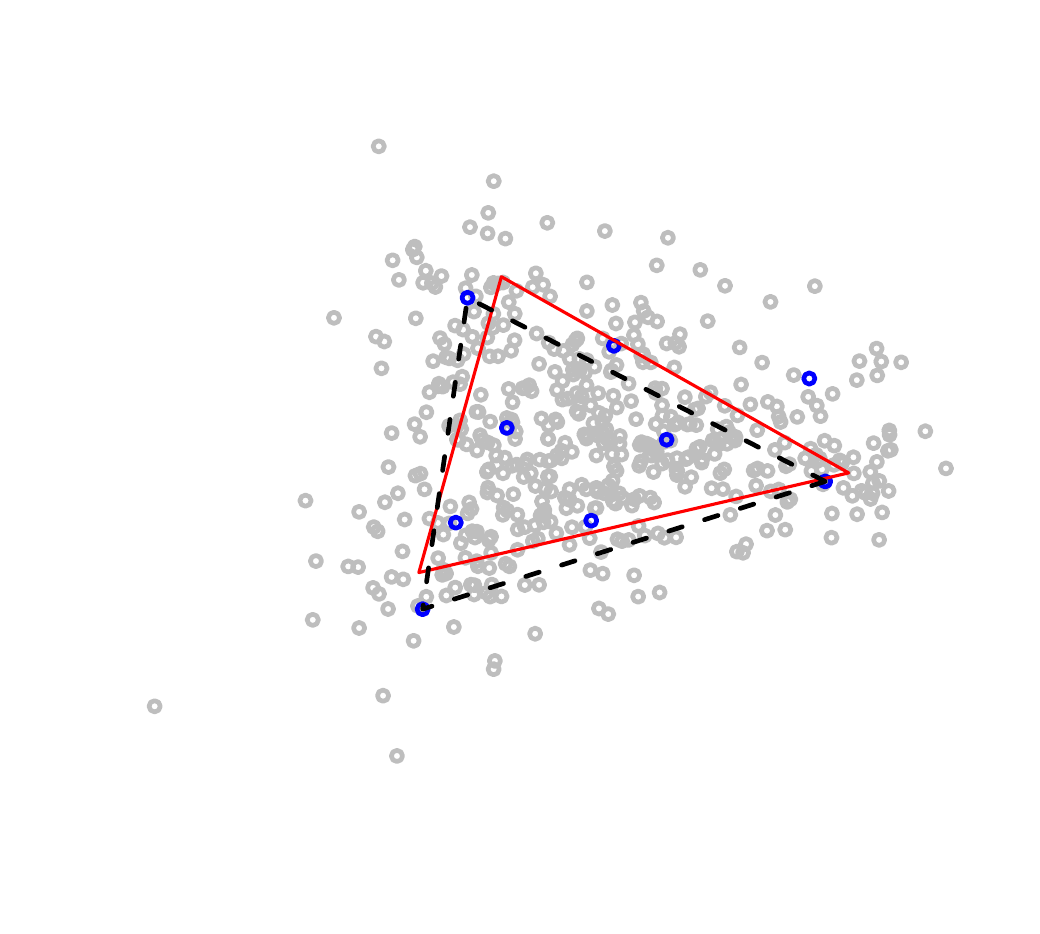}
\caption{Illustration of the Vertex Hunting step. From left to right, $L=7,8,9$. Although the local cluster centers (blue points) are different, the estimated $2$-simplex (dashed black) changes very little, and it approximates the IS (solid red) well. } \label{fig:VHillust}
\end{figure}

{\bf Experiment 7: Tuning parameter selection}.   
We first study the choice of the tuning parameter $L$ in Mixed-SCORE. We aim to see (i) how the estimation errors change for a range of $L$, and (ii) how the adaptive choice $\hat{L}^*_n(A)$ in \eqref{chooseL} performs. Fix $(x, \rho, z)=(0.4, 0.2, 5)$ and let $n_0$ range in $\{60, 80, 100\}$. For each setting, we run Mixed-SCORE with $L\in\{4,5,\cdots,9\}$ and $\hat{L}_n^*(A)$. The results are displayed in Figure~\ref{fig:choice-k}. First, when there are relatively few mixed nodes (e.g., $n_0=100$), small values of $L$ yield good performance; but as the number of mixed nodes going up, we favor larger values of $L$; these match our theoretical results (Lemmas~\ref{lem:VH1}-\ref{lem:VH2}). Second, under the circumstances of a moderate number of mixed nodes (e.g., $n_0=60,80$), for a range of $L$ (e.g., $L\in\{7,8,9\}$), the statistical errors of Mixed-SCORE are similar, and $\hat{L}^*_n(A)$ falls in this range with high probability. Figure~\ref{fig:VHillust} shows the estimated $2$-simplex in one repetition ($n_0=80$), and the simplex changes very little when $L$ falls in a range.

\spacingset{1}
\spacingset{1.45}
\spacingset{1}
\spacingset{1.45}

\begin{figure}[htb]
\centering
\includegraphics[width=.48\textwidth, height=.35\textwidth]{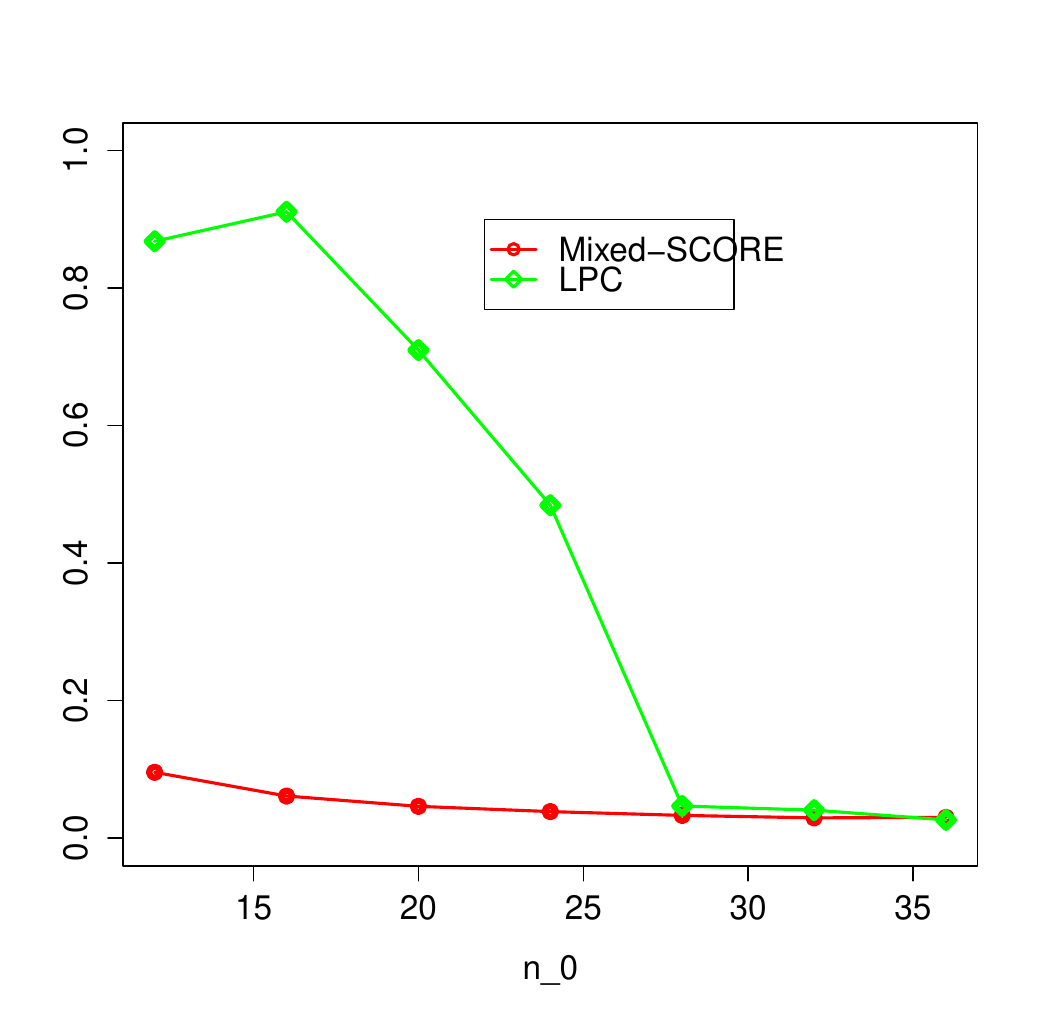}
\caption{Estimation errors of Mixed-SCORE and LPC ($y$-axis: $n^{-1}\sum_{i=1}^n\|\hat{\pi}_i-\pi_i\|^2$).}\label{fig:comparison2}
\end{figure}

{\bf Experiment 8: Comparison with latent space approach}. We compare 
Mixed-SCORE with the Bayesian method based on LPC \cite{handcock2007model} (we use the R package {\it latentnet}). In this experiment, we fix $n=120$, $K=3$, $(x, \rho, z)=(0.4, 0.3, 5)$, and let $n_0$ range in $\{12, 16, 20, \cdots, 32, 36\}$ (so the number of mixed nodes in each group decreases from $21$ to $3$). The results are displayed in Figure~\ref{fig:comparison2}. We find that, when the fraction of mixed nodes is comparably small, LPC has a perfect performance; however, as the fraction of mixed nodes increases to more than $40\%$, the performance of LPC deteriorates rapidly; one reason is that, when $n_0$ is not very large, LPC often estimates the PMF of all the nodes as the same. In contrast, the performance of Mixed-SCORE is quite stable. In terms of computing time, Mixed-SCORE takes only seconds for one repetition while LPC takes $>20$ minutes (both measured in R). 

\spacingset{1}
\spacingset{1.45}

\section{More Real Data Results}\label{sec:Real-moreResults}

We present additional results for the trade networks. First, we plot the rows of $\hat{R}$ for the GOS network (see Figure~\ref{subfig:GUS} for a comparison). Recall that edges in the GOS network indicate significant over-estimation of trade flows in the initial gravity model. This embedding is not as informative as the embedding we obtained for the GUS network. One interesting observation is that countries with high GDPs tend to cluster together and countries with low GDPs tend to cluster together. 

\spacingset{1}
\begin{figure}[htb!] 
\centering
\includegraphics[width=.49\textwidth, height=.42\textwidth]{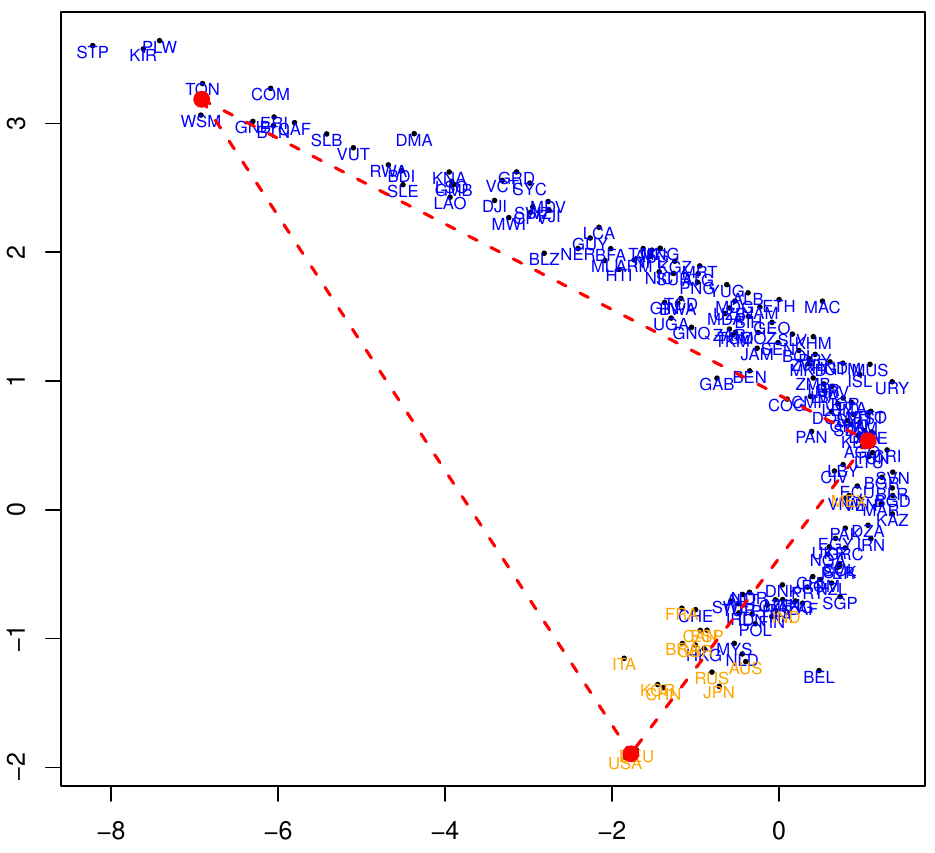}
\caption{Rows of $\hat{R}$ for the GOS network after fitting a gravity model. We set $K=3$ in Mixed-SCORE, so the Ideal Simplex is a triangle. Each $\hat{r}_i$ corresponds to a country, whose ISO3 code is shown (orange color: top 15 countries with highest GDPs). In each plot, the dashed triangle is the estimated simplex from SVS with $L=40$. We note that although each $r_i$ is in the Ideal Simplex, some $\hat{r}_i$'s can be outside the estimated simplex due to noise corruption.} 
\label{fig:GOS} 
\end{figure}
\spacingset{1.45}

Next, we present the estimated mixed memberthips of representative countries in the trade in service (TIS) network. 

\spacingset{1}
\begin{table}[hbt!]
\centering
\caption{The estimated $\hat{\pi}_i$ for the 10 countries with largest total service exports. By Figure~\ref{subfig:TSI}, the three communities are interpreted as `North Africa', `Southeast Asia' and `South/Central Europe'.} \label{tb:trade}
\scalebox{.96}{
\begin{tabular}{l | r | r | ccc}
\hline
Economy & \shortstack{Service\\export} & degree &  $\hat{\pi}_i(1)$ & $\hat{\pi}_i(2)$ & $\hat{\pi}_i(3)$\\
\hline
    USA            &      3,998,419 & 45 & 0.128 & 0.424 & 0.448\\
    UK                &      1,914,255 & 34 & 0.202   &  0.319 &   0.479\\
    Germany       & 1,534,393 & 29 & 0.348 & 0.215 & 0.436\\
    France    & 1,354,407 & 26 & 0.243 &  0.193  &  0.564\\
    China    & 1,146,845 &  14 & 0.130  &  0.606   &  0.264\\
    Netherlands     &  1,064,165 & 19 & 0.218 &  0.215 &   0.567 \\
    Japan     & 882,650 &  17 &  0.124 & 0.611 &  0.265\\
    India     & 865,543   & 6 & 0.033  &   0.598  &   0.369\\
    Singapore     & 830,975 & 20 & 0.313 &  0.554 &   0.134\\
    Ireland     & 811,105 & 12 & 0.144 & 0.269  & 0.586\\
\hline
\end{tabular}}
\end{table}
\spacingset{1.45}

We also present additional results for the citee network. The following table shows those ``high-degree and relatively pure" nodes in each of the three communities.

\spacingset{1}
\begin{table}[htb!]  
\caption{\small Estimated PMF of the $100$ nodes with the highest degrees in the Citee network, among which only the $12$ purist nodes in each community are reported.} \label{tb:membership1} 
\vspace{5pt}

\hspace*{-2cm}
\scalebox{.6}{
\begin{tabular}{l|l| lll || l|l| lll || l|l| lll}
\hline
Name & Deg. & MulTest & SpatNon & VarSelect & Name & Deg. & MulTest & SpatNon & VarSelect & Name & Deg. & MulTest & SpatNon & VarSelect\\
\hline
Felix Abramovich & 366 & 0.943 & 0 & 0.057 & Peter Muller & 429 & 0.326 & 0.613 & 0.061 & Lixing Zhu	& 432 & 0.121 & 0 & 0.879 \\
Joseph Romano & 377 & 0.868 & 0 & 0.132 & Jeffrey Morris & 452 & 0.146 & 0.519 & 0.335 & Zhiliang Ying & 382 & 0.107 & 0.027 & 0.866\\
Sara van de Geer & 372 & 0.834 & 0 & 0.166 & Michael Jordan & 383 & 0.321 & 0.495 & 0.184 & Zhezhen Jin & 361 & 0.134 & 0 & 0.866 \\
Yoav Benjamini & 478 & 0.821 & 0 & 0.179 & Mahlet Tadesse	& 383 & 0.373	& 0.493 & 0.134 & Dennis Cook & 424 & 0.253 & 0 & 0.747\\
David Donoho & 484 & 0.819 & 0 & 0.181 & Naijun Sha & 383 & 0.373 & 0.493 & 0.134 & Wenbin Lu & 405 & 0.255 & 0 & 0.745 \\
Christopher Genovese & 521 & 0.810 & 0 & 0.190 & Michael Stein & 379 & 0.093 & 0.449 & 0.458 & Dan Yu Lin & 527 & 0.257 & 0 & 0.743\\
Larry Wasserman & 535 & 0.800 & 0 & 0.200 & Adrian Raftery & 413 & 0.175 & 0.446 & 0.379 & Donglin Zeng & 489 & 0.270 & 0 & 0.730 \\
Jon Wellner & 387 & 0.798 & 0.05 & 0.152 & Robert Kohn & 429 & 0.310 & 0.428 & 0.262 & Gerda Claeskens & 404 & 0.247 & 0.033 & 0.720\\
Alexandre Tsybakov & 521 & 0.784 & 0 & 0.216 & George Casella & 430 & 0.303 & 0.425 & 0.271 & Yingcun Xia & 358 & 0.302 & 0 & 0.698 \\
Jiashun Jin & 441 & 0.780 & 0 & 0.220 & Marina Vannucci & 571 & 0.304 & 0.418 & 0.278 & Naisyin Wang & 586 & 0.283 & 0.043 & 0.674\\
Yingying Fan & 410 & 0.741 & 0 & 0.259 & Bernard Silverman & 577 & 0.514 & 0.395 & 0.091 & Hua Liang & 509 & 0.334 & 0 & 0.666 \\
John Storey & 544 & 0.737 & 0 & 0.263 & Catherine Sugar & 501 & 0.450 & 0.360 & 0.190 & Wolfgang Karl Hardle & 456 & 0.343 & 0 & 0.657\\
\hline
\end{tabular}
}
\end{table}
\spacingset{1.45}

\section{Using Mixed-SCORE for the Estimation of $\Omega$} \label{sec:EstimateOmega}

In Remark 9 of Section~\ref{subsec:data-tradeGoods}, we mentioned that Mixed-SCORE can be used to estimate $\Omega$, where we let $\hat{\Omega}=\hat{\Theta}\hat{\Pi}\hat{P}\hat{\Pi}'\hat{\Theta}$ by using $\hat{\Pi}$ from Mixed-SCORE and $(\hat{\Theta},\hat{P})$ in Section~\ref{subsec:Refitting}. Alternatively, we may also estimate $\Omega$ by the standard PCA, where $\hat\Omega=\sum_{k=1}^K\hat\lambda_k\hat\xi_k\hat\xi_k$. The following simulation results suggest that the $\hat{\Omega}$ by Mixed-SCORE is much better than the $\hat{\Omega}$ by standard PCA.

\begin{table}[htb]
\centering
\scalebox{.9}{
\begin{tabular}{l|ccc}\hline
Parameters & $\hat\Omega$=$\sum_{k=1}^K\hat\lambda_k\hat\xi_k\hat\xi_k$  & Mixed-SCORE  \\
\hline
$\theta_i^{-1}\sim\text{Unif}(5,10)$, $\alpha_1$=$(.6, .2, .2)$, $\alpha_2$=$(.3, .4, .3)$ & 78.84 &46.63  \\
\hline
$\theta_i^{-1}\sim\text{Unif}(5,10)$, $\alpha_1$=$(.4, .2, .4)$, $\alpha_2$=$(.2, .6, .2)$ & 78.78  &44.43  \\
\hline
$\theta_i^{-1}\sim\text{Unif}(5,10)$, $\alpha_1$=$(.4, .2, .4)$, $\alpha_2$=$(.1, .8, .1)$ & 80.65  &44.84  \\
\hline
$\theta_i\sim\text{Unif}(0.05, 0.2)$, $\alpha_1$=$(.4, .2, .4)$, $\alpha_2$=$(.2, .6, .2)$ & 71.83  &44.31  \\ 
\hline
$\theta_i\sim\text{Unif}(0.05,0.2)$, $\alpha_1$=$(.6, .2, .2)$, $\alpha_2$=$(.3, .4, .3)$ & 71.73  &38.86  \\
\hline
\end{tabular}}
\caption{Comparison of the Frobenius errors of estimating $\Omega$ based on 100 repetitions. Settings: $K=3$,  $n=540$; There are $n/6$ pure nodes for each community, and the $\pi_i$'s of the remaining nodes are i.i.d. drawn from a mixture distribution $0.5\, \mathrm{Dirichlet}(\alpha_1)+0.5\, \mathrm{Dirichlet}(\alpha_2)$. The diagonals of $P$ are $1$ and off-diagonals are $0.3$.}\label{tb:CompareHatOmega}
\end{table}

\end{document}